\documentclass[12pt,letterpaper]{article}
\usepackage{tikz}
\usepackage{graphicx} 

 \usepackage[skins,theorems]{tcolorbox}
\tcbset{highlight math style={enhanced,
  colframe=red,colback=white,arc=0pt,boxrule=1pt}}
  \usepackage[bookmarksopen, bookmarksnumbered, bookmarksopenlevel=2]{hyperref}
  \usepackage{tikz}
  \usepackage{tikz-3dplot}
 \usetikzlibrary{calc}
 \usetikzlibrary{decorations} %
 \usepackage[UKenglish]{babel}
 \usepackage[toc,page]{appendix}
 \usepackage{amsmath}
 \usepackage{amssymb}
 \usepackage{enumerate}
 \usepackage{graphicx,color}
 \usepackage{makecell}
 \usepackage{hhline}
\usepackage[width=1.1\textwidth,font=footnotesize,labelfont=bf]{caption}
\usepackage{cite}
\usepackage[vcentermath]{youngtab}
\usepackage{geometry}
\usepackage{slashed}
\usepackage{color}
\usepackage{stackrel}
\usepackage{tikz-cd} 
\usepackage{epsfig}
\usepackage{float}
\usepackage{subcaption}
\usepackage{mathtools}
\usepackage{cancel} 
\usepackage{multirow}
\usepackage{tabularx}
\usepackage{mathtools}

\newcolumntype{D}{>{\centering\arraybackslash}X}
\newcolumntype{L}{>{$}l<{$}}
\newcolumntype{R}{>{$}r<{$}}
\newcolumntype{C}{>{$}c<{$}}
\usepackage{IEEEtrantools}
\usepackage{makecell}

\usepackage{empheq}
\usepackage{arydshln}

\newcommand{\bqa}{\begin{eqnarray}}
\newcommand{\eqa}{\end{eqnarray}}

\newenvironment{eqn}{\begin{equation}\begin{aligned}}{\end{aligned}\end{equation}\noindent}
\newenvironment{eqn*}{\begin{equation*}\begin{aligned}}{\end{aligned}\end{equation*}\noindent}
\hypersetup{
    pdftitle={},
    pdfauthor={},
    pdfsubject={}
}
\numberwithin{equation}{section}
\numberwithin{table}{section}

\makeatletter

\newcommand{\be}{\begin{equation}}
\newcommand{\ee}{\end{equation}}
\newcommand{\beq}{\begin{equation}}
\newcommand{\eeq}{\end{equation}}
\newcommand{\ba}{\begin{aligned}}
\newcommand{\ea}{\end{aligned}}

\newcommand{\bea}{\begin{eqnarray}}
\newcommand{\eea}{\end{eqnarray}}

\newcommand\bi{\begin{itemize}}
\newcommand\ei{\end{itemize}}

\renewcommand{\b}{{\beta}}

\def\unit{{1\kern-.65ex {\rm l}}}
\def\1{{1\kern-.65ex {\rm l}}}

\def\bbR{{\mathbb{R}}}

\newcount\hour \newcount\minute
\hour=\time \divide \hour by 60
\minute=\time
\def\now{%
\ifnum \hour<13
  \ifnum \hour=0 \advance \hour by 12 \number\hour:\else \number\hour:\fi%
     \ifnum \minute<10 0\fi%
     \number\minute%
\ A.M.%
\else \advance \hour by -12 \number\hour:%
  \ifnum \minute<10 0\fi%
  \number\minute%
  \ P.M.%
\fi%
}

\makeatother

\begin{document}

\vspace*{.6cm}
\begin{center}
{\Large
Axion-Scalar Systems and Dynamical Distances}

\vspace{.6cm}
\end{center}

\vspace{0.35cm}
\begin{center}
 Thomas W.~Grimm$^{1,2}$, Damian van de Heisteeg$^3$, Filippo Revello$^{1,4}$
\end{center}

\vspace{1cm}
\begin{center} 
\vspace{0.5cm} 
\emph{$^{1}$
Institute for Theoretical Physics, Utrecht University,\\ 
Princetonplein 5, 3584 CC Utrecht, 
The Netherlands}\\[.2cm]

\emph{$^2$\, Center of Mathematical Sciences and Applications,\\
Harvard University, Cambridge, MA 02138, USA}\\[.2cm]

\emph{$^{3}$Jefferson Physical Laboratory,\\
Harvard University, Cambridge, MA 02138, USA
}\\[.2cm]

\emph{$^{4}$Instituut voor Theoretische Fysica \& Leuven Gravity Institute, KU Leuven,\\ Celestijnenlaan 200D, B-3001 Leuven, Belgium}\\[.2cm]

\vspace{0.3cm}
\end{center}

\vspace{01cm}


\begin{abstract}
\noindent
We study the cosmology of axion-scalar pairs, coupled by a hyperbolic field space metric and with a string-motivated rational scalar potential. Borrowing tools from the theory of dynamical systems, we are able to classify all late-time trajectories and extract physical properties of the asymptotic solutions.
These results suggest a Dynamical Distance Conjecture: along the physical (possibly non-geodesic) trajectories, towers of states become exponentially light as a function of the traversed field-space distance. We further rule out possible counterexamples with wildly oscillating solutions. The considered axion-scalar systems are realized in F-theory compactifications, where the axion–scalar pair is a complex-structure modulus and four-form fluxes induce the asymptotic potentials. We also provide a complete Hodge-theoretic classification of all one-modulus asymptotic potentials of this type.
\end{abstract}

\vfill

\newpage

\tableofcontents

\setcounter{page}{1}

\newpage
\section{Introduction}

String cosmology \cite{Cicoli:2023opf} has served as a key source of inspiration for many ideas and conjectures concerning the constraints that Quantum Gravity may place on consistent effective field theories. These insights are part of the Swampland program (see \cite{Palti:2019pca,vanBeest:2021lhn} for reviews). A notable example is the study of trans-Planckian field displacements $(\Delta \Phi \gg M_P)$ in lower-dimensional effective field theories (EFTs) arising from String Theory. On the one hand, such large field ranges are a hallmark of large-field inflationary models, which predict an observable scalar-to-tensor ratio in the spectrum of primordial gravitational waves \cite{Baumann:2014nda}. On the other hand, trans-Planckian displacements are considered problematic in string-theoretic EFTs, as they typically lead to the emergence of light towers of states that invalidate the EFT description \cite{Ooguri:2006in}. This is a well-known instance of phenomenologically relevant models that are apparently consistent from an IR perspective, but are potentially 
in tension with UV principles. More generally, this approach has lead to the development of various conjectures, the Swampland conjectures, which offer hope of connecting String Theory and general principles of Quantum Gravity to observational data. 

A fundamental obstacle in making  connections between the Swampland conjectures and cosmological observations is the fact that the former are most often tested or  formulated in regimes which are not, strictly speaking, those relevant for cosmology. Indeed, most of them either concern and constrain vacua of the theory or adiabatic transitions between them, while cosmology has to do with \emph{dynamical}, \emph{time-dependent} backgrounds
which look and behave very different from the vacuum state(s) of a theory.\footnote{For completeness, let us also mention a few notable exceptions: the Transplanckian Censorship Conjecture (TCC) \cite{Bedroya:2019snp,Bedroya:2019tba,Payeur:2024kyy} and EFT constraints on variations of the species scale \cite{Hebecker:2018vxz,Scalisi:2018eaz, vandeHeisteeg:2023uxj, Bedroya:2025ris}, the Festina Lente (FL) bound \cite{Montero:2019ekk,Montero:2021otb}, and the application of entropy bounds to constrain scale separation for dynamical backgrounds \cite{Andriot:2025cyi}.}
To close this gap 
it is highly desirable to extend the validity of the conjectures to cosmologically relevant backgrounds, taking dynamical aspects into account.
In particular, a dynamical formulation is absent for the celebrated Distance Conjecture \cite{Ooguri:2006in,Ooguri:2018wrx}, one of the pillars of program. The distance conjecture asserts that infinite distance limits in moduli space are always accompanied by towers of light states, signaling the breakdown of the EFT description. At the quantitative level, such towers are expected to become exponentially light in the geodesic distance as one moves from a point $P$ to a point $Q$ that is located close to an infinite distance boundary of the field space. The tower masses are then conjectured to behave as
\begin{equation}\label{eq:odc}
    m_t \sim e^{- \alpha_d d(P,Q)}\, ,
\end{equation}
where $\alpha_d$ is an $\mathcal{O}(1)$ coefficient, and the distance is measured in units of $M_P$.  In its original formulation, the distance conjecture applies to exact moduli spaces, where the scalar potential vanishes everywhere and the moduli are genuine flat directions. This version is also the one that is most supported from top-down examples in String Theory, as the higher degree of supersymmetry allows for more computational control \cite{Klaewer:2016kiy,Baume:2016psm,Grimm:2018ohb,Blumenhagen:2018nts,Grimm:2018cpv,Corvilain:2018lgw,Marchesano:2019ifh,Grimm:2019wtx,Font:2019cxq,Lee:2019xtm,Lee:2019wij,Gendler:2020dfp} (see also \cite{Palti:2019pca,vanBeest:2021lhn}). 

In physical situations of interest, the moduli have to be stabilised with a non-zero scalar potential. As part of the Refined Distance Conjecture \cite{Klaewer:2016kiy,Baume:2016psm}, it was postulated that the same behaviour should hold in the presence of a potential, at least in regions where the latter is not too steep (for example along ``valleys") \cite{Calderon-Infante:2020dhm}. Such an interpretation also raises the question of how the usual notion of moduli space distance should be generalised with a potential, and various proposals have appeared in the literature \cite{Landete:2018kqf,Schimmrigk:2018gch, Lust:2019zwm, Kehagias:2019akr,Shiu:2022oti, Li:2023gtt, Basile:2023rvm, Shiu:2023bay, Palti:2024voy, Mohseni:2024njl,Debusschere:2024rmi,Demulder:2024glx,  Palti:2025ydz}. As mentioned above, the prototypical application of the (refined) distance conjecture is to rule out models of inflation characterised by super-Planckian field displacements, which are appealing from a phenomenological perspective. Notice that in the context of String Theory, the most developed examples of large field inflation typically involve so called axion monodromy models \cite{Silverstein:2008sg,McAllister:2008hb}, where axion field ranges can be large due to winding many times around their no longer compact domain. In this work, we will precisely look to constrain the presence of similar dynamics along trajectories approaching the boundary of moduli space, and taking the axions into account.\footnote{See \cite{Lanza:2024uis} for a recent analysis of similar phenomena moving away from the boundary of moduli space.}

A tightly related issue is that the distance conjecture is formulated for adiabatic motion in moduli space, and in terms of the distance measured along a geodesic. Such a picture does not take into account dynamical aspects such as time-dependent backgrounds and/or field gradients, which will necessarily arise in presence of a scalar potential. Moreover, while the actual dynamical trajectories coincide with geodesics in the absence of a potential, it is not necessarily true when the latter is taken into account. In principle, this allows trans-planckian scalar field excursions without the light tower(s) predicted by the distance conjecture, if the distance of the geodesic connecting the two points is short enough.\footnote{In \cite{Buratti:2018xjt}, it was claimed that infinitely extended paths might arise from an infinite distance limit involving winding axions in a Klebanov-Strassler throat. However, the example relies on a consistent truncation rather than a \emph{bona fide} EFT, and may be outside the scope of the conjecture. } A concrete goal of this work is to understand where similar phenomena may occur on trajectories flowing towards the boundary of moduli space. In particular, one might hope this could help to shed light on possible dynamical generalisations of the distance conjecture.  

This paper is structured as follows. Section \ref{sec:as} gives a general introduction to the low-energy dynamics of saxions and axions, and their relation to the distance conjecture in a dynamical setting. Section \ref{sec:Cosm_solutions} contains an in-depth classification of all the possible late-time, cosmological solutions for an axion-saxion pair with a hyperbolic field-space metric and an arbitrary polynomial potential. While the analysis itself is quite intricate, relying on various results from the dynamical systems theory, the results are summarised concisely in Section \ref{sec:phim}, together with a discussion of their physical significance and of relevant examples. Section \ref{F-theory_embedding} provides a detailed analysis of how the models that we have considered can be embedded in String Theory, and how they arise as asymptotic limits in F-theory compactifications. This includes a classification of all possible scalar potentials that can be realised asymptotically for a single, complex-structure modulus.   Finally, we conclude in Section \ref{sec:con}.

\section{Axion-Scalar Systems and a Distance Conjecture}\label{sec:as}

In this section we introduce the four-dimensional effective theory that we will study throughout this work and highlight our main motivation that arises from a dynamical distance conjecture. To begin with, we introduce in section~\ref{sec:gen_action+potential} a system of a scalar and an axion with a non-trivial scalar potential. When demanding that such a model can be embedded into a UV complete theory of quantum gravity, we expect various constraints to restrict the effective theory. In Section \ref{sc:ddc}, we introduce one such condition, which states that dynamical distances cannot parametrically exceed the geodesic distances. We explain how it extends the claims of the original distance conjecture \cite{Ooguri:2006in} to non-geodesic paths determined by the equations of motions. In Section \ref{ssc:ds}, we introduce the dynamical system reformulation of the equations of motion that will be used in the rest of the paper.

\subsection{Axion-Scalar system} \label{sec:gen_action+potential}

Throughout this work we will focus our attention to studying the dynamics 
of a pair of scalars $(s,a)$ that are coupled to Einstein gravity via the 
$d$-dimensional effective action 
\begin{equation}\label{eq:action}
S= M_{P,d}^2\int d^d x \, \sqrt{- g}  \Big\{ \tfrac{1}{2}\mathcal{R} - \tfrac{1}{2} G_{ss} \partial_\mu s \partial^\mu s  - \tfrac{1}{2} G_{aa} \partial_\mu a \partial^\mu a - V\Big\}\ ,
\end{equation}
where $V$ is a scalar potential. 
A key assertion is that the scalar $a$ describes an axion-like particle. In practice, this means that when considering $V=0$, this field has a classical shift symmetry $a \rightarrow a + c$, for some constant $c$. This implies that $G_{aa}$ and $G_{ss}$ are independent of $a$. In this generality the couplings of the scalars $s,a$ are not related. To proceed we will make pick a well-motivated toy model constraining the form of $G_{aa}(s),G_{ss}(s)$, and $V(a,s)$. Let us stress that our choice of model is expected to be severely constrained if one insists on a concrete UV embedding. This holds true both for the geometry of the axion-scalar field space and the choice of scalar potential. We will come back to these constraints at various points below.

\paragraph{Field-space metric.} In this work, we will mostly constrain ourselves to study scalars $a,s \in \bbR$ parametrizing a two-dimensional moduli space in the region $s \gg 1$. In this asymptotic region we approximate the moduli space metric to be the hyperbolic metric of the upper-half plane, i.e.~we consider
\beq \label{hyper_metric}
 G_{aa}=G_{ss} = \frac{C}{s^2} \ ,
\eeq
where $C>0$ is a constant.
This expression is generally corrected by sub-leading contributions but gives a valid leading term unless 
one considers the case $C=0$, which we will exclude from the following analysis.\footnote{Note that cases with $C=0$ can arise within the string motivated examples. Such situations need to be treated separately but this will be beyond the scope of this work.} 

To control the sub-leading corrections, one needs to study the embedding of this model in an underlying more fundamental theory. If this theory has $\mathcal{N}=1$ supersymmetry, we can exploit the fact that the metric is actually K\"ahler and corrections will correct the K\"ahler potential. Introducing $\Phi= s + i a$ one checks that the metric \eqref{hyper_metric} is derived from the K\"ahler potential $K= - 2C \, \text{log}(\Phi + \bar \Phi)$, when taking $G_{I\bar J}= 2 \partial_{\Phi} \partial_{\bar \Phi} K$. The corrections naturally take the 
form 
\beq \label{complexK}
   K= - \, \text{log}\big[(\Phi + \bar \Phi)^{2C} + f(\Phi,\bar \Phi)\big]\ . 
\eeq
When embedding such settings into string theory in Section \ref{F-theory_embedding}, we will be able to say more about the expected form of $f(\Phi,\bar \Phi)$. The following discussion of the equations of motion assumes that $f(\Phi,\bar \Phi) \approx 0$. 

\paragraph{Scalar potential.} In addition to the kinetic terms, we also need to specify a class of scalar potentials. Here again, we restrict our attention to models that are well-motivated from a UV point of view. Concretely, we choose to consider 
\begin{equation}\label{eq:fullp_e}
  V(s,a) = \frac{1}{s^{\lambda}} \sum_{n=0}^N \frac{1}{s^n} P_n \left( \frac{a}{s}\right)\ ,
\end{equation}
where the $P_n$ are arbitrary polynomials. From a bottom-up perspective, this expression might look rather convoluted. Up to an overall factor or $s^{-\lambda}$ it merely states that we are considering a general polynomial potential that can be formed using $a$ and $1/s$.
The chosen expression \eqref{eq:fullp_e} turns out to fit well with the embedding of this model into F-theory, as we will discuss in Section~\ref{F-theory_embedding}.
In our explicit study of classical trajectories in this potential, we will need to impose further restrictions on the $P_n$ that are motivated by the UV embedding. Generally, we highlight that we demand 
\begin{equation}\label{eq:resV}
    V(a,s) \geq 0\ ,  \qquad V(a,s)\ \xrightarrow{\  s\rightarrow \infty\ }\ 0 \ , 
\end{equation}
in the considered region of the field space.

\paragraph{Equations of motion.} We next determine the equations 
of motion, restricting ourselves to solutions described by an FLRW metric with Hubble rate $H(t)$.
Let us denote the scalars collectively as $\phi^i = (s,a)$. 
The first Friedmann equation is then given by
\begin{equation}\label{eq:f1}
\frac{(d-1)(d-2)}{2}H^2 =\frac{1}{2}G_{i j}\partial_{\mu} \phi^i \partial^{\mu} \phi^j+V(\phi^i)\ .
\end{equation}
The equations of motion for (homogeneous) scalar field profiles are
\begin{equation}\label{eq:eom}
\ddot{\phi}^i+ \Gamma^{i}_{j\,k} \dot{\phi}^j \dot{\phi}^k+(d-1)H\dot{\phi}^i+\partial^i V =0,
\end{equation}
where all the contractions and the Christoffel symbols are defined with respect to the real metric $G_{ij}$. 
With the hyperbolic metric \eqref{hyper_metric}, the non-vanishing Christoffel symbols are $\Gamma^s_{a\,a}=-\Gamma^a_{s\,a}= -\Gamma^s_{s\,s} = \frac{1}{s}$.
Then, the equations of motion can be recast as
\begin{equation}\label{eq:am}
\ddot{a}-\frac{2 \dot{a} \dot{s}}{s} + (d-1) H \dot{a}+ \frac{s^2}{C}\partial_a V=0
\end{equation}
\begin{equation}\label{eq:sm}
\ddot{s}-\frac{\dot{s}^2}{s}+\frac{\dot{a}^2}{s} + (d-1) H \dot{s}+  \frac{s^2}{C} \partial_s V=0
\end{equation}
\begin{equation}\label{eq:H}
(d-1)(d-2)H^2=  C \frac{\dot{s}^2+\dot{a}^2}{s^2} +2V.
\end{equation}
Using the constraint \eqref{eq:H} to eliminate one variable, the above equations turn into a non-linear, second-order system of two differential equations. In the following, we will find it convenient to reformulate it into a system of first-order equations, so that techniques from the theory of dynamical systems can be used to analyse the asymptotic structure of its solutions (see Section \ref{sec:Cosm_solutions} for details).

\subsection{Non-geodesic solutions and the distance conjecture}\label{sc:ddc}
A cornerstone of the swampland program is the distance conjecture, which postulates the appearance of (exponentially) light towers of states when approaching a boundary point of the moduli space which is at infinite geodesic distance. As such, the conjecture is formulated for adiabatic field displacements, and makes statements only about trajectories that are geodesics. However, realistic cosmological applications involve time-dependent field configurations and non-geodesic trajectories sourced by the presence of scalar potentials. 

Extending the distance conjecture to such settings, it is natural to ask how the distance between two points in field space should be measured. A logical first choice is to just consider the length of the dynamical trajectory traced out by the fields, considered for example in \cite{Landete:2018kqf}. Denoting the dynamical trajectory between two points $P,Q$ by $\gamma$, the length is computed as
\begin{equation}\label{eq:ddc}
    \Delta_{\gamma}(P,Q) = \int_{\gamma} {\rm{d}} \tau \sqrt{G_{ij} \,  \dot{\phi}^i \dot{\phi}^{j} }.
\end{equation}
One of our main goals is to investigate whether this dynamical length $\Delta_{\gamma}(P,Q)$ can become parametrically larger than the original geodesic length $\Delta_{\rm geod}(P,Q)$ that one would consider without a scalar potential. For instance, in hyperbolic moduli spaces we know that $\Delta_{\rm geod}(P,Q)$ scales logarithmically with the saxionic field $s$, but in the presence of a scalar potential the dynamical distance $\Delta_{\gamma}(P,Q)$ could potentially scale at an even faster rate. Concordance with the distance conjecture would then require the presence of a new tower of states that becomes light exponentially in $\Delta_{\gamma}(P,Q)$ instead of $\Delta_{\rm geod}(P,Q)$, which would therefore be parametrically lighter than the original tower. One of the main take-aways for the potentials considered in this work is that such additional towers are not required, since $\Delta_{\gamma}(P,Q)$ never becomes parametrically larger than $\Delta_{\rm geod}(P,Q)$ for any of the dynamical trajectories. In other words, this suggests that one can formulate a \textit{Dynamical Distance Conjecture} in which \eqref{eq:odc} for the masses is replaced by
\begin{equation}
    m_t \sim e^{- \alpha_d \Delta_{\gamma}(P,Q)}\, ,
\end{equation}
when approaching an infinite distance boundary.

While we only took the kinetic terms for the scalars into account in defining the distance \eqref{eq:ddc} above, in general the scalar potential $V(\phi)$ should also be incorporated in the distance measure. This is especially important in the context of the exponential fall-off of the tower mass, since the normalization of the distance measure is directly related to the coefficient in the exponent. There have been proposed various notions of generalized distances including these effects in the literature, see for instance~\cite{Schimmrigk:2018gch, Lust:2019zwm, Kehagias:2019akr,Calderon-Infante:2020dhm,Shiu:2022oti, Li:2023gtt, Basile:2023rvm, Shiu:2023bay, Palti:2024voy, Mohseni:2024njl,Debusschere:2024rmi,Demulder:2024glx,  Palti:2025ydz}. As an example, consider for instance the distance measure proposed in \cite{Debusschere:2024rmi}
\begin{eqn}
    \Delta_{V}(P,Q) = \int_{\gamma} {\rm{d}} \tau \sqrt{G_{ij} \,  \dot{\phi}^i \dot{\phi}^j + 2 V(\phi)},
\end{eqn}still computed along the actual dynamical trajectory. For our purposes, the inclusion of $V(\phi)$ in the distance measure does not make a particular difference, since the kinetic and potential energies scale similarly asymptotically, so its only effect would be to modify the $\mathcal{O}(1)$ factor in the decay rate of the tower. Since our focus lies on ruling out parametric deviations from the original geodesic distances, we restrict our attention to \eqref{eq:ddc} for the dynamical distance.

Let us now specialize the discussion again to the main focus of this work, axion-scalar systems. We first recall the situation without a scalar potential. In this case the metric is hyperbolic \eqref{hyper_metric}, which means that the usual distance conjecture predicts the presence of a tower whose mass scales as
\begin{eqn}\label{eq:mtw}
    m_t \sim s^{-\alpha} \quad \quad \text{with}  \quad  \alpha \sim \mathcal{O}(1).
\end{eqn}We also know that geodesic trajectories for which $s \to \infty$ are given by straight lines along the imaginary axis of the upper-half plane. 

\begin{figure}[h!]
\centering
 \includegraphics[width=0.4\textwidth]{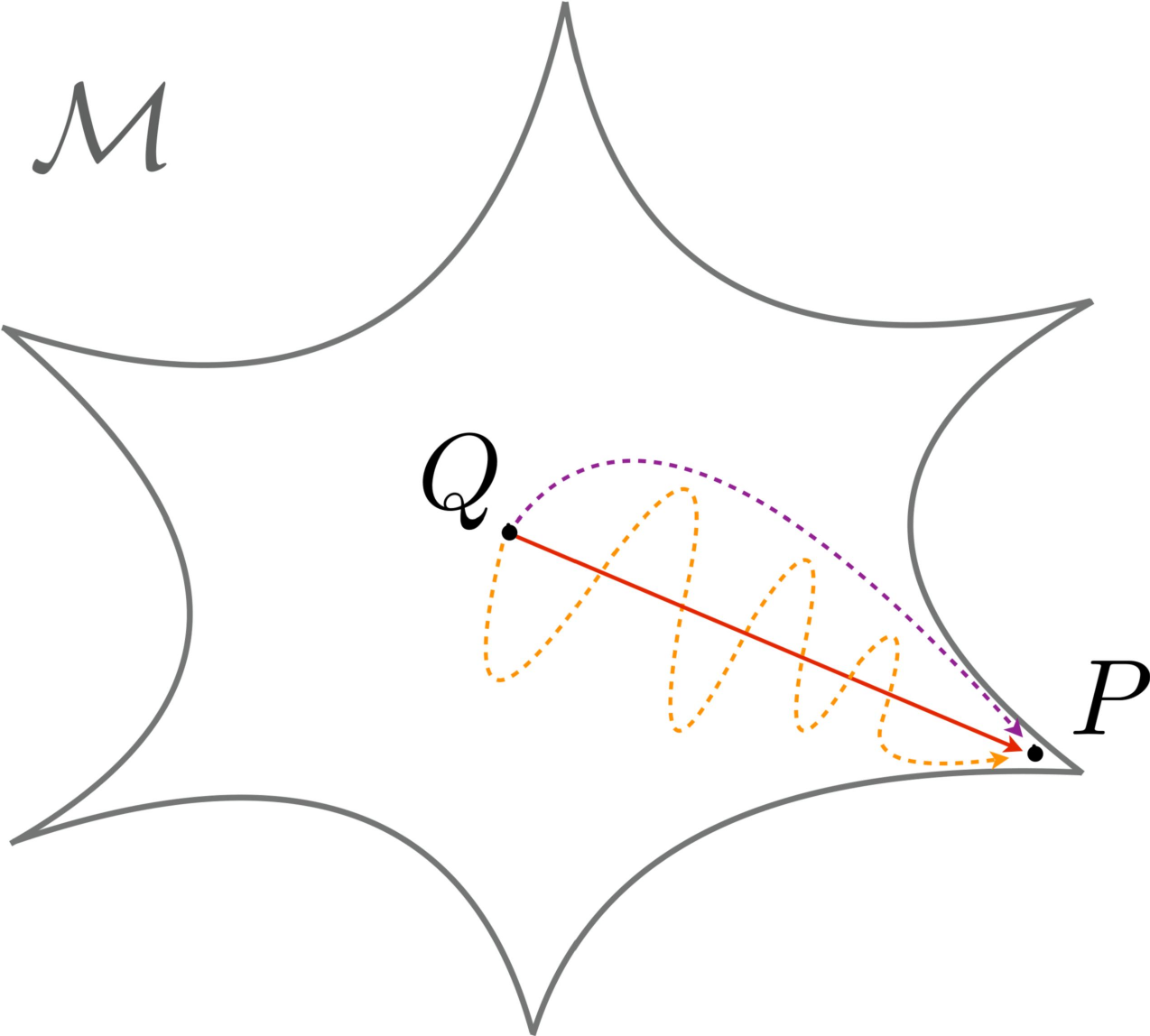}
\caption{Schematic depiction of different dynamical trajectories connecting a point $Q$ in the bulk of moduli space $\mathcal{M}$ to a point $P$ on the boundary. The red, solid line is a geodesic between the two points, while the dashed orange and purple lines represent oscillating and non-oscillating trajectories respectively (as described in the text).}
\label{fig:traj}
\end{figure}

Let us compare this situation to the case with a scalar potential. The question we want to investigate is the extent to which the dynamics can alter the trajectories relative to the original straight-line geodesics. We assume that the same tower of states is present with the same mass scale \eqref{eq:mtw} as before. From the perspective of the distance conjecture, we find that as long as the dynamical length $\Delta_{\gamma}(P,Q)$ scales linearly with the original geodesic length $\Delta_{\rm geod}(P,Q)$, then the same tower suffices to satisfy the distance conjecture; for faster rates, a new tower of states would need to be identified. Since we do not expect the presence of a scalar potential to introduce such a tower, we interpret this as a bound on what the trajectories can look like. To make this more precise, let us first write the distance of a dynamical trajectory as
\begin{eqn}\label{eq:d1}
    \Delta_{\gamma}(P,Q) = \sqrt{C} \int_{s(P)}^{s(Q)}  \frac{{\rm{d}} s}{s} \sqrt{1+\left( \frac{{\rm{d}}a}{{\rm{d}}s}\right)^2 },
\end{eqn}where we implicitly assume the parametrization $a\equiv a(s)$ along the trajectory $\gamma$ and the saxion $s(t)$ to be monotonic along the trajectory. Before considering dynamics, we can first ask what the valleys of the potential look like. In this context, the study of axion backreaction on saxion trajectories in moduli space \cite{Baume:2016psm,Klaewer:2016kiy,Buratti:2018xjt,Calderon-Infante:2020dhm, Grimm:2020ouv} already suggests that this does not lead to parametrically larger distances. For the scalar potentials  considered in this work, this follows from their polynomial dependence on $a/s$ given in \eqref{eq:fullp_e}, since this implies that $a$ scales linearly with $s$, with the slope set by the zero of the polynomial. One of the main goals of this work is to generalise this picture to a fully dynamical setting, including for example cosmological scaling solutions.\footnote{See \cite{Debusschere:2024rmi} for a discussion on this last point.} Considering such dynamical trajectories, the aim is to prove that the axion-to-saxion velocity ratio is always bounded from above as
\begin{eqn}\label{eq:bnd}
    \left| \frac{{\rm{d}}a}{{\rm{d}}s} \right| \leq K \quad \quad \text{as} \quad t \rightarrow + \infty,
\end{eqn}so that $\Delta_\gamma$ grows at most logarithmically with $s$, and hence scales linearly with $\Delta_{\rm geod}$.  If $s(t),a(t)$ are monotonic functions, the maximal allowed deviation is precisely the case where the saxion and the axion have the same scaling, $s \sim a$ as $t \rightarrow \infty$. Dynamical solutions realising this scaling can arise even in the absence of a potential for the axion, saturating the bound with a constant \cite{Sonner:2006yn,Cicoli:2020cfj,Cicoli:2020noz,Brinkmann:2022oxy,Russo:2022pgo,Revello:2023hro,Shiu:2024sbe}. The fact that the axion can never grow faster than the saxion is a consequence of the specific form taken by the kinetic term, as well as the polynomial dependence of the scalar potential on $a/s$. In general, however, highly oscillatory trajectories can in principle violate the bound even if the axion is subdominant with respect to the saxion. A schematic depiction of these different kinds of trajectories is shown in Figure \ref{fig:traj}. To show that \eqref{eq:bnd} is valid, it will not be enough to use the generic scaling properties of the kinetic and potential terms mentioned above, but it will be necessary to refer to the detailed classification of asymptotic potentials. This analysis is the subject of the next section \ref{sec:Cosm_solutions}.

\subsection{Dynamical system approach}\label{ssc:ds}

In this subsection, we introduce a reformulation of the equation of motions \eqref{eq:am}-\eqref{eq:H} as an autonomous system (following the seminal works \cite{Copeland:1997et,Ferreira:1997hj}), which will be used throughout the rest of the paper. In particular, this will allow us to use various techniques from the theory of dynamical systems to study asymptotic solutions. To give a flavour of the mathematical problems that we will be addressing in Section \ref{sec:Cosm_solutions}, we can already show how the reformulation works in a simple case. For ease of presentation, we specialise to the case where the sum \eqref{eq:fullp} contains a single term and $P(a,s) = s^{-\lambda} P(a/s) \equiv \tilde{V}(s) P(a/s)$.\footnote{Recall the general form of the potential \ref{eq:fullp_e}. This case will be analysed in detail in section \ref{sec:Cosm_solutions}.}

\noindent In order to anticipate the conversion of Eqs \eqref{eq:am}-\eqref{eq:H} into a dynamical system, we define the following variables
\begin{equation}\label{eq:newvar}
    x =  \frac{\dot{s}}{ \alpha H s} \quad \quad y = \frac{\dot{a}}{\alpha H s} \quad \quad z = \frac{1}{H}\sqrt{\frac{2\tilde{V}(s)}{(d-1)(d-2)}} \quad \quad w=\frac{a}{s},
\end{equation}
where \begin{eqn}
    \alpha = \sqrt{\frac{(d-1)(d-2)}{C}}.
\end{eqn}From the Hubble constraint, they are not all independent, and satisfy the relation
\begin{eqn}\label{eq:csH}
    x^2+y^2+z^2  P(w) =1.
\end{eqn}Conveniently, the $\varepsilon$ parameter relevant to accelerated expansion can be expressed as
\begin{equation}
    \varepsilon \equiv -\frac{\dot{H}}{H^2}= (d-1)(x^2+y^2).
\end{equation} 
With these definitions, \eqref{eq:am}-\eqref{eq:H} can be converted to
\begin{equation}
\left \{
\begin{aligned}
\frac{d x}{dN}&=- \alpha\, y^2 -\left(1- x^2-y^2\right) \left[(d-1)x- \frac{\alpha}{2} \left( \lambda + \frac{w \, \partial_w P(w)}{P(w)}\right) \right]\\
\frac{d y}{dN}&=  \alpha\,  x y-\left(1- x^2-y^2\right) \left[(d-1)y+\frac{\alpha}{2} \frac{\partial_w P(w)}{P(w)}\right]\\
\frac{d w}{dN}&= \alpha (y- wx),\\
\end{aligned} \right. 
\end{equation}
with $N$ the logarithm of the FLRW scale factor $a_s(t)$, $N= \log a_s(t)$. This is a highly non-linear, autonomous system of three variables, and our goal will be to classify all of the late-time solutions. Importantly, we will not limit ourselves to a local (perturbative) stability check of the fixed points, but rather perform a global analysis of all asymptotic trajectories.

\noindent The results will then be used to discuss the connection to the distance conjecture, as well as phenomenological implications. Notice how, in this language, it becomes easy to address the kind of questions discussed in the previous section \ref{sc:ddc}. To ensure that the dynamical distance \eqref{eq:ddc} is not parametrically larger than the geodesic distance, for example, it is sufficient (but not necessary) to verify the condition given by Eq.\eqref{eq:bnd}. The latter amounts to showing that $|y/x|$ is bounded as $t \rightarrow \infty$. Viceversa, if $|y/x| \rightarrow + \infty$, the conjecture is falsified.\footnote{If $|y/x|$ has unbounded oscillations, both options are possible in principle.}

\section{Cosmological solutions} \label{sec:Cosm_solutions}

In this section we study the cosmological solutions to the axion-scalar theories introduced in section \ref{sec:gen_action+potential}.
We are motivated by two main goals. The first one, purely theoretical, is to establish whether the dynamical form of the distance conjecture, introduced in \eqref{eq:ddc}, holds for all infinite distance, one-modulus limits. The second one is to understand more in detail the different kinds of cosmological phases that can arise in this setting, with more of an eye to phenomenological applications. Indeed, recent work \cite{Conlon:2022pnx,Apers:2022cyl,Revello:2023hro,Apers:2024ffe,Andriot:2024jsh,Conlon:2024uob,Apers:2024dtn,Andriot:2024sif,Revello:2024gwa,Brunelli:2025ems,Andriot:2025cyi,Ghoshal:2025tlk,SanchezGonzalez:2025uco,Mosny:2025cyd} has highlighted the cosmological consequences of string-theory motivated, exotic epochs such as kination, dominated by the kinetic energy of a scalar field. It would be interesting to understand whether such epochs are a common feature of string cosmologies, and if any other possibilities can occur. 
Moreover, there has been renewed interest in the idea that the observed accelerated expansion of the universe might be driven by rolling moduli in the asymptotic regions of moduli space 
\cite{Cicoli:2018kdo,Cicoli:2020cfj,Cicoli:2020noz,Cicoli:2021fsd,Brinkmann:2022oxy,Calderon-Infante:2022nxb,Shiu:2023nph,Shiu:2023fhb,Andriot:2023wvg,Shiu:2024sbe,Andriot:2024jsh,Andriot:2024sif,Rahimy:2025iyj,Licciardello:2025fhx}. However, we will not address this possibility in detail in the present work.

As anticipated in the previous section, we will consider a potential of the form
\begin{eqn}\label{eq:fullp}
  V(s,a) = \frac{1}{s^{\lambda}} \sum_{n=0}^N \frac{1}{s^n} P_n \left( \frac{a}{s}\right) \equiv \tilde{V}(s) \sum_{n=0}^{N} \frac{1}{s^n} P_n \left( \frac{a}{s}\right),
\end{eqn}where $\lambda >0$ and the $P_n$ are polynomials specific to the case under consideration. In the string theory realization of this potential, each term in the sum comes from a single ``sector"---turning on a particular combination of flux quanta---within the asymptotic approximation used in Section \ref{F-theory_embedding} known as the $\rm{sl}(2)$ approximation. Notice that \eqref{eq:fullp} encompasses a much more general class of potentials than those arising from the geometry of asymptotic limits. In particular, it is the most general polynomial potential in terms of $a$ and $1/s$. Whenever necessary, we will also make additional assumptions on the nature of the polynomials $P_n(w)$'s of the axion-to-saxion ratios $w=a/s$. As already stated in \eqref{eq:resV}, an important physical requirement is for the potential to die off as $s \rightarrow \infty $ (for fixed values of the axion $a$). Moreover, the overall potential is required to be positive definite by construction when it is coming from F-theory, although individual terms in \eqref{eq:fullp} need not be positive (as can be seen in explicit examples). Finally, another general property implied by the classification in Section \ref{F-theory_embedding} is that the degrees of the polynomials $P_n(w)$ are bounded by
\begin{eqn}
    {\rm{deg}} \left( P_n\right) + \lambda + n \leq 4,
\end{eqn}as $1/s^4$ is the most singular term that can appear in the overall potential.\footnote{An exception is when some of the leading order terms are zero, and one can consider higher-order corrections (such as $\alpha'$ corrections) which may contain higher powers of $\frac{1}{s}$.} For simplicity, and to illustrate some of the necessary techniques, in Section \ref{ssc:sise} we will begin by tackling the case of a single sector (term) in the sum \eqref{eq:fullp}, before delving into the general case in Section \ref{ssec:generalpolynomial}.\\

\subsection{Analysis of the single-sector potential}\label{ssc:sise}
Having set up the general parametrization of the dynamical system given by \eqref{eq:am}-\eqref{eq:H}, we now specialize the scalar potential \eqref{eq:fullp} to a simpler form. 
While this case is considerably more tractable than the fully general polynomial potential, it already captures many of the essential qualitative features and  ingredients of the general setting. We refer to Section \ref{ssec:generalpolynomial} for an extension of the analysis 
to the case of a general polynomial potential. 

The scalar potential we simplify to is given by a single sector
\begin{equation}\label{eq:onesec}
    V(s,a) = \frac{1}{s^{\lambda}} P \left( \frac{a}{s}\right)\equiv \tilde{V}(s)  P \left( \frac{a}{s}\right),
\end{equation}and where the polynomial $P(w)$ is positive definite on the real line. This corresponds to the case where only one term is present in \eqref{eq:fullp}. Such a potential can be realised when only the fluxes corresponding to a specific sector are turned on as described in Section \ref{F-theory_embedding}, or it can provide the leading order approximation to \eqref{eq:fullp} in the asymptotic limit $s\rightarrow \infty$ if $P \left( w \right)$ does not approach zero along the trajectory.\footnote{As we will see later, this is not always the case in explicit examples, and one must in that case resort to the general case.} In terms of the new variables $x,y,w$ introduced in \eqref{eq:newvar}, the equations of motion can be rewritten as an autonomous, non-linear dynamical system of the first order with only three equations, thus eliminating one degree of freedom. In particular, the system of differential equations \eqref{eq:am}-\eqref{eq:H} can be recast as 
\begin{equation}\label{eq:sysN}
\left \{
\begin{aligned}
\frac{d x}{dN}&=- \alpha\, y^2 -\left(1- x^2-y^2\right) \left[(d-1)x- \frac{\alpha}{2} \left( \lambda + \frac{w \, \partial_w P(w)}{P(w)}\right) \right]\\
\frac{d y}{dN}&=  \alpha\,  x y-\left(1- x^2-y^2\right) \left[(d-1)y+\frac{\alpha}{2} \frac{\partial_w P(w)}{P(w)}\right]\\
\frac{d w}{dN}&= \alpha (y- wx),\\
\end{aligned} \right. 
\end{equation}
where $N \equiv \log a_s$ (and $a_s(t)$ is the scale factor). Classifying the asymptotic solutions of this differential equation system is the main subject of the next subsections;
the analysis is performed in Section \ref{ssc:sise}, and a summary is also provided in Section \ref{ssec:summary}.

Before proceeding, we perform sanity checks on \eqref{eq:sysN}, introducing along the way quantities useful for classifying asymptotic solutions. We first ask whether \eqref{eq:sysN} is regular, i.e.~whether all first derivatives remain finite. This is nontrivial due to the appearance of 
$P(w)$ in denominators, whose zeroes may signal singularities.
To do that we introduce the  function $z$, defined in \eqref{eq:newvar}, which is redundant due to the constraint \eqref{eq:csH}.
In terms of $x,y,z$ and $w$ the system \eqref{eq:sysN} can indeed be reformulated in a manifestly regular manner
\begin{equation}\label{eq:sysNz}
\left \{
\begin{aligned}
\frac{d x}{dN}&=- \alpha y^2 - (d-1)x z^2 P(w)+ \alpha \frac{z^2}{2} \left[ \lambda P(w) + w \, \partial_w P(w) \right] \\
\frac{d y}{dN}&= \,\,\,\, \alpha  x y- (d-1)y z^2 P(w)-\alpha \frac{z^2}{2} \partial_w P(w)\\
\frac{d z}{dN}&=z \left[(d-1)(x^2+y^2)-\frac{\lambda x}{2} \alpha \right],\\
\frac{d w}{dN}&=\alpha (y- wx)\ .\\
\end{aligned} \right. 
\end{equation}
Note that this system now has a polynomial dependence on $(x,y,z,w)$ on the right-hand side and hence no singularity unless these variables blow up.   

There is another way to see regularity of the system \eqref{eq:sysN} in variables suitable for our later classification. Let us define
\beq
   S \equiv x^2+y^2\ , \qquad  T \equiv x+y w\ , \qquad T_\lambda \equiv \frac{\alpha \lambda}{2}\ .  
\eeq
Using \eqref{eq:sysN} the first derivatives take the form 
\begin{eqnarray}\label{eq:dT}
\frac{d T}{dN} &=& -(1-S) (d-1) \left[ T- T_{\lambda} \right]\ ,\\
\label{eq:dS}
   \frac{d S}{dN}&=& -(1-S) \left[2(d-1)S+\frac{d w}{dN} \frac{\partial_w P(w)}{P(w)} - \alpha \lambda x \right]\ ,\\
\label{eq:dw}
    w'^2 &\equiv&\left(\frac{d w}{dN} \right)^2 = \alpha^2 \left[ S(1+w^2)-T^2 \right]\ ,
\end{eqnarray}
where $x(S,T,\omega)$ is given by  
   $x = T \mp w \sqrt{S(1+w^2)-T^2}/(1+w^2)$ with the sign of the second term chosen to be opposite to the one of $w'$.

This form of the differential system has various advantages. Firstly, we note that   \eqref{eq:dT} suggests that in the asymptotic limit 
\beq 
T\rightarrow T_\lambda \quad \text{for} \quad N\rightarrow \infty\ ,
\eeq 
as we will justify more rigorously in a moment.
Secondly, we can formally integrate \eqref{eq:dS} to find 
\begin{eqn}\label{eq:int2}
   \frac{P(w(N))}{P(w_i)} \left(\frac{1+w(N)^2}{1+w_i^2}\right)^{\frac{\lambda}{2}} =\frac{1-S(N)}{1-S_i} e^{- 2(d-1) [I_1(N)+I_2(N)]},
\end{eqn}where 
\begin{eqn}\label{eq:I12}
    I_1(N) = \frac{1}{\alpha^2} \int_{N_i}^N {\rm{d}}\hat{N} \,  \frac{w'^2}{1+w^2} \quad \quad \text{and} \quad \quad I_2(N) =  \int_{N_i}^N {\rm{d}}\hat{N} \,  \frac{T(T-T_{\lambda})}{1+w^2},
\end{eqn}while $S_i$ and $w_i$ are the values of $S$ and $w$ at the (arbitrarily  large) initial time $N_i$.\footnote{Note that they are not initial conditions that one may choose freely, but rather some unknown asymptotic value determined by the original initial conditions.}
This expression gives another way to see that the system \eqref{eq:sysN} is regular, since the combination $(1-S)/P(w)$ is finite for any value of $N$. In particular, the asymptotic behavior of the integral $I_1(N)$ as $N \to \infty$ will be used to distinguish between different types of asymptotic solutions.
With these ingredients at our disposal, we can now perform the general analysis of the single-sector potential. As our starting point, recall the first-order differential equation \eqref{eq:dT} satisfied by $T$. Due to its form, the variable $T$ is monotonous on either side of $T= T_{\lambda}$, so it can never cross the line $T=T_{\lambda}$. This divides the phase space into two invariant subspaces. To describe these subspaces more precisely, let us consider the Lyapunov function 
\begin{eqn}
    \mathcal{L}= \left((d-1) T- \frac{\alpha \lambda}{2}  \right)^2,
\end{eqn}which satisfies
\begin{eqn}
    \mathcal{\dot{L}}=-(1-S)(d-1)\left[(d-1) T- \frac{ \alpha \lambda}{2}  \right]^2 \leq 0. 
\end{eqn}
Since $ \mathcal{L}$ is also positive definite, the level sets
\begin{eqn}
\Omega_c = \Big\{(x,y,w)  \quad \text{such that }\quad \mathcal{L}(x,y,w) \leq c \Big\}    \, .
\end{eqn}are positively invariant and compact. By La Salle's invariance principle, $T$ will then approach the largest invariant set contained within $\Omega_0$, which is defined as
\begin{equation}
    \Omega_0 = \Big\{ (x,y,w) \quad \text{such that} \quad S=1 \text{ or }T= T_{\lambda}\Big\}\, .
\end{equation}
Let us consider each of the asymptotic cases $S=1$ and $T =T_{\lambda}$ separately.

We begin with the case where $T \rightarrow T_{\lambda}$ and $S$ is not converging to 1. In order to classify all asymptotic solutions in this case, it is helpful to look at the asymptotic behavior of the integrals $I_1(N)$ and $I_2(N)$ defined in \eqref{eq:I12}, since through \eqref{eq:int2} this would lead to constraints on $P(w)$ asymptotically. We immediately see that the integrand of $I_1(N)$ is positive, but for $I_2(N)$ this depends on whether $T$ is above or below $T=T_{\lambda}$. To this end, it is helpful to note based on \eqref{eq:dT} that $T$ can never cross $T=T_{\lambda}$, since this first-order differential equation ensures that $T>T_{\lambda}$ or $T< T_{\lambda}$ will always be preserved by time evolution. In other words, for $T>T_{\lambda}$ the integrand of $I_2(N)$ is always positive, while for $T<T_{\lambda}$ it is negative. In either case, we can show that $I_2(N)$ is uniformly bounded in $N$ and converges as $N \to \infty$. This can be seen from rewriting the integral through \eqref{eq:dT} and bounding it as follows
\begin{equation}\label{eq:boundint}
    \begin{aligned}
        \big| I_2(N) \big| &= \Bigg|-\int_{N_i}^N d\hat{N} \frac{ T T'}{(d-1)(1+w^2)(1-S)}\Bigg|  \\
        &\leq \Bigg| -\frac{T^2}{1-S}\bigg|_{N_i}^N + \int_{N_i}^N d\hat{N} \, T^2 \frac{d}{dN} \left(\frac{1}{1-S}\right) \Bigg| \\
        & \leq \Bigg| \frac{T_{\lambda}^2}{1-S(N)} -\frac{T_i^2}{1-S_i} - T_{\lambda}^2\Bigg| \frac{1}{1-S_i}- \frac{1}{1-S(N)}\Bigg| \Bigg|\, .
    \end{aligned}
\end{equation}
In the second line we dropped factors of $d-1$ and $1+w^2$ that are irrelevant for boundedness of the integral, and used integration by parts. In the third line we evaluated the first term at the boundaries, substituting the asymptotic value $T_{\lambda}$ for $T(N)$, while for the second term we bounded the integrand from the monotonicity of $T(N)$, and performed the remaining integral. Notice that, since $S\nrightarrow 1$, there exists an infinite subsequence $\{N_j\}$ (with $N_j \rightarrow \infty)$ and $S_{\rm{min}} <1$ such that $S(N_j)= S_{\rm{min}}$.
Along that subsequence, the integral is therefore bounded from \eqref{eq:boundint}, since the denominators proportional to $1-S(N)$ cannot diverge. Finally, since $I_2(N)$ is also a monotonous function, we conclude that it is bounded everywhere as $N \rightarrow \infty$, and the integral $I_2(N)$ converges. At this point, we can distinguish two cases, based on the convergence properties of $I_1(N)$. These are detailed in subsections \ref{ssec:oscillating} and \ref{ssec:properfixed}.

\subsubsection{Oscillating solutions}\label{ssec:oscillating}
If $I_1$ diverges as $N \rightarrow \infty$, by \eqref{eq:int2} $P(w) \rightarrow 0$, so that $w$ converges to $w_0$, an absolute minimum of $P(w)$. Notice that this can only happen if the potential does have an absolute minimum, and if said minimum is attained for a value of $w$ satisfying $w_0^2 \geq T_{\lambda}^2-1$. Then, two of the variables ($T$ and $w)$ will asymptotically converge to a constant value, but that is not true for the remaining variable $S$. Indeed, the system \eqref{eq:sysN} admits no fixed point where $w= w_0$, so that $S$ and $w'$ will keep oscillating indefinitely (subject to the constraint \eqref{eq:dw}). This gives rise to what we will refer to as ``oscillating" solutions, where the trajectory approaches a one-dimensional locus in phase space rather than a point. 

In this case, we can also write down an (approximate) asymptotic solution as follows. If $P(w_0)=0$, the potential close to the minimum can be Taylor expanded as $P(w)= f^2 (w-w_0)^{2p} + \mathcal{O}\left( (w-w_0)^{2p+1}\right)$, with $p$ integer and where $f$ is an unimportant constant. With this form of $P(w)$, the equations of motion for $S$ and $T$ can be formally integrated as
\begin{eqn}\label{eq:exexp}
   \frac{(w-w_0)^{2p}}{(w_i-w_0)^{2p}} \left(\frac{1+w^2}{1+w_i^2}\right)^{\frac{\lambda}{2}} =  \frac{1-S}{1-S_i} {\text{Exp}} \left( {-2(d-1)  \int_{N_i}^N {\rm d}\hat{N} S} \right)  
\end{eqn}
and
\begin{eqn}
    T =T_{\lambda}+(T_i-T_{\lambda}) {\text{Exp}} \left( {-(d-1)  \int_{N_i}^N {\rm d}\hat{N} (1-S)} \right),
\end{eqn}where the index $i$ refers to quantities evaluated at some initial time $N=N_i$. From the above, we see how the evolution of the system is governed by the average of $S$, defined as
\begin{eqn}
    \bar{S} \equiv \frac{1}{N-N_i} \int_{N_i}^N {\rm{d}}\hat{N} S(\hat{N}), \quad \quad \quad 0 \leq \bar{S} \leq 1.
\end{eqn}In the case of increasingly fast oscillations, it is natural to expect that $\bar{S}$ will approach a constant: we can make this assumption, and then verify it a posteriori. This leaves us with a single undetermined function, $w(N)$, for which we can take the ansatz 
\begin{eqn}\label{eq:ansatzw}
    w(N)= w_0+ C_3(w_i-w_0) e^{-\gamma (N-N_i)} F \left( C_1 e^{\gamma (N-N_i)} +C_2\right),
    \end{eqn}and where $C_1,C_2,C_3$ are integration constants. Neglecting exponentially suppressed terms, the equation of motion for $w$ \eqref{eq:dw} together with the ansatz \eqref{eq:ansatzw} give the differential equation
\begin{eqn}\label{eq:dF}
    F'^2 = 1-F^{2p},
\end{eqn}if the following conditions are satisfied:\footnote{Since these are asymptotic solutions, only valid as $N\rightarrow \infty$, the initial time $N_i$ also has to be sufficiently large. As $N_i \rightarrow \infty$, $C_3 \rightarrow 1$ from \eqref{eq:condA}, so that \eqref{eq:condC} can always be satisfied for the form $F$ given by \eqref{eq:wF}.}
\begin{subequations}
\begin{eqnarray}
& C_3= \left[ \frac{1}{1-S_i} \frac{(1+w_0^2)-T^2_{\lambda}}{1+w_0^2} \left(\frac{1+w_0^2}{1+w_i^2}\right)^{\frac{\lambda}{2}} \right]^{\frac{1}{2n}} \label{eq:condA} \\
&  C_1 = \frac{\alpha}{\gamma C_3} \frac{\sqrt{(1+w_0^2)-T_{\lambda}^2}}{|w_i-w_0|} \\
& C_3 F(C_1+C_2)=1 \label{eq:condC} \\
&  \gamma= \frac{(d-1) \bar{S}}{p}. 
\end{eqnarray}
\end{subequations}
For $p=1$, the equation \eqref{eq:dF} is solved by simple trigonometric functions. For arbitrary $n$, a general solution is given by the generalized trigonometric functions \cite{EDMUNDS201247}. The latter are defined as
\begin{equation}
\sin _{k, l}^{-1}(x)\equiv \int_0^x \frac{{\rm{d}} t}{\left(1-t^l\right)^{1 / k}}, \quad 0 \leq x \leq 1, \quad \quad \cos _{k, l}(x) \equiv \frac{{\rm{d}}}{{\rm{d}}x} \sin _{k, l}(x),
\end{equation}
and satisfy the relation
\begin{eqn}
    \left( \sin_{k, l}(x) \right)^{l}+\left( \cos_{k, l}(x) \right)^{k}=1.
\end{eqn}Just like the ordinary trigonometric functions, they are periodic with a semi-period $\pi_{k,l}$ satisfying
\begin{equation}
\pi_{k, l}:=2 \int_0^1\left(1-t^l\right)^{-1 / k} \mathrm{~d} t = \frac{2}{l} \frac{\Gamma \left( \frac{k-1}{k}\right) \Gamma \left( \frac{1}{l}\right)}{\Gamma \left( \frac{k-1}{k}+\frac{1}{l}\right)}.
\end{equation}
The asymptotic solution to the equation of motion for $w$ is then given by \eqref{eq:ansatzw}, where
\begin{eqn}\label{eq:wF}
    F(x) = \sin_{2,2p} (x),
\end{eqn}while 
for $S$ it becomes
\begin{eqn}\label{eq:SF}
    S(N)= \left(1-\frac{T_{\lambda}^2}{1+w_0^2}\right) F^2 \left( C_1 e^{\gamma (N-N_i)} +C_2\right)+ \frac{T_{\lambda}^2}{1+w_0^2}.
\end{eqn}This shows \emph{a posteriori} how it is well justified to assume $\bar{S}$ averages to a constant, and in particular
\begin{eqn}
    \bar{S}= \frac{1}{p+1} \left(p+\frac{T_{\lambda}^2}{1+w_0^2} \right),
\end{eqn}where we have used
\begin{eqn}
     \frac{2}{\pi_{2,2p}} \int_{0}^{\frac{\pi_{2,2p}}{2}} {\rm{d}t} \cos_{2, 2p}(t)^2 = \frac{2}{\pi_{2,2p}}\int_{0}^1 {\rm{d}t} \sqrt{1-t^{2p}}= \frac{p}{p+1}.
\end{eqn}Finally, it follows that $T$ is well approximated by
\begin{eqn}\label{eq:solt}
    T =T_{\lambda}+(T_0-T_{\lambda}) e^{-\frac{d-1}{p+1}\left( 1-\frac{T^2_{\lambda}}{1+w_0^2}\right)(N-N_i)}.
\end{eqn}

\subsubsection{Proper fixed points}\label{ssec:properfixed}
To describe the remaining possibility, let us take a step back to \eqref{eq:I12} and the surrounding discussion. If the integral $I_1(N)$ converges, its argument is a uniformly continuous function of $N$, as its second derivative is bounded. This follows from the expression
\begin{eqn}\label{eq:dwdN}
    \begin{aligned}
    \frac{{\rm d}}{{\rm d}N} \left( \frac{w'^2}{1+w^2}\right) = & -2 \frac{w w'^3}{(1+w^2)^2} +\frac{2 w'}{1+w^2} \Bigg[  \alpha^2 w S-(1-S)(d-1)w' 
    \\ & - 
    \frac{\alpha^2}{2}(1-S)\left(\lambda w+ (1+w^2)\frac{\partial_wP(w)}{P(w)} \right) \Bigg], 
\end{aligned}
\end{eqn}where the only potentially unbounded term, proportional to $(1-S)/P(w)$, is finite from Eq. \eqref{eq:int2}. Using Barbalat's lemma, we can conclude that the integrand vanishes asymptotically, and thus $w' \rightarrow 0$. From the convergence of the integral and also using \eqref{eq:dw}, both $S$ and $w$ have a finite limit
\begin{eqn}
    w \longrightarrow \bar{w} \quad \quad \quad \quad S \longrightarrow \bar{S}=T_{\lambda}^2/(1+\bar{w}^2).
\end{eqn}
In terms of $x$ and $y$, the expressions read
\begin{eqn}
    \bar{x}= \frac{\lambda \alpha}{2(d-1)}  \frac{1}{1+\bar{w}^2} \quad \quad \quad \quad \bar{y}= \frac{\lambda \alpha}{2(d-1)}  \frac{\bar{w}}{1+\bar{w}^2}.
\end{eqn}The above equations describe standard fixed points, whose stability can be studied perturbatively by a local linearization of the system. Their location is fully specified by a solution to the equation 
\begin{eqn}\label{eq:fp}
    \frac{\partial_w P(\overline{w})}{P(\overline{w})}- \frac{2 T_{\lambda} \, \overline{w}}{\alpha (1+\overline{w}^2)} \left[ \frac{\alpha}{1+\overline{w}^2-T_{\lambda}^2}-(d-1)\right]=0.
\end{eqn}In order for the solution to be valid, the constraint \eqref{eq:csH} has to be respected, translating to
\begin{eqn}
    \bar{S}=\bar{x}^2+\bar{y}^2 < 1, \quad \quad \bar{w}^2 > \frac{\lambda^2 (d-2)}{4C(d-1)}.
\end{eqn}Incidentally, the condition for accelerated expansion only differs by a numerical factor, \emph{i.e.}
\begin{eqn}
   \bar{S}= \bar{x}^2+\bar{y}^2 < \frac{1}{d-1}, \quad \quad \bar{w}^2 > \frac{\lambda^2 (d-2)}{4C}.
\end{eqn}If $P(w)$ has no minima for $w^2 > T_{\lambda}^2-1$, $I_1(N)$ cannot diverge and a fixed point has to exist. Reassuringly, in this case \eqref{eq:fp} has at least two solutions, one for both positive and negative $w$. This can be seen from continuity of the LHS: for $w\rightarrow \pm\sqrt{w^2-1}^{\pm} $ the LHS goes to $\mp \infty$, while for $w$ going to $\pm \infty$ it approaches zero from above (below).

\subsubsection{Improper fixed point (and kination)}\label{ssc:ifp}

If $S=1$, the constraint equation \eqref{eq:csH} implies $P(w)=0$, so one must further impose $w'=0$ to lie within an invariant subset of $\Omega_0$. Therefore, the largest invariant set is the point specified by $S=1,T^2=1+w_0^2$. Alternatively, one can see this from the fact that $T$ must necessarily converge to a constant, being a monotonous and bounded function. If $S \rightarrow 1$, then $T^2 \rightarrow 1+w_0^2$ from \eqref{eq:dw}, since $w'^2$ cannot converge to a positive constant. We call this type of solution an improper fixed point, since all variables converge to a constant value but $w=w_0$ is outside the domain where the system \eqref{eq:sysN} is defined. For this reason, it can be analyzed more conveniently using the reformulation \eqref{eq:sysNz}. It is easy to see that improper fixed points are characterized by $x=\pm 1, y=0$, and $w=w_0=0$, and can only exist for potentials satisfying $P(0)=0$. Physically, they correspond to the situation where the saxion is kinating, and its energy density is dominating the evolution. Notice that since we have introduced an additional variable, a fixed point of the original system \eqref{eq:sysN} may not necessarily correspond to a fixed point \eqref{eq:sysNz}, as $z$ can depend non-trivially on time. If $T_{\lambda} >1$, however, $z$ is driven to zero if $x\rightarrow \pm 1, y=0$, so the improper fixed point must necessarily be a fixed point of \eqref{eq:sysNz}, with $z=0$. For a system of the form $x'_i=f_i(x_j)$, the (linear) stability of a fixed point $\bar{x}$ can be evaluated from the eigenvalues of the Jacobian $J \equiv \partial_{x_j} f_i \lvert_{x=\bar{x}}$. In this case, one can easily see from the computation of the Jacobian (evaluated at $x=\pm1, y=w=z=0$) that there is always a positive eigenvalue 
\begin{eqn}
   J \big \lvert _{x= \pm1,y=w=z=0}= \left(
\begin{array}{cccc}
 0 & 0 & 0 & 0\\
 0 & \pm \alpha  & 0 & 0 \\
 0 & 0 &(d-1) \mp \frac{\lambda \alpha}{2}  & 0 \\
 0 & \alpha  & 0 & \mp \alpha  \\
\end{array}
\right),
\end{eqn}and such improper fixed points are never stable. On the other hand, if $T_{\lambda} <1$, there is a general (although approximate) solution to the system given in the paragraph \ref{ssec:oscillating} on oscillating solutions. For any choice of the integration constants, $S \nrightarrow 1$. We conclude that in the single-sector case, solutions with $S \rightarrow 1$ are never an attractor, and $T$ will always converge to
\begin{eqn}
   T \rightarrow  T_{\lambda} \equiv \frac{\alpha \lambda}{2(d-1)} = \frac{\lambda}{2} \sqrt{\frac{d-2}{C(d-1)}}.
\end{eqn}As we will see later, this is to be contrasted to the multi-sector case, where $S \rightarrow 1$ and $T \rightarrow \sqrt{1+w_0^2}$ can be realized asymptotically.

As a short aside, let us also notice that if $P(0) > 0$, there still exists a (proper) kinating fixed point, given by $x=\pm1,y=0$ and $w=0$. From the above analysis, it should always be unstable. Again, this can be seen explicitly from the Jacobian
\begin{eqn}
   J \big \lvert _{x= \pm1,y=w=0}= \left(
\begin{array}{ccc}
 \mp \alpha  \lambda +2 d-2 & 0 & 0 \\
 \pm \frac{\alpha  P'(0)}{P(0)} & \pm \alpha  & 0 \\
 0 & \alpha  & \mp \alpha  \\
\end{array}
\right),
\end{eqn}which always has at least one positive eigenvalue. 

\subsection{Analysis of the general polynomial potential}\label{ssec:generalpolynomial}

It is clear that the oscillating solutions discussed in Section \ref{ssc:sise} may break down asymptotically if sub-leading corrections to the scalar potential are present. For this reason, we now consider the more general potential \eqref{eq:fullp}, which we report here again for ease of presentation:
\begin{eqn}\label{eq:fullp2}
  V(w,s) = \frac{1}{s^{\lambda}} \sum_{n=0}^N \frac{P_n \left( w\right)}{s^n} .
\end{eqn}For instance, if for a given $m$
\begin{eqn}\label{eq:cond}
\frac{T_{\lambda}^2}{1+w_0^2} \left[m \alpha -\frac{d-1}{p(p+1)} \right]< \frac{d-1 }{p+1},
\end{eqn}the leading order term $P_0(w)= f^2 (w-w_0)^{2p}+  \mathcal{O}\left((w-w_0)^{2p+1}\right)$ becomes subdominant with respect to the leading correction $v^m P_m(w)$ along the trajectories defined by Eqs \eqref{eq:ansatzw}, \eqref{eq:wF} and \eqref{eq:SF}.\footnote{Assuming $P_m(w_0)>0$.} Even if \eqref{eq:cond} were not satisfied, along an oscillating solution $w \rightarrow w_0$ by crossing the line $w=w_0$ infinitely many times, and one might worry that close enough to those points the sub-leading correction would also dominate over the leading term, potentially spoiling the solution. On the other hand, if $P_0(w) > \varepsilon $ asymptotically (for some $\varepsilon >0$), all sub-leading terms will suppressed by powers of $s$, and it will not be necessary to take them into account. 

When the potential takes the form \eqref{eq:fullp2}, no accidental simplifications occur and the equations of motion can be reformulated as a first order, dynamical system of four equations (rather than three). While this can be trivially done by simply defining two new variables corresponding to $\dot{s}$ and $\dot{a}$, it will be convenient to write the system in a form that is reminiscent of the single sector case, to see if any of the results obtained before can be straightforwardly generalized. Upon further defining $v \equiv 1/s$, the Hubble constraint becomes
\begin{eqn}\label{eq:zms}
    x^2+y^2+z^2 \sum_{n=0}^N v^n P_n(w) =1,
\end{eqn}
and the equations of motion can be recast as
\begin{equation}\label{eq:sysNg}
\left \{
\begin{aligned}
\frac{d x}{dN}&=\,- \alpha\, y^2 -\left(1- x^2-y^2\right) \left[(d-1)x- \frac{\alpha}{2} \left( \lambda + \frac{ \, \sum_{n=0}^N v^{n} \left( w\partial_w P_n(w)+n P_n(w) \right)}{\sum_{n=0}^N v^{n} P_n(w)}\right) \right]\\
\frac{d y}{dN}&=  + \alpha\,  x y-\left(1- x^2-y^2\right) \left[(d-1)y+\frac{\alpha}{2} \frac{ \, \sum_{n=0}^N v^{n} \partial_w P_n(w)}{\sum_{n=0}^N  v^{n} P_n(w)} \right]\\
\frac{d w}{dN}&= \alpha (y- wx)\\
\frac{d v}{dN}&= -\alpha vx.\\
\end{aligned} \right. 
\end{equation}
In this form, the system \eqref{eq:sysNg} can be analyzed with similar tools to the ones used in the previous section. For the reasons we have just stated, it suffices to analyze the cases where $w \rightarrow w_0$, with $P_0(w_0)=0$. Otherwise, we may refer to the single-sector classification.

An in-depth analysis is performed in the next subsection, showing how $w' \rightarrow 0$ and no oscillating solutions can exist asymptotically, under the assumption that the overall potential never vanishes exactly. The latter amounts to requiring the existence of at least one term in the above sum such that $P_n(w_0) > 0$, for any $w_0$ satisfying $P_0(w_0)=0$. Moreover, we also show how the variable $T$ still converges to a constant, determining only two possible asymptotic solutions. The first one is characterised by
\begin{eqn}\label{eq:fa1}
    T \rightarrow T_{\lambda+m} \equiv \frac{\lambda+m}{2(d-1)} \quad \quad \quad \quad S \rightarrow \frac{T_{\lambda+m}^2}{1+w_0^2},
\end{eqn}where $m$ is the lowest integer such that $P_m(w_0) >0$. Effectively, this behaves similarly to the fixed point solutions discussed previously, although the asymptotic value of $T$ is now determined by the next-to-leading term rather than the leading one. If $T_{\lambda+m}^2 > 1+w_0^2$, however, this solution is not viable ($S \leq 1)$, and the only other possibility is
\begin{eqn}\label{eq:fa2}
    T  \rightarrow \sqrt{1+w_0^2} \quad \quad \quad \quad S \rightarrow 1.
\end{eqn}Unlike the oscillating solutions, these do not violate the dynamical version of the distance conjecture discussed in this paper, following the discussion around \eqref{eq:dads}. 

\subsubsection{General remarks}

In analogy to the previous section, the quantity $T$ can be shown to obey
\begin{eqn}\label{eq:dTQ}
\frac{dT}{dN}= -(1-S) \left[(d-1) T- \frac{\alpha}{2} \left( \lambda + \frac{\sum_{n=0}^N n v^{n} P_n(w)}{\sum_{n=0}^N v^n P_n(w)}\right) \right],
\end{eqn}\noindent
and Equation \eqref{eq:int2} can be generalized to 
\begin{eqn}\label{eq:int3}
   \frac{\sum_{n=0}^N  v^{n} P_n(w)}{\sum_{n=0}^N  v^{n} P_n(w_i)} \left(\frac{1+w^2}{1+w_i^2}\right)^{\frac{\lambda}{2}} =\frac{1-S}{1-S_i} {\rm{Exp}} \left(-2(d-1) \int_{N_i}^N {\rm{d}}\hat{N} \,    S- \frac{T T_{\lambda}}{1+w^2} \right).
\end{eqn}Let us now assume that, for a given potential, there exist constants $P_-,P_+$ such that the following inequality applies asymptotically to any trajectory:
\begin{eqn}
    P_{-} \leq Q(v,w) \leq P_+, \quad \quad \text{with} \quad \quad Q(v,w) \equiv \frac{\sum_{n=0}^N n v^{n} P_n(w)}{\sum_{n=0}^N v^n P_n(w)}.
\end{eqn}
Then, one can define the Lyapunov functions
\begin{eqn}
    \mathcal{L}_{\pm} \equiv  \left[(d-1) T- \frac{\alpha}{2} ( \lambda + P_{\pm}) \right]^2.
\end{eqn}
Their derivatives satisfy
\begin{equation}\label{eq:dt2}
\frac{d  \mathcal{L}_{\pm}}{dN}=-2(1-S) \mathcal{L}_{\pm} -  \frac{\alpha(1-S)}{2}(P_{\pm}-Q)\left[(d-1)T -\frac{\alpha}{2}(\lambda + P_{\pm})\right],
\end{equation}and in particular they are always negative outside the interval $\left[\lambda+P_{-} ,\lambda+P_{+}\right]$. Therefore, by La Salle's invariance principle, $T$ will asymptotically converge to the strip $\left[\lambda+ P_-,\lambda + P_+\right]$, unless $S \rightarrow 1$. More precisely, this can be shown by applying the theorem to the compact sets resulting from the intersection of the half-spaces $\mathcal{P}_{\pm}= \left\{ T \lessgtr P_{\pm}  \right\}$ with the level sets of $\mathcal{L}_{\pm}$.

\noindent To infer properties of the solution, it is therefore essential to establish bounds on the quantity $Q(v,w)$. Here, we will exploit the fact that $v\rightarrow 0$ for any trajectory probing the infinite distance limit. If the attractor is located in the region $P_0(w_0) > 0$ (more generally, if $P(w) > \varepsilon $ for a fixed $\varepsilon >0$ asymptotically),
\begin{eqn}
    Q(v,w) \rightarrow 0 \quad \quad \text{and} \quad \quad T \rightarrow \frac{\lambda \alpha}{2(d-1)}. 
\end{eqn}Then, the sub-leading terms in the potential can be neglected, and one can use the classification of Section \eqref{ssc:sise}.

In the more interesting case where $P_0(w_0) \rightarrow 0$ asymptotically, sub-leading corrections in $1/s$ can play a crucial role, precisely because the (naively) leading order term is becoming arbitrarily small. In terms of the quantities defined above,
\begin{eqn}\label{eq:Q}
    Q(v,w) = \frac{m v^m P_m(w)}{P_0(w)+v^m P_m(w)} + \mathcal{O}(v),
\end{eqn}\noindent
where $m$ is the smallest integer such that $P_m(w_0) \neq 0$. In order for the potential to be always positive definite, $P_m(\bar{w}) > 0$. Therefore, $ 0 \leq Q(v,w) \leq m$ and $T $ will converge to the strip
\begin{eqn}\label{eq:intv}
    T \rightarrow \left[T_{\lambda}, T_{\lambda}+T_m\right], \quad \quad \text{with} \quad \quad T_m=\frac{m \alpha}{2(d-1)}.
\end{eqn}\noindent
We now aim to prove that, unless $S \rightarrow 1$, $T$ will actually converge to the maximum value allowed in the interval \eqref{eq:intv}.

\subsubsection{Convergence of $T$}\label{ssc:Tconv}

From \eqref{eq:dTQ}-\eqref{eq:Q}, we can now understand how the presence of additional terms affects the solutions to the single-sector potential when $P_0(w) \rightarrow 0$.
As mentioned at the beginning of this section, corrections to the potential can dominate over the leading term if \eqref{eq:cond} is satisfied. In that case, $Q \rightarrow m$ from \eqref{eq:Q}, and it would appear that $T \rightarrow T_m + T_{\lambda}$ rather than $T\rightarrow T_{\lambda}$. 

To proceed further, is convenient to rewrite the system \eqref{eq:sysNg} as a second-order differential equation for $w(N)$,
\begin{eqn}
\begin{split}\label{eq:w2}
   & w'' = \alpha w \left[\alpha S-(d-1)(1-S)\left( T_{\lambda}+\frac{\alpha Q}{2(d-1)} \right) \right] \\
    & -(1-S)(d-1)w'-\frac{\alpha^2}{2}(1-S)(1+w^2) \frac{ \, \sum_{n=0}^N v^{n} \partial_w P_n(w)}{\sum_{n=0}^N  v^{n} P_n(w)}.
    \end{split}
\end{eqn}Notice that all the terms on the RHS are bounded except the last one, which is the key to understanding the asymptotic behavior of the system. \footnote{If $w''(N)$ were bounded, we could immediately conclude using Barbalat's Lemma that $w'\rightarrow 0$ if $w\rightarrow \bar{w}$, immediately ruling out oscillating solutions.} In a large $s$ (small $v$) expansion, the dominating $1/s$ correction to the potential is given by
\begin{eqn}
    P(w) = \left[ P_0(w)+ v^m P_m(w) \right] +\mathcal{O}(v (w-w_0)^2)+\mathcal{O}(v^{m+1}) 
\end{eqn}with $m$ the smallest integer for which $P_m(\bar{w}) \neq 0$ as in \eqref{eq:Q}. In particular, the combination appearing in the RHS of \eqref{eq:w2} can be re-expressed as
\begin{eqn}
    \frac{ \,\sum_{n=0}^N v^{n} \partial_w P_n(w)}{\sum_{n=0}^N  v^{n} P_n(w)} = \frac{p(1- \frac{Q}{m})}{w-w_0}  + \mathcal{O}(Q(w-w_0)),
\end{eqn}where $p$ is defined by the Taylor expansion of the potential close to the minimum of $P_0$, \emph{i.e.} $P(w)= f^2 (w-w_0)^{2p} + \mathcal{O}\left( (w-w_0)^{2p+1}\right)$. This results in
\begin{eqn}\label{eq:w3}
    w''= - \frac{ p \alpha^2(1- \frac{Q}{m})}{(w-w_0)}(1-S)(1+w^2)+ f(S,w,w'),
\end{eqn}where $f(S,w,w')$ is a bounded function, whose precise expression is not relevant for the present discussion. \footnote{In any case it can be read off from \eqref{eq:w2}.} Intuitively, this tells us that $w$ will oscillate very fast as $w\rightarrow w_0$. Every time the line $w=w_0$ is crossed, either $S$ or $Q/m$ have to be exactly equal to $1$. Therefore, one might suspect $T\rightarrow T_{\lambda}+T_m$ or $S \rightarrow 1$ to be the only two asymptotic possibilities.

To formalize such an intuition, we first prove that $T$ will always converge to a constant for the system \eqref{eq:sysNg}. If $w \nrightarrow w_0$, the leading term $P_0(w)$ dominates at large $s$, and one can apply the classification of the previous section, with a single term in the potential. Therefore, we only have to consider solutions with $w\rightarrow w_0$, where $P(w_0)=0$. Let us now proceed by contradiction, and assume that $T$ does not converge to a constant along such solutions. Since $T$ is bounded, we can define the sequence $\left\{N_i^-,N_i^+ \right\}$, where $N_i^+ < N_i^- $ denote the location of consecutive local maxima and minima for $T$ respectively, with $T_i^{\pm} \equiv T(N_i^{\pm})$, $\liminf_{i \rightarrow \infty }T_i^- < \limsup_{i \rightarrow \infty } T_i^+$ and $T_i^{\pm} \in \left[ T_{\lambda},T_{\lambda}+T_m\right]$ from \eqref{eq:intv}. From \eqref{eq:w3}, we have that for any arbitrary times $N_1$ and $N_2$, 
\begin{eqn}\label{eq:boundQ}
\begin{split}
\int_{N_1}^{N_{2}}  {\rm {d}}\hat{N}\left(1-\frac{Q}{m}\right)& (1-S)  \leq \\ & \Bigg \lvert w'(N_{2}) - w'(N_{1}) - \int_{N_1}^{N_{2}} {\rm {d}} \hat{N} f(w,w',S) \Bigg \lvert  \times 
 \underset{N \in [N_1,N_2] }{{\rm Max}} \big |w(N)-w_0 \big |,
\end{split}
\end{eqn}where we have suppressed some irrelevant constants. Then, if $\lvert N_i^+-N_i^- \lvert$ is bounded from above as $i \rightarrow \infty$, we can use the inequality \eqref{eq:boundQ} to show how for any $\varepsilon >0$ one can find a value $\bar{i}$ such that
\begin{eqn}\label{eq:Tpm}
\begin{split}
    T(N_i^-)-T(N_i^+) = (d-1)\int_{N_i^+}^{N_i^-} {\rm {d}} \hat{N}
    (1-S)\left(T_{\lambda}+T_m -T\right)\\-T_m \int_{N_i^+}^{N_i^-} {\rm {d}} \hat{N} (1-S) \left(1-\frac{Q}{m}\right) >- \varepsilon
\end{split}
\end{eqn}for any $i > \bar{i}$. The above equation is essentially an integrated version of \eqref{eq:dt2}, and we have used the fact that the first integral is positive while the second one is bounded by \eqref{eq:boundQ}, as well as the fact that $w(N) \rightarrow w_0$. We conclude that $T_i^+-T_i^- \rightarrow 0 $ and $T$ converges to a constant. The remaining case, where $\lvert N_i^+-N_i^- \lvert$ is not bounded from above, can be excluded on the basis of the following reasoning. Let us denote $N_i^+ <\tilde{N}_i < N_i^-$ as the time for which $T(\tilde{N}_i) = (T_i^+ + T_i^-)/2$. Since $\frac{{\rm d} T}{ {\rm d} N}$ is bounded, $N_i^--\tilde{N}_i  > K$, where $K$ is a positive constant. One can then consider the sequence of integrals given by
\begin{eqn}
    I_i \equiv \int_{N_i^- - K}^{N_i^-}  {\rm {d}}\hat{N}\left(1-\frac{Q}{m}\right)& (1-S),
\end{eqn}where $I_i \rightarrow 0$ from \eqref{eq:boundQ}. Moreover, since $T' <0$ between $N_i^- - K $ and $N_i^-$, 
\begin{equation}
    \frac{Q}{m} < \frac{T_i^++T_i^--T_{\lambda}}{2 T_m} < 1 - \delta_1
\end{equation}asymptotically (at least along a subsequence), for some constant $\delta_1 >0$. Therefore,
\begin{eqn}
   I_{S,i} \equiv  \int_{N_i^--K}^{N_i^-}  {\rm {d}} \hat{N}& (1-S) \rightarrow 0.
\end{eqn}This implies the existence of a subsequence $I_{S,j(i)} \in I_{S,i} $ where $S\rightarrow 1$ almost everywhere, \footnote{Convergence in measure implies convergence almost everywhere along a subsequence.} and since $S$ is continuous $S\rightarrow 1$ pointwise. From \eqref{eq:dw}, we can then deduce that $(T_i^++T_i^-)^2 \leq 4 (1+w_0^2)$, and hence $w'^2 = S(1+w^2)-T^2 \rightarrow (1+w_0^2)-T^2 > \delta_2 > 0 $ for any $N \in (\tilde{N}_i,N_i^-]$.\footnote{All of these inequalities are to be understood as valid asymptotically.} Finally,
\begin{equation}\label{eq:diffw}
    \big \lvert w(N_i^-)-w(N_i^--K) \big\rvert =    \Bigg \lvert \int_{N_i^--K}^{N_i^-} {\rm {d}}\hat{N} w' \Bigg \rvert >  \alpha \int_{N_i^--K}^{N_i^-}  {\rm {d}}\hat{N} \sqrt{\left(\frac{T_i^++T_i^-}{2}\right)^2-T^2}
\end{equation}
asymptotically. Since the RHS is uniformly bounded by some positive constant, this contradicts the initial hypothesis $w \rightarrow w_0$. This completes the proof that $T$ converges to a constant value, $T \rightarrow T_{\infty}$.

By slightly tweaking the same reasoning, it is also possible to show $T_{\infty}=T_{\lambda}+ T_m$, unless $S \rightarrow 1$. Let us consider the same equation as in \eqref{eq:Tpm}, with the integrals now taken between (now arbitrary) times $N_i$ and $N_i+ \Delta N$, with $\Delta N$ constant. The $\left\{ N_i\right\} $ can be any sequence $N_i \rightarrow \infty$, and $\Delta N$ can be arbitrarily large (but finite). Then, 
\begin{eqn}\label{eq:Tpm}
\begin{split}
  & (d-1)\int_{N}^{N_i + \Delta N} {\rm {d}} \hat{N}  
    (1-S)\left(T_{\lambda}+T_m -T\right) = \\ 
    &\qquad \qquad T_m \int_{N_i}^{N+ \Delta N} {\rm {d}} \hat{N} (1-S) \left(1-\frac{Q}{m}\right) + T(N+\Delta N) - T(N) \rightarrow 0. 
\end{split}
\end{eqn}If $T_{\infty} < T_{\lambda}+T_m$, it follows (as before) that there exists a subsequence of intervals $\left[ N_{j(i)}, N_{j(i)}+\Delta N \right ] $ where $S \rightarrow 1 $ pointwise. Within such intervals $w'^2 >0$, and this leads to a contradiction (as in \eqref{eq:diffw}) unless $T \rightarrow \sqrt{1+\bar{w}}^2$.

Finally, this also shows how there can be no asymptotic oscillating solutions in the presence of corrections lifting the degeneracy of the potential (\emph{i.e.} such that the potential can never be zero exactly). One can rewrite \eqref{eq:dt2} as
\begin{eqn}\label{eq:dTfin}
\begin{aligned}
    & \frac{{\rm{d}}}{{\rm{d}}N}\left[T-\frac{m}{p}\frac{w'(w-\bar{w})}{ (1+w^2)} \right]= -(1-S)(d-1) \left[T -T_{\lambda}-T_m \right]\\ & -\frac{m}{p}\frac{w'^2}{ (1+w^2)} + g(S,w,w')(w-w_0),
    \end{aligned}
\end{eqn}where $g(S,w,w')$ is again bounded. From this expression, it follows that $w'\rightarrow 0$, since the first term on the RHS is also converging to zero.

\subsubsection{Improper fixed point and kination}\label{ssc:ifpk}

From the analysis in the previous section, we deduce that in the multi-sector case the system will eventually approach a fixed point. If $T \nrightarrow T_{\lambda+m}$, one must necessarily have $S \rightarrow 1$. As in the single-sector case, this corresponds to improper fixed points, where $w=v=0$ and the system \eqref{eq:sysNg} is not well defined. As before, this can only happen if $P_0(w)$ has an absolute minimum at $w_0=0$, and the variables converge to $x= \pm 1, y=w=v=0$. To analyze their  stability, we can proceed as in Section \ref{ssc:ifp}, and reformulate the system with the additional variable $z$, defined in \eqref{eq:zms}.  
\begin{equation}\label{eq:sysNgz}
\left \{
\begin{aligned}
\frac{d x}{dN}&=\,- \alpha\, y^2 - z^2 (d-1)  \left(x-T_{\lambda} \right)\sum_{n=0}^N  v^{n} P_n(w)
+ \frac{\alpha z^2}{2}\sum_{n=0}^N v^{n} \left[w\partial_w P_n(w)+n P_n(w) \right] \\
\frac{d y}{dN}&=  + \alpha\,  x y- z^2 (d-1) y \sum_{n=0}^N  v^{n} P_n(w) - \frac{\alpha z^2}{2}\sum_{n=0}^N v^{n} \partial_w P_n(w)\\
\frac{d w}{dN}&= \alpha (y- wx)\\
\frac{d v}{dN}&= -\alpha vx.\\
\frac{d z}{dN}&= z \left[(d-1)(x^2+y^2)-\frac{\lambda x}{2} \alpha \right] \\
\end{aligned} \right. 
\end{equation}If $T_{\lambda} > 1$, $z\rightarrow 0$ when approaching the improper fixed points. Therefore, the latter can be viewed as the (now) proper fixed points $x= \pm1, y=w=v=z=0$ of \eqref{eq:sysNgz}. The corresponding Jacobian is
\begin{eqn}
   J \big \lvert _{x= \pm1,y=w=v=z=0}= \left(
\begin{array}{ccccc}
 0 & 0 & 0 & 0 & 0\\
 0 & \pm \alpha  & 0 & 0 & 0 \\
 0 & \alpha   & \mp \alpha & 0 & 0 \\
 0 & 0 & 0   & \mp \alpha & 0  \\
 0 & 0 & 0 & 0 &(d-1) \mp \frac{\lambda \alpha}{2} \\
\end{array}
\right),
\end{eqn}so that they are never stable. If $T_{\lambda} < 1$, however, the analysis does not apply, as it is possible to have improper fixed points of \eqref{eq:sysNg} which are not a fixed point of \eqref{eq:sysNgz}, since $z$ does not converge to a constant value. From the numerical analysis presented in the next section, we will indeed find examples of asymptotic solutions where $S \rightarrow 1$.

\section{Physical implications}\label{sec:phim}
In this section we discuss the physical implications of the cosmological solutions found in the previous Section \ref{sec:Cosm_solutions}. We studied dynamical solutions for the axion-scalar systems with a polynomial potential for the axions. In the first subsection \ref{ssec:summary} we summarize our classification of solutions for both the single- and multi-sector potential, including a remarkable class of solutions that oscillates indefinitely in the former case. We then discuss in subsection \ref{ssec:dynamicaldistances} the implications for the length of these dynamical trajectories, highlighting in particular the consequences of oscillating solutions. We discuss in subsection \ref{ssc:ctrex} a particular example out of this class, which appears to have a dynamical distance that grows parametrically faster than the geodesic distance, but we explain how corrections to the potential or the decay of the oscillating field would resolve this issue in a more realistic scenario.

\subsection{Summary of asymptotic solutions}\label{ssec:summary}

Let us begin by reviewing our classification of asymptotic cosmological solutions for the axion-saxion scalar system, both in the single and multi-sector cases. The interested reader may refer to Section \ref{sec:Cosm_solutions} for a thorough analysis, containing detailed proofs of our statements. Here, we will simply provide a summary of our results, with a particular emphasis of those aspects that are most relevant to the connection between dynamics and the Swampland Distance Conjecture

\subsubsection{Single sector}

The single-sector system is given by \eqref{eq:sysN}, and it corresponds to a scalar potential given by a single axion polynomial $V(s,a) = s^{-\lambda} P(a/s)$. It is analysed in detail in subsection \ref{ssc:sise}. Table \ref{tab:sum} provides an overview of the main characteristics of each type of asymptotic solution.

\begin{table}[h]
\centering
\begin{tabular}{ |l|c|c|c|c| }
 \hline
 Asymp. solution & Nec. condition & $T$ & $S$ & $w$ \\ 
\hline \hline 
 Fixed point  & $\bar{w}^2 > T_{\lambda}^2-1$ & $T_{\lambda}$ & $\frac{T_{\lambda}^2}{1+\bar{w}^2}$ &  $\bar{w}$ solving \eqref{eq:fp} \\
 \hline
  Oscillating & $w_0^2 > T_{\lambda}^2-1$ & $T_{\lambda}$ & osc. &  $w_0$ solving $P(w_0)=0$ \\
 \hline
\end{tabular}
\caption{Classification of all possible asymptotic solutions to \eqref{eq:sysN} based on the asymptotic values of the variables $T=x+yw$, $S=x^2+y^2$, and $w$, where $w=a/s$ is the axion-saxion ratio and $x,y$ are the saxion and axion velocities respectively.}
\label{tab:sum}
\end{table}

We first summarize the criteria that underlie the classification before we get into the specifics of each case. Recall that the original differential system \eqref{eq:sysN} was formulated in terms of the saxion and axion velocities $x,y$ and the axion-to-saxion ratio $w$. For the purposes of the classification, we readily introduced more suitable combinations of variables $S=x^2+y^2$ and $T=x+yw$ to replace the velocities $x,y$. The classification then hinges on the first-order differential equation satisfied by $T$. Our analysis of Section \ref{ssc:sise} tells us that asymptotic solutions all satisfy $T'(N)=0$ asymptotically. From the structure of the differential equation, it then follows straightforwardly that either $S=1$ or $T=T_{\lambda}$. The case of $S=1$ corresponds to an improper fixed point, while for $T= T_{\lambda}$ we find both an oscillating or a proper fixed point solution. We can separate between the latter two cases by looking at whether a certain integral, denoted as $I_1(N)$,\footnote{For a precise definition, see Eq. \eqref{eq:I12}.} diverges or converges asymptotically as $N \to \infty$. Below we elaborate on each of these cases.

Before we do so, let us comment on the limiting value $T \to \bar T = \alpha \lambda/2$. It is suggestive to rephrase this asymptotic value as follows
\begin{equation}\label{eq:kineticlimit1}
   2 G_{I \bar{J}}  \frac{{\rm d}}{{\rm d}N} \left( \Phi^I \bar{\Phi}^{\bar{J}}\right) = \frac{C}{s^2} \frac{{\rm d}}{{\rm d}N} \left( s^2+a^2\right) \rightarrow \frac{\lambda (d-2) }{2} \, ,
\end{equation}
where  we have switched to the complex notation $G_{I \bar{J}} \equiv 2 \partial_{\Phi_I} \partial_{\bar{\Phi}_J} K$ with $K$ given in~\eqref{complexK}. The term on the LHS of \eqref{eq:kineticlimit1} is reminiscent of the kinetic term in the action, although we are lacking a more precise physical understanding. For instance, the structure of $T$ does not match precisely with that of the kinetic term, since here there is only a single derivative which acts on the sum $s^2+a^2$. We find that this asymptotic behavior \eqref{eq:kineticlimit1} also persists in the case of a general scalar potential with multiple terms in subsection~\ref{ssec:generalpolynomial}.

\paragraph{Proper fixed point.} We first consider proper fixed point solutions, where all three variables asymptote to fixed values. As mentioned above, we found that in this case the combination $T=x+yw$ approaches its fixed value $T=T_{\lambda}$ and the integral $I_1(N)$ converges. It is from this convergence of $I_1(N)$ that we can see that the other variables approach fixed values. Upon closer inspection of its integrand,
we conclude the derivative $w'(N) \to 0$ must vanish asymptotically. On the one hand, this tells that $w$ is driven to some fixed point $\bar{w}$.\footnote{The precise value of $\bar{w}$ follows through our detailed analysis from \eqref{eq:fp}.} On the other hand, the fact that asymptotically $w'(N)=0$ implies that $S$ approaches the fixed value $\bar{S}=T_{\lambda}^2/(1+\bar{w}^2)$. Thus we learn that all three variables $T,w,S$ are driven to particular fixed values, where $T$ in particular approaches the universal one $T_{\lambda}=\alpha\lambda/2$, while $\bar{w}$ (and thus also $\bar{S}$) depend on further details of the scalar polynomial $P(w)$.

\paragraph{Oscillating.} We next consider the case of an oscillating solution, where the axion velocity $y$ does not converge to a fixed value asymptotically, but instead it rather surprisingly oscillates indefinitely. We have depicted a numerical solution of this type in Figure \ref{fig:osc}. Due to this remarkable behavior, this case poses a risk where the dynamical distance is parametrically larger than the geodesic distance. We investigate this aspect in subsection \ref{ssec:dynamicaldistances}, where we find that this parametrical separation does \textit{not} happen (apart from one outlier case, that we discuss separately in subsection \ref{ssc:ctrex}). For now, let us focus on how this oscillating solution fits into the classification. Recall that again the combination of variables $T=x+yw$ approaches the fixed value $T=T_{\lambda}$, but now the integral $I_1(N)$ diverges as $N \to \infty$. To see what the asymptotic solution behaves like, it is useful to recall the relation of $I_1(N)$ with the polynomial $P(w(N))$ appearing in the scalar potential. Namely, the fact that $I_1(N)$ diverges implies that $P(w(N))$ must vanish asymptotically, so the axion-to-saxion ratio $w$ approaches one of its zeroes $w=w_0$. However, its derivative $w'(N)$ does not go to zero fast enough, which allows for room to spare in \eqref{eq:dw} such that the remaining variable $S=x^2+y^2$ does not converge. Rather, $S$ will keep oscillating indefinitely through its constituent $y$, the axion velocity. We have given an explicit description of this asymptotic solution in
subsection~\ref{ssec:oscillating}. 

\subsubsection{Multi-sector} 

Similar techniques allow us to classify the form of the asymptotic solutions for the multi-sector potential, as discussed in subsection \ref{ssec:generalpolynomial}. In particular, we carried out the analysis for a potential of the form
\begin{eqn}\label{eq:gp2}
    V(s,a) = \frac{1}{s^\lambda} \sum_{n=0}^N \frac{1}{s^n} P_n \left(\frac{a}{s}\right),
\end{eqn}involving a finite number of axion-saxion polynomials weighed by different powers of the saxion. This expression is more general than and includes all of the potentials derived from the F-theory construction in Section \ref{F-theory_embedding}, up to non-perturbative corrections. Let us also stress the fact that we only consider solutions extending all the way to the boundary of moduli space, where $s(N) \rightarrow \infty$ as $N \rightarrow \infty$. This excludes, for example, cases where the field is stabilized in an eventual minimum of the potential. Because of the different powers of $s$ appearing in \eqref{eq:gp2}, the potential can be effectively approximated as single-term if $P_0(a/s)$ is lower bounded along asymptotic trajectories where $s \rightarrow \infty$. In that case, the sub-leading terms can safely be neglected, and one can use the classification of the previous section. From Table \ref{tab:sum}, the only remaining option is a fixed point.\footnote{Since $P_0(w) \nrightarrow 0$.}

\noindent The other possibility occurs when $P_0(w) \rightarrow 0$. This may only arise if there exists $w_0$ such that $P_0(w_0)=0$, and $w \rightarrow w_0$. In this case, apparently sub-leading terms in the potential compete with the leading order term, since the corresponding polynomial vanishes asymptotically. Therefore, the classification in the previous subsection is no longer viable. To derive the form of possible asymptotic solutions, we again used the fact that the quantity $T$ obeys a simple equation of motion. With a slightly more convoluted reasoning than the one presented in the previous section, it is possible to prove that $T$ still converges to a constant, and $w' \rightarrow 0$. We refer to such possibilities as improper fixed points, because all variables converge to a fixed value which is however not included in the domain where the system \eqref{eq:sysNg} is defined ($w\rightarrow w_0$ and $v\rightarrow 0$). From our analysis, only two such possibilities can arise (as well as standard fixed points). This is shown in Table \ref{tab:sum2} below.

\begin{table}[h]
\centering
\begin{tabular}{|l|c|c|c|c|}
 \hline
 Asymp. solution & Nec. condition & $T$ & $S$ & $w$ \\ 
\hline \hline 
 Fixed point  & $\bar{w}^2 > T_{\lambda}^2-1$ & $T_{\lambda}$ & $\frac{T_{\lambda}^2}{1+\bar{w}^2}$ &  $\bar{w}$ solving \eqref{eq:fp} \\
 \hline
Imp. fixed point & $w_0^2 > T_{\lambda+m}^2-1$ & $T_{\lambda+m}$ & $T_{\lambda+m}^2/(1+w_0^2$) &   $w_0$ $|$ $P(w_0)=0$ \\
 \hline
Kin. fixed point & $P_0(0)=0$, $T_{\lambda} <1,$ & $1$ & $1$ &  $0$ \\
 \hline
\end{tabular}
\caption{Classification of all possible asymptotic solutions to \eqref{eq:sysNg} based on the asymptotic values of the variables $T=x+yw$, $S=x^2+y^2$, and $w$, where $w=a/s$ is the axion-saxion ratio and $x,y$ are the saxion and axion velocities respectively.}
\label{tab:sum2}
\end{table}

\paragraph{Improper fixed point.} 
The first possibility is that $T \rightarrow T_{\lambda+m}$, where $m$ is the smallest integer such that $P_m(w_0) >0$. In order to be viable, it requires $T^2_{\lambda+m} < (1+w_0^2)$, since $S \rightarrow T_{\lambda+m}^2/(1+w_0^2)$. The fact that this is the only possibility (except $S \rightarrow 1$) can be understood intuitively from the differential equation for $T$
but is more formally proven in subsection \ref{ssc:Tconv}. 
Effectively, the would-be-leading term $P_0(w)$ vanishes quicker than the sub-leading correction $v^m P_m(w)$ asymptotically, and it can be neglected. Moreover, the fact that $w' \rightarrow 0$ is also shown. A numerical example is presented in Figure \ref{fig:subl2}. It is also evident from the plot how the trajectory initially approaches one of the oscillating solution described in the previous subsections, before the effect of higher-order corrections eventually takes over and it converges to a fixed point.

\paragraph{Improper, kinating fixed point.} Finally, there is one possibility that can arise only in a very specific situation, when $w_0=0$ and $T_{\lambda}^2 <1 $. As in the previous case, its existence can be understood from the equation for $T$,
as well as the fact that $w' \rightarrow 0$. It is characterized by $S=x^2+y^2=S \rightarrow 1$ asymptotically ($x \rightarrow 1$, $y\rightarrow 0$). Physically, it corresponds to saxion kination, where the kinetic energy of the saxion dominates over the scalar potential. It is discussed more in detail in subsection \ref{ssc:ifpk}, and a numerical example is presented in Figure \ref{fig:subl1}.

\subsection{Consequences for dynamical distances}\label{ssec:dynamicaldistances}
These results can now be used to address a dynamical formulation of the distance conjecture, as discussed in Section \ref{sc:ddc}. Within the specific context of this paper - one-modulus asymptotic limits - we can now answer the question of whether the distance conjecture is obeyed in terms of the traversed distance $\Delta$ defined in Eq. \eqref{eq:d1}. In particular, the latter is controlled by the ratio
\begin{eqn}\label{eq:dads}
    \frac{{\rm d}a}{{\rm d}s} = \frac{y}{x}= \frac{T w \pm \sqrt{S(1+w^2)-T^2}}{T \mp w \sqrt{S(1+w^2)-T^2}}.
\end{eqn}Following the discussion around
\eqref{eq:bnd}, if this ratio is bounded the dynamical distance conjecture is guaranteed to hold. On the other hand, if it diverges ``uniformly" in time (\emph{i.e.} there exists a time $N_M$ such that $|y/x| >M$ for any $N> N_M$), the conjecture will not be satisfied.

The first thing to notice is that if the system converges to a fixed point, both proper or improper, the square root terms in \eqref{eq:dads} (proportional to $w'\rightarrow 0$) vanish. Therefore, the ratio $\frac{{\rm d}a}{{\rm d}s}$ approaches the constant value $\overline{w}$, satisfying the proposed version of the conjecture. In particular, since the only options in Table \ref{tab:sum2} are fixed points, this implies that a putative violation can only occur when the potential can be effectively approximated as single-sector, as in Table \ref{tab:sum}. The only possibility left to examine is that of an oscillating solution, where $T \rightarrow T_{\lambda}$ and $w$ flows to a minimum of the potential ($w \rightarrow w_0, P(w_0)=0$). If $T_{\lambda} \neq 0 $ and the condition $|w_0|< T_{\lambda}$ is satisfied, the ratio above is still bounded, and again no violation occurs. In particular, the latter is true for all of the string theoretic examples examined in Section \ref{F-theory_embedding} with $T_{\lambda} \neq 0$. On the other hand, for $T_{\lambda} =0$ the question becomes slightly more subtle. If there are no fixed points, the only possibility is for $w \rightarrow w_0$. In that case, the ratio above becomes $\frac{{\rm d}a}{{\rm d}s} = 1/ w$, and diverges if $w \rightarrow 0$, leading to a violation of the dynamical distance conjecture discussed above if and only if $w_0=0$. Therefore, violations do occur for potentials characterized by
\begin{eqn}\label{eq:vl}
   \lambda =0, \quad \quad \quad  P(w)= w^{2m} Q(w) \quad \quad \text{with} \quad m \geq 1
\end{eqn}and where $Q(w)$ is another polynomial. Notice how, by our previous analysis, we really require \eqref{eq:vl} to be the full contribution to the potential, with no higher order terms (unless they also vanish for $w=0$). The latter would result in an asymptotic value for $T$ larger than zero, curbing the dangerous oscillating behaviour. Can examples of this kind arise in any asymptotic limit, and for some particular flux choice? Using the classification in Section \ref{F-theory_embedding}, we see that there are two cases when this can happen.

The first one concerns a singularity of the type $\mathrm{III}_{0,0}$, with a particular combination of fluxes turned on. In the notation of \eqref{eq:III00}, if one sets $g_1=g_2=g_5=g_6=0$ the potential becomes
\begin{eqn}
    P(w)= (g_3^2+g_4^2)w^2 \quad \quad \quad \lambda=0,
\end{eqn}clearly of the form \eqref{eq:vl}. The second option arises for a type \emph{V}$_{1,1}$ limit, with the flux choice $g_1=g_2=g_4=g_5=0$ in \eqref{eq:V11}
\begin{eqn}\label{eq:fc}
    P(w)=3 g_3^2 w^2(1+w^2) \quad \quad \quad \lambda=0.
\end{eqn}This limit type is more commonly known as the Large Complex Structure (LCS) limit for a CY 4-fold, and some explicit examples will be presented later.

In the rest of this section, we will nevertheless try to explain why these potentials do not necessarily give rise to a counterexample to the dynamical formulation of the distance conjecture, and how the apparent pathology can be cured. In particular, we will discuss two possible solutions: the effect of higher order (e.g. $\alpha'$) corrections to the scalar potential, and a more physical mechanism involving the decay of the oscillating field, somewhat reminiscent of well-known, cosmologically relevant scenarios such as reheating.

\subsubsection{Obstructions to parametric separation of dynamical distances}
\label{ssc:ctrex}
\begin{figure}[h!]
\centering
\begin{subfigure}[b]{0.49\textwidth}
    \centering
  \includegraphics[width=\textwidth]{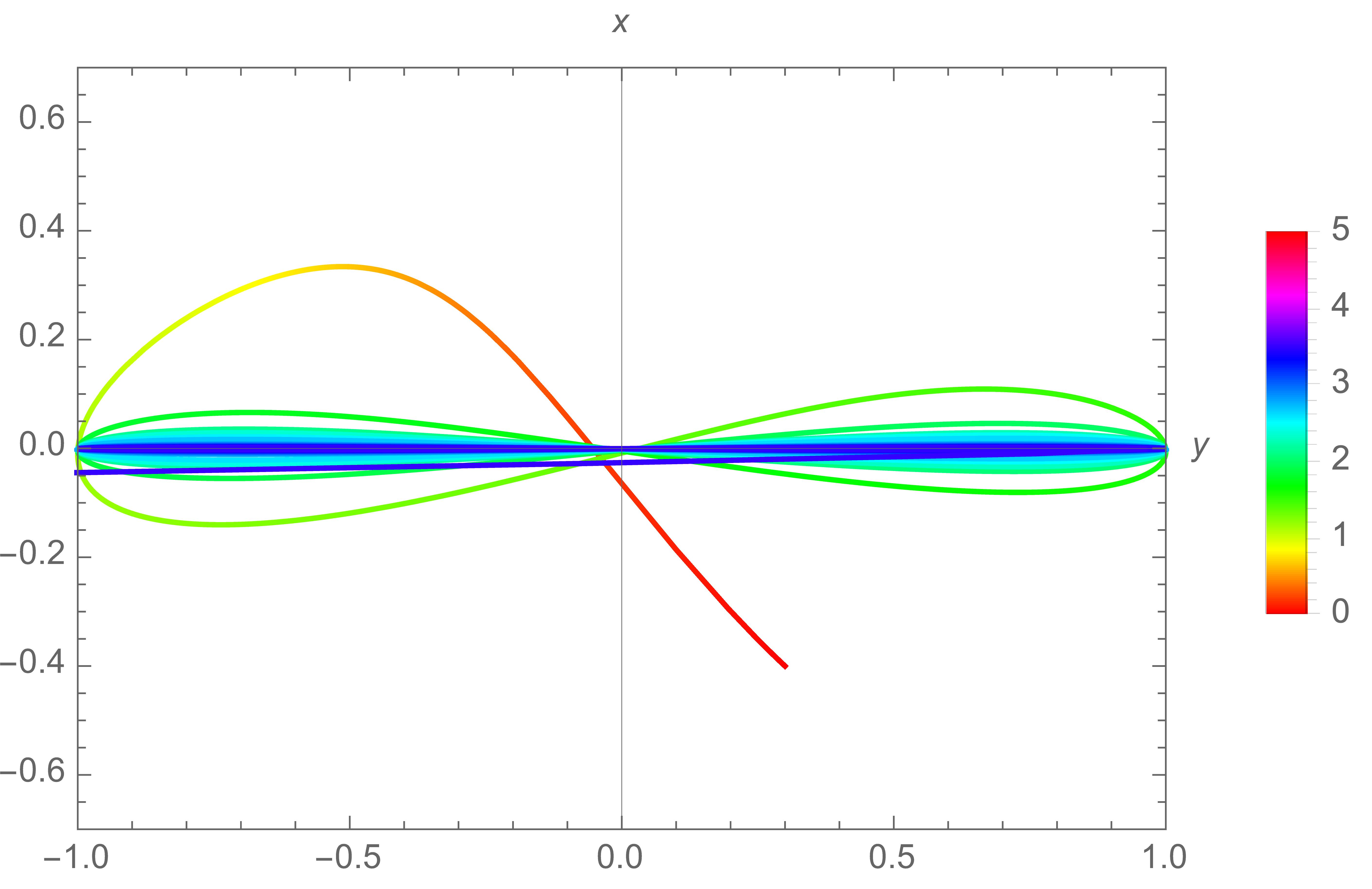}
  \label{fig:phase1}
  \caption{}
\end{subfigure}
\begin{subfigure}[b]{0.49\textwidth}
  \centering
  \includegraphics[width=\textwidth]{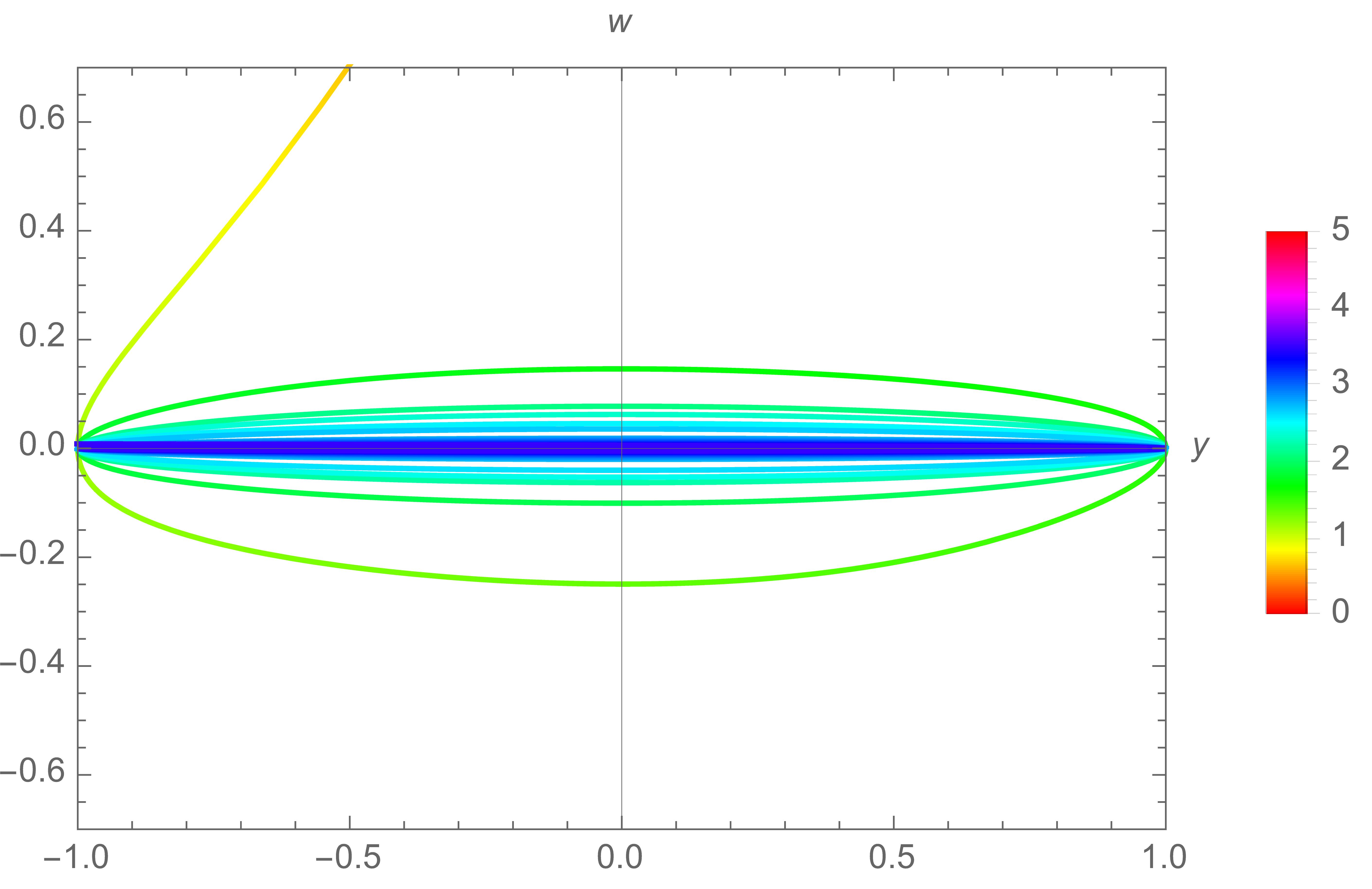}
\label{fig:phase2}
\caption{}
\end{subfigure}
\caption{Numerical solution to the system \eqref{eq:sysC} plotted in phase space, exhibiting the segment-like attractor described in the main text. The left and right panel depict the $y-x$ and $y-w$ planes respectively, with color varying as a function of time. As in the previous figure, we have chosen initial conditions specified by $x(0)=-0.4,y(0)=0.3,w(0)=1$, and the evolutions is between $N=0$ and $N=3.5$.}
\label{fig:phase}
\end{figure}

As stated above, potentials of the form \eqref{eq:vl} can apparently be realized within the classification outlined in Section \ref{F-theory_embedding}. Let us analyze the first example, arising from the well known asymptotic limit of Large Complex Structure, described as a singularity of type $\mathrm{V}_{1,1}$ (see also Appendix \ref{ssec:V}). Schematically, it is described by a potential 
\begin{eqn}
    P(w)= 3 g^2 w^2(1+w^2) \quad \quad \quad \lambda=0,
\end{eqn}and a kinetic term where the constant $C$ takes the value $C=2$. The resulting dynamical system is
\begin{equation}\label{eq:sysC}
\left \{
\begin{aligned}
\frac{d x}{dN}&=- \alpha\, y^2 -\left(1- x^2-y^2\right) \left[(d-1)x- \alpha \frac{2+3w^2}{1+w^2} \right]\\
\frac{d y}{dN}&=  \alpha\,  x y-\left(1- x^2-y^2\right) \left[(d-1)y+\frac{\alpha}{w} \frac{2+3w^2}{1+w^2} \right]\\
\frac{d w}{dN}&= \alpha (y- wx),\\
\end{aligned} \right.
\end{equation}
with $\alpha = \sqrt{3}$. Let us now try to understand the asymptotic solutions in detail. From the analysis in the previous section, we know that $T$ will converge to $T_{\lambda}=0$. Moreover, there are no, non-trivial fixed points, so $w$ will approach the minimum of the potential (located at $w=0$) asymptotically. We then expect the remaining variable - in this case $y$ - to exhibit an oscillatory behavior. As shown in Figure \ref{fig:phase}, the attractor locus is not a point, but rather a fixed segment stretching between $y=\pm1$, and located at $x=w=0$.
\begin{figure}[h!]
\centering
\begin{subfigure}[b]{0.49\textwidth}
    \centering
  \includegraphics[width=\textwidth]{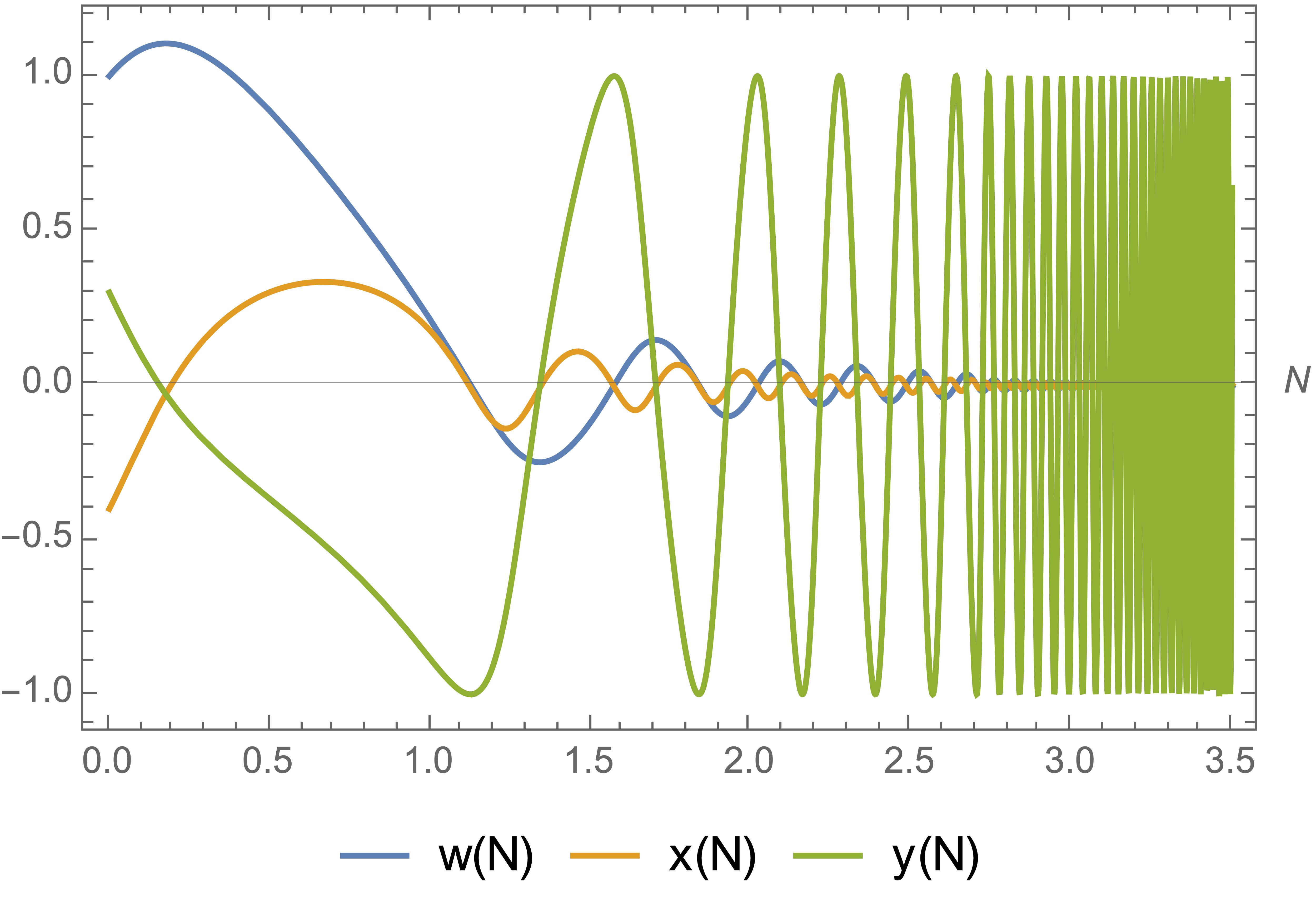}
  \label{fig:osc1}
  \caption{}
\end{subfigure}
\begin{subfigure}[b]{0.49\textwidth}
  \centering
  \includegraphics[width=\textwidth]{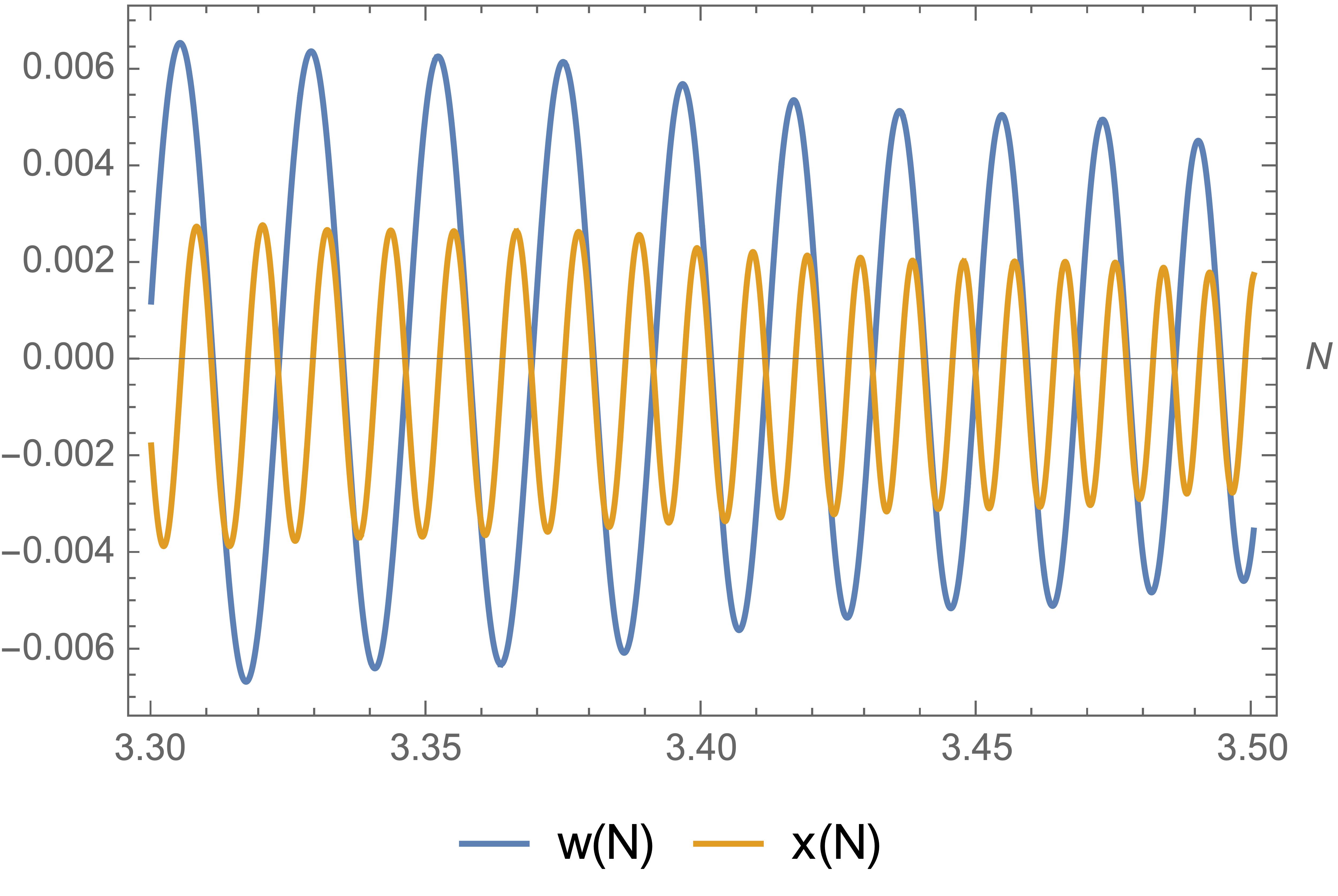}
\label{fig:osc2}
\caption{}
\end{subfigure}
\caption{Numerical solution to the system \eqref{eq:sysC}, with initial conditions specified by $x(0)=-0.4,y(0)=0.3,w(0)=1$. The left panel shows the full evolution from $N=0$ to $N=3.5$, while the right panel zooms in on the interval $[3.3,3.5]$, and only shows $w(N),x(N)$ for clarity. The qualitative features of the plot are described in the text, and can be understood analytically.}
\label{fig:osc}
\end{figure}

Such oscillatory features are visible in numerical solutions to the system, such as those shown in Fig. \ref{fig:osc}. Moreover, approximate, asymptotic solutions to the system are discussed in \eqref{eq:ansatzw}-\eqref{eq:solt}. Specializing to this particular example (and approximating $P(w) \simeq 3 g^2 w^2$ at small $w$), they become
\begin{eqn}\label{eq:asymp} 
\begin{split}
    & x(N) =  \left[ C_1-\frac{\alpha}{d-1}  \sin \left(2 e^{ \frac{3}{2} (N-N_0)}  +2 \theta\right) \right]e^{-\frac{3}{2} (N-N_0)} \\ & y(N) = \cos \left( e^{ \frac{3}{2}(N-N_0)}+\theta \right) \\ & w(N) =  \frac{2 \alpha}{3} e^{- \frac{3}{2} (N-N_0)} \sin \left( e^{\frac{3}{2} (N-N_0)} +\theta \right),
    \end{split}
\end{eqn}where $C_1,\theta$ and $N_0$ are integration constants. Notice how, for $N \rightarrow \infty$, this is exactly a period of matter domination, with a Hubble rate $H= \frac{2}{3 t}$. Indeed, the solution for the axion is exactly that of a field oscillating in a quadratic potential once we switch back to coordinate time:
\begin{equation}\label{eq:areh}
    a(t) = \frac{a(t_0) t_0}{t} \sin (m_a t),
\end{equation}
with $a_0,t_0$ being initial constants and $m_a$ the axion mass.

From the above, the equation for $x$ can be integrated to show that the saxion comes to a stop during its motion, exponentially fast in $N$. Physically, this is quite surprising, as there is still a runaway direction for $s$ which does not get stabilized. Although we will not discuss this further, this constitutes a rare example of moduli that are fixed even without a minimum in their scalar potential, thanks to Hubble friction (see also \cite{Tonioni:2024huw}). On the other hand, it is easy to see how the axion travels an infinite distance as $t \rightarrow \infty$, diverging as the harmonic series. 

We can now return to our main conceptual point, the implications for the distance conjecture in a dynamical sense. From the discussion above, it would seem that the oscillating solutions may provide a counterexample to the dynamical version of the distance conjecture, discussed in subsection \ref{sc:ddc}. This is because the distance measured along the trajectory grows parametrically faster than the geodesic one. In particular, the latter reaches a finite value, suggesting that the towers of states predicted by the usual distance conjecture do not have to become light in this limit. However, as we have just seen, the length of the trajectory in the full axion-saxion subspaces is divergent. 

A possible resolution is given by the fact that, in a realistic set-up, one does not expect that the oscillations will last eternally. In a physical situation, they will eventually be damped by the dissipation of the axion's kinetic energy into other sources, for example through decays of the condensate to other particles. To convince ourselves that this should be the case, it is instructive to compare the situation to a more familiar one. Once the saxion field is frozen, the evolution is essentially identical to that of a scalar (axion) field oscillating around the minimum of a Minkowski vacuum. Although the field would also appear to be traveling an infinite distance in that case, this certainly does not imply that Minkowski vacua are in tension with the distance conjecture, or (even worse) that they should belong the Swampland. This example is not only academic, but has an important phenomenological incarnation in the context of reheating (see \cite{Allahverdi:2010xz,Amin:2014eta,Lozanov:2019jxc} for a review). In the vanilla scenario of perturbative reheating,\footnote{Non-perturbative effects or Bose condensation can seldom be neglected, but in any case they would tend to enhance the decay rate.} mediated by perturbative decays (or scattering processes) of quanta in the oscillating field,  the oscillations of a scalar field around its minimum are exponentially suppressed in terms of a decay rate $\Gamma$. In our case, \eqref{eq:areh} would be modified to
\begin{equation}
    a(t) =\frac{a(t_0) t_0 e^{- \frac{\Gamma_{a} t}{2}}}{t} \sin (m_a t),
\end{equation}
with $\Gamma_a$ the decay rate of the axion condensate. Such an exponential behaviour would cause the field-space distance converge quickly enough to avoid any issues with the dynamical distance conjecture. In typical scenarios arising from string compactifications \cite{Cicoli:2023opf}, it is usually a saxion that oscillates, and it decays to its partner axions which are approximately massless. In this case the opposite process might happen, with the production of light saxions.\footnote{A direct decay $a\rightarrow s\, s $ is not allowed by the couplings in the lagrangian. However, the process $a \, a \rightarrow s \, s$ is allowed. Although inefficient at the perturbative level $(\Gamma \ll H)$, parametric resonance can be very efficient, as in toy models of chaotic inflation.} Moreover, any other model-dependent coupling to light particles could constitute a viable decay channel, for example massless gauge bosons.

Let us stress that the examples exhibiting this behavior also appear in the classification of section \ref{F-theory_embedding}. However, this does not guarantee that they arise concretely in an actual compactification. While the classification of Section \ref{F-theory_embedding} includes all possible scalar potentials that can be obtained from one-modulus asymptotic limits using Hodge theory, it could be that one case is never geometrically realized as compactification data (manifold, fluxes). In particular, this is true for the $\mathrm{III}_{0,0}$ singularity type example mentioned above, for which no explicit embedding is known to the authors. However, this may only be used to rule out certain specific cases, as the other problematic limit of type \emph{V}$_{1,1}$, corresponding to the LCS point, is known to arise in explicit compactifications. See Section \ref{ref:classification} for a more details on where geometric realizations of the limits can be found.

Finally, it is possible that sub-leading corrections may intervene to spoil this pathologic behavior. Indeed, the leading order approximation \eqref{eq:onesec} to the full scalar potential fails precisely when $w \rightarrow w_0$, as sub-leading corrections to the scalar potential (suppressed by powers of $1/s$) may potentially become relevant when the leading term vanishes. These corrections include both those arising from the large-$s$ expansion of the tree-level, flux scalar potentials (such as those parametrized by the sum over $n$ in \eqref{eq:fullp}) and genuine stringy effects such as $\alpha'$ corrections. While the former can be often ``switched off" by a judicious choice of fluxes, the latter are more general and thus difficult to avoid. \footnote{Unless a cancellation occurs for symmetry reasons.} Indeed, we will show in the next subsection that corrections to the scalar potential which do not vanish for $w=w_0$ will in general cause the oscillating solutions with $T \rightarrow 0$ to be unstable.

\subsubsection{The effect of corrections - an explicit example}\label{sssec:correction}

As a concrete example, we can consider the potential discussed in Section~\ref{ssc:ctrex}, which in the absence of a mechanism dampening the oscillations would have lead to a violation of the dynamical distance conjecture. However, we now include additional corrections to the scalar potential, to show that they would also make the oscillating solution unstable and drive the system towards a late-time solution that does not violate the conjecture. In the case of the LCS point, sub-leading terms in the potential generically appear as a consequence of $\alpha'$ corrections to the K\"ahler potential. The form of such corrections is well-known, also for fourfolds \cite{CaboBizet:2014ovf,Gerhardus:2016iot,Cota:2017aal, vandeHeisteeg:2024lsa}.  In the language of Section \ref{sec:emb_string}, they correspond to a K\"ahler potential of the form 
\begin{eqn}
    K = -\log \left(\frac{2}{3}s^4 -4 s\xi \right),
\end{eqn}where the parameter $\xi$ captures the correction. In geometric examples, it is related to the third Chern class $c_3 (Y_4)$ and intersection number $\kappa$ of the mirror Calabi-Yau fourfold as
\begin{equation}
    \xi = \frac{\zeta(3) c_3(Y_4)}{(2\pi)^3 \kappa}\, ,
\end{equation}
where $\zeta$ denotes the Riemann zeta function. With the particular flux choice leading to \eqref{eq:fc}, this translates into a correction to the potential
\begin{eqn}
\begin{aligned}
    V(s,a) &= g_3^2 \frac{3 \left(a^2+s^2\right) \left(4 a^2 s^6+12 a^2 \xi  s^3+9 \xi ^2 \left(a^2+9
   s^2\right)\right)}{4 \left(s^3-3 \xi \right)^2 \left(s^4+6 \xi  s\right)} \\
   &= g_3^2 \left(a^2+s^2\right) \left(\frac{3 a^2  \left(s^3+3 \xi\right)}{s^7}+ \frac{27 \xi ^2  \left(13 a^2+9 s^2\right)}{4 s^{10}}+  \mathcal{O}\left( (\tfrac{\xi}{s^3})^{3}\right)\right)\, ,  
\end{aligned}
\end{eqn}where in the second line we expanded in $\xi/s^3$. We notice in particular the presence of a term in the potential that does not carry powers of $a$, and hence does not vanish in the limit $a \rightarrow 0$. This is exactly the kind of correction discussed in Section \ref{ssec:generalpolynomial}. In the language of Eq. \eqref{eq:fullp}, it can be recast as 
\begin{eqn}\label{eq:corralpha'}
    V(s,a) \supset   \frac{243 g_3^3 \xi^2}{ s^6} \quad \quad \longrightarrow \quad \quad m=6, \quad \quad P_6(w) = \frac{243}{4}\xi^2 g_3^2.
\end{eqn}By the analysis in Section \ref{ssec:generalpolynomial} (see also Table \ref{tab:sum2}), such a term prevents the existence of the oscillating solutions.

\begin{figure}[h!]
\centering
 \includegraphics[width=0.65\textwidth]{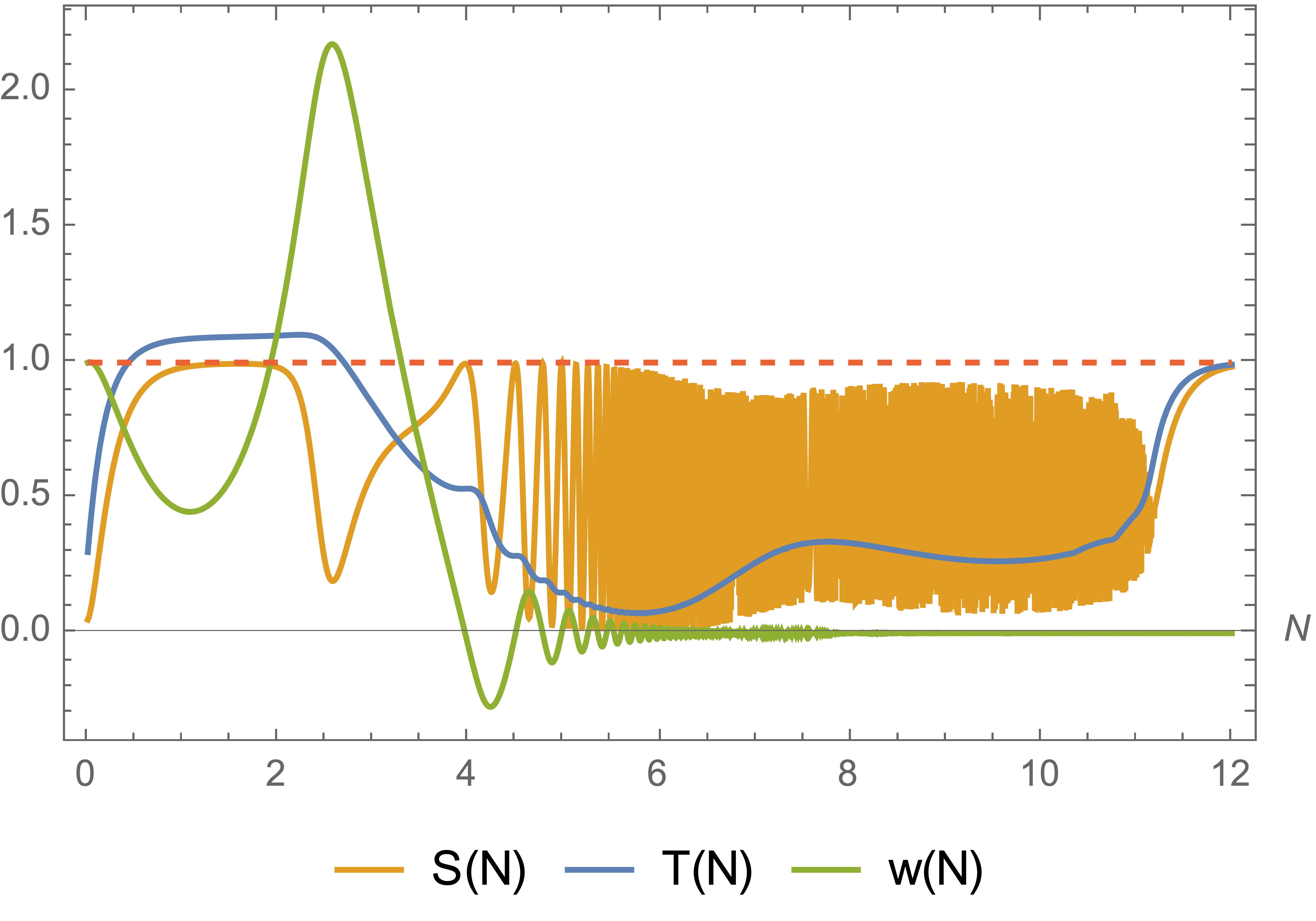}
\caption{Numerical solution of the system \eqref{eq:sysNg} with the potentials \eqref{eq:sp1}. The initial conditions are specified by $x(0)=0.1,y(0)=0.2,w(0)=1,v(0)=20$, and the evolutions is between $N=0$ and $N=12$. The dotted line denotes the asymptotic values of $T$, equal to $1$.}
\label{fig:subl1}
\end{figure}

\noindent We can take this further, and explicitly consider two cases, with potentials of the form \eqref{eq:fc} (with $g_3=1$) plus corrections:
\begin{eqn}\label{eq:sp2}
   \lambda =0 \quad \quad C=2 \quad \quad P_0(w) = 3  w^2 (1+w^2) \quad \quad P_{3}(w) = 1 \quad (m=3),
\end{eqn}and
\begin{equation}\label{eq:sp1}
   \lambda =0 \quad \quad C=3 \quad \quad P_0(w) = 3 w^2 (1+w^2) \quad \quad P_6(w) = 1 \quad (m=6).
\end{equation}
The first potential has been chosen for purposes of illustration only, while the second one corresponds to the $\alpha'$ correction in \eqref{eq:corralpha'}.
\begin{figure}[h!]
\centering
\includegraphics[width=0.65\textwidth]{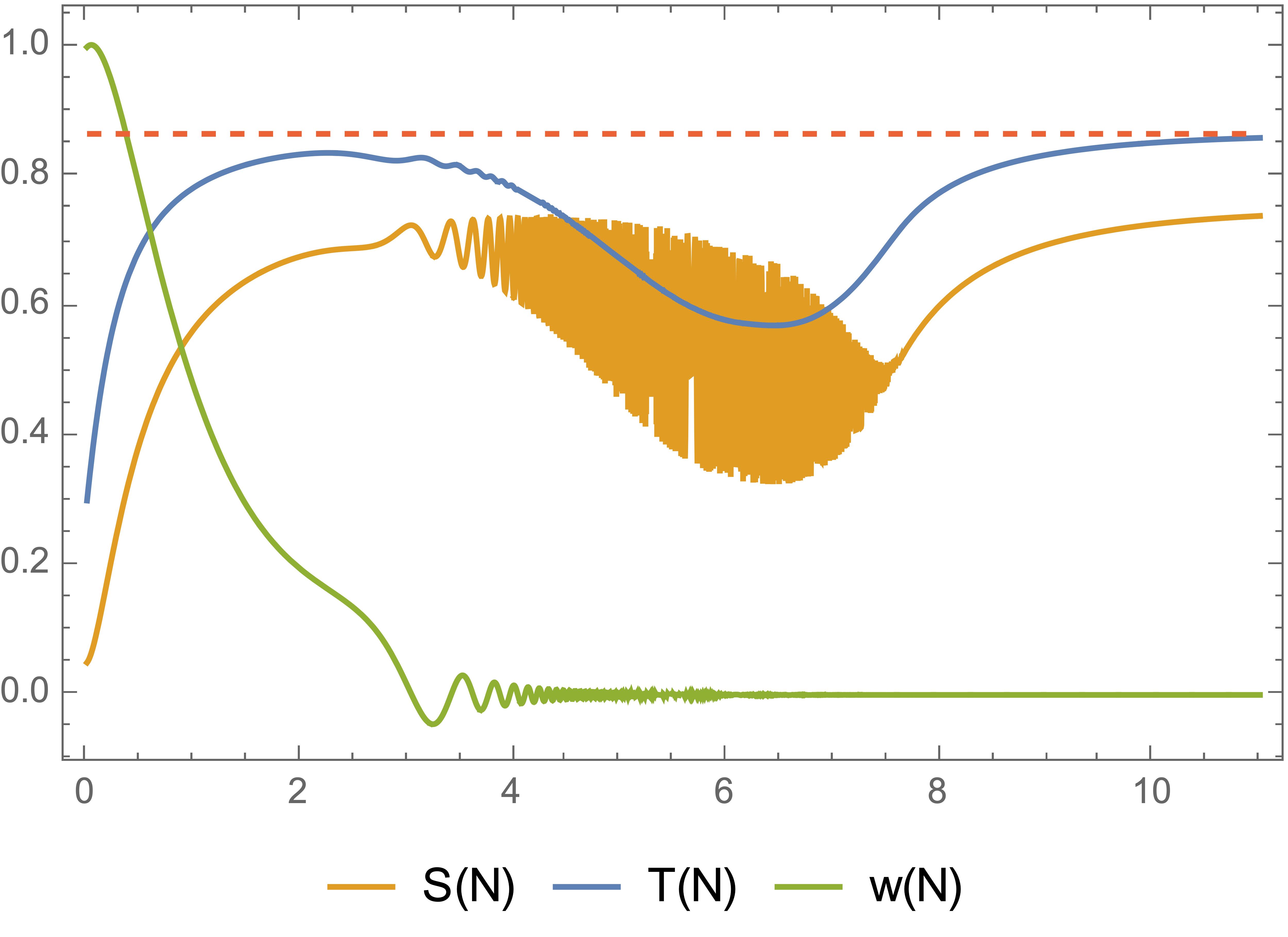}
\caption{Numerical solution of the system \eqref{eq:sysNg} with the potential \eqref{eq:sp2}. The initial conditions are specified by $x(0)=0.1,y(0)=0.2,w(0)=1,v(0)=20$, and the evolutions is between $N=0$ and $N=11$. The dotted line denotes the asymptotic value of $T$, equal to $T_m$.}
\label{fig:subl2}
\end{figure}
As discussed in Section \ref{ssec:generalpolynomial}, the ansatz given in Section~\ref{ssc:ctrex} is not an asymptotic solution once these sub-leading corrections have been taken into account, and in particular $T$ can no longer asymptote to zero. One can also see from Table \ref{tab:sum2} that only fixed points are allowed in this case, which are always compatible with the dynamical version of the distance conjecture according to the discussion in \ref{ssec:dynamicaldistances}. We can verify these statements by solving the equations of motion numerically, as plotted in Figures \ref{fig:subl2} and \ref{fig:subl1}.

In both instances, the solution converges to a fixed point, with $w'=0$. In the first case, $T \rightarrow T_{\lambda+m}=T_m$, as expected on the basis of the considerations outlined before. In the second case, this cannot happen, as $T_m^2 >1$. \footnote{Eq. \eqref{eq:dw} would then imply $S>1$, which is inconsistent.} Then, both $T$ and $S$ converge to $1$, corresponding to pure kination in the asymptotic future. These two cases are exactly the two possibilities discussed at beginning of this section, corresponding to Eqs \eqref{eq:fa1} and \eqref{eq:fa2}, and also the second and third entries of Table \ref{ssec:dynamicaldistances}.  From a phenomenological perspective, it is quite curious to see a (future) attractor corresponding to a kinating modulus.\footnote{However, we expect the inclusion of radiation or matter would modify this conclusion, as they redshift slower than kinetic energy \cite{Copeland:1997et}.} Typically, when axion flat directions are present, kination is always an unstable fixed point \cite{Cicoli:2020cfj,Cicoli:2020noz,Cicoli:2023opf,Revello:2023hro}, and an attractor only in the past \cite{Andriot:2024sif}. 

\section{F-theory embedding and classification of scalar potentials \label{F-theory_embedding}}

In this section we describe the embedding of the axion-scalar potentials in string compactifications. To be precise, we consider F-theory compactifications on Calabi--Yau fourfolds with fluxes. In Section~\ref{sec:emb_string} lay out the general 4d $\mathcal{N}=1$ supergravity framework that underlies this setting. We then specialize in Section~\ref{ref:classification} to F-theory compactifications with a single complex structure modulus, i.e.~$h^{3,1}=1$, and classify all possible asymptotic scalar potentials for the axion-scalar pair. We describe perturbative and non-perturbative corrections to these scalar potentials in Section~\ref{sec:corrections}.

\subsection{Motivation from string theory} \label{sec:emb_string}
The compactification of F-theory on an elliptic Calabi--Yau fourfold with four-form fluxes yields a four-dimensional $\mathcal{N}=1$ supergravity theory. Equivalently, one can think of this setup as a Type IIB Calabi--Yau orientifold with RR and NS-NS three-form fluxes. The advantage of the F-theory picture is that it enables us to go beyond the usual weak Type IIB string-coupling regime and access other asymptotic limits in the $\mathcal{N}=1$ field space. The action of this 4d supergravity theory is given by
\begin{equation}
S= M_{4}^2\int d^4 x \, \sqrt{- g}  \Big[ \tfrac12 \mathcal{R} -K_{I \bar{J}} \, \partial_{\mu} \Phi^I \partial^{\mu} \bar{\Phi}^{\bar{J}}  - V(\Phi, \bar{\Phi} )\Big],
\end{equation}
where $M_4$ denotes the four-dimensional Planck mass. The $\Phi^I$ are complex scalars, and the scalar potential is denoted by $V(\Phi,\bar{\Phi})$. Following the usual supergravity formalism, the metric on this field space is derived from the K\"ahler potential as $K_{I\bar J}= \partial_I \partial_{\bar{J}} K$ and relates with 
 $G_{I\bar J}$ used in \eqref{eq:action} as $G_{I\bar J} = 2 K_{I\bar J}$. 
The scalar potential is obtained through the standard $\mathcal{N}=1$ formula from the superpotential $W$ and $K$ as
\begin{equation}
V(\Phi, \bar{\Phi} )=e^K\left(\sum_{\Phi_I, \Phi_J} K^{I \bar{J}} D_I W D_{\bar{J}} \bar{W}-3|W|^2\right).
\end{equation}
where the sum runs over all moduli $\Phi_I,\Phi_J$. The explicit expression for $K$ and $W$, or equivalently $V$, depends on the choice of compactification manifold and the region in field space. The complex scalars $\Phi_I$ are given by the $h^{1,1}$ K\"ahler moduli and $h^{3,1}$ complex structure moduli. The K\"ahler potential $K$ is given by
\begin{equation}\label{eq:K}
    K = -2\log \mathcal{V}_b - \log \int_{Y_4} \Omega \wedge \bar\Omega\, ,
\end{equation}
where the base volume $\mathcal{V}_b$ in the first term captures the tree-level K\"ahler potential for the complex coordinates $T^\alpha$. The second term is the K\"ahler potential for the complex structure moduli. We emphasize that the volume $\mathcal{V}_b$ may depend on some of the complex structure moduli of $Y_4$. For instance, in the Type IIB orientifold limit it depends on the axio-dilaton, which is one of the complex structure moduli of the Calabi--Yau fourfold; we refer to \cite{Grimm:2019ixq} for a detailed discussion. For the purposes of this work we include a suitable extra term for the complex structure moduli in the asymptotic K\"ahler potential \eqref{eq:Kasymp} to incorporate this effect.

\paragraph{K\"ahler moduli.} Let us first discuss the K\"ahler moduli, even though we focus mostly on the complex structure moduli throughout this work. As one of these K\"ahler moduli parametrizes the volume of the elliptic fiber, and since the F-theory limit sends this to zero size, only $h^{1,1}-1$ K\"ahler moduli $T^\alpha$ remain. The superpotential only depends on these K\"ahler moduli through non-perturbative corrections. Ignoring these terms as they are suppressed, the no-scale property $K^{\alpha \bar\beta}K_\alpha K_{\bar \beta}=3$ tells us that the scalar potential reduces to \cite{Haack:2001jz}
\begin{equation}\label{eq:Vnoscale}
    V = e^K K^{i \bar j}D_i W \overline{D_j W}\, .
\end{equation}
Even though the superpotential $W$ does not depend on the K\"ahler moduli in this approximation, the scalar potential still depends on these $T^\alpha$ through the volume factor $(\mathcal{V}_b)^{-2}$ coming from $e^K$. This dependence can be of great importance in assessing whether cosmological scenarios such as accelerated expansion can happen in string theory: while this has shown to be possible in \cite{Calderon-Infante:2022nxb} when restricting to asymptotic regimes in the complex structure moduli space, it has been pointed out in \cite{Hebecker:2023qke} that the K\"ahler moduli contribution to the slope of $V$ always causes $|\nabla V|/V \geq \sqrt{2}$, and hence do not allow for accelerated expansion \cite{Rudelius:2021oaz, Rudelius:2021azq}. The focus of this work is to study axion-scalar dynamics in the complex structure moduli sector. While the K\"ahler moduli could affect the asymptotic behavior of our cosmological solutions, we find that the complex structure moduli by themselves already yield self-consistent solutions which do not violate the dynamical distance conjecture. Alternatively, one could try to embed these models in non-geometrical backgrounds without any K\"ahler moduli, such as the Type IIB orientifolds of Landau-Ginzburg models pioneered in \cite{Becker:2006ks}. These models were recently revisited to construct the first examples in string theory of fully-stabilized four-dimensional Minkowski vacua \cite{Rajaguru:2024emw, Becker:2024ayh, Chen:2025rkb} and de Sitter saddle points \cite{Chen:2025rkb}.

\paragraph{Superpotential.} After this short digression on the K\"ahler moduli, let us now return to the scalar potential \eqref{eq:Vnoscale}. We consider F-theory compactifications with fluxes, and the four-form flux $G_4$ generates the flux superpotential \cite{Gukov:1999ya}
\begin{equation}
    W = \int_{Y_4} G_4 \wedge \Omega\, .
\end{equation}
The dependence of the flux superpotential on the complex structure moduli comes from the holomorphic $(4,0)$-form $\Omega(\Phi)$. The quantization of the fluxes requires $G_4 \in H^4_{\rm p}(Y_4,\mathbb{Z})$. The fluxes should satisfy the tadpole cancellation condition \cite{Sethi:1996es}
\begin{equation}
    \frac{1}{2}\int_{Y_4} G_4 \wedge G_4 + N_{\rm D3} = \frac{\chi(Y_4)}{24}\, , 
\end{equation}
where $N_{\rm D3}$ denotes the number of mobile D3-branes and $\chi(Y_4)$ the Euler characteristic of $Y_4$. Since this work focuses on general asymptotic regimes in complex structure moduli space, and not any particular models, we will not work out the quantization of the fluxes or the tadpole cancellation condition in detail.

\paragraph{Periods.} Let us now describe the moduli dependence of the flux superpotential, both from general considerations and in asymptotic limits. In order to make the moduli dependence explicit, it is convenient to expand the holomorphic $(4,0)$-form $\Omega$ in a basis of so-called period functions $\Pi^{\mathcal{I}}(\Phi)$. Formally, they are obtained from integrating $\Omega$ over an integral basis of four-cycles $\Gamma_I \in H_{4}(Y_4,\mathbb{Z})$ as
\begin{equation}
    \Pi^I = \int_{\Gamma_I} \Omega \, , \qquad \Omega(\Phi) = \Pi^I(\Phi) \gamma_I\, .
\end{equation}
where $\gamma_I \in H^4_p(Y_4,\mathbb{Z})$ denotes the Poincar\'e dual basis of four-cycles. In practice, the periods $\Pi^I(\Phi)$ are obtained as solutions to a system of linear ordinary differential equations known as the Picard-Fuchs equation, see for instance \cite{vandeHeisteeg:2024lsa} for a recent exposition (supplemented with pedagogical notebooks) of the Calabi--Yau fourfolds with the simplest moduli space $\mathcal{M}_{\rm cs}=\mathbb{P}^1-\{0,1,\infty\}$.

\paragraph{Asymptotic periods.} In this work, we can forego such techniques as we do not need the global dependence of the functions $\Pi^I(\Phi)$ on the complex structure moduli, but just their asymptotic dependence near a boundary in moduli space. In this setting there are general approximation methods available \cite{Schmid, CattaniDeligneKaplan} coming from asymptotic Hodge theory, that have been widely used in the study of string compactifications \cite{Grimm:2018ohb, Grimm:2018cpv, Corvilain:2018lgw, Grimm:2019wtx,Grimm:2019bey, Font:2019cxq, Grimm:2019ixq, Grimm:2020cda, Gendler:2020dfp, Bastian:2020egp, Grimm:2020ouv, Grimm:2021ikg, Bastian:2021eom, Bastian:2021hpc, Palti:2021ubp, Grimm:2021ckh, Bakker:2021uqw, Grimm:2021vpn, Grana:2022dfw, Bastian:2023shf, Grimm:2023lrf, Grimm:2024fip}. For a detailed review we refer the reader to \cite{vandeHeisteeg:2022gsp, Monnee:2024gsq}. It allows us to parametrize any boundary as a limit $\Phi \to i \infty$, where we can decompose this complex field as
\begin{equation}
    \Phi = a+i s\, ,
\end{equation}
such that the ``saxionic" scalar $s \to \infty$ parametrizes the boundary limit. Alternatively, this boundary can be parametrized as the point $e^{2\pi i \Phi}=0$. From this presentation, it is apparent that circling the boundary corresponds to $a \to a +1$, which is why this real field is interpreted as an axion. The period vector admits an asymptotic expansion in terms of these coordinates as 
\begin{equation}\label{eq:piasymp}
    \mathbf{\Pi}(a,s) = e^{(a+is)N} \left( \mathbf{a}_0 + e^{2\pi i (a+is)} \mathbf{a}_1 + \ldots \right)\, ,
\end{equation}
which is known as the nilpotent orbit approximation \cite{Schmid}. The matrix $N$ here is nilpotent $N^d\neq 0$, $N^{d+1}=0$ for some integer, with $d \leq 4$ for Calabi--Yau fourfolds. When circling the boundary $a\to a+1$ this induces a monodromy transformation $\mathbf{\Pi}(a+1,s)=  e^N \mathbf{\Pi}(a,s)$.\footnote{Monodromies can be decomposed into a semisimple part of finite order $T_{ss}$ and a unipotent part $T_u = e^N$. In this work we only consider the unipotent part of infinite order, and assume that the finite order part $T_{ss}$ has been removed by a coordinate redefinition. }The expansion in $e^{-2\pi s}$ can be understood in physical terms as an instanton expansion. In fact, when the boundary is a large complex structure point, these exponential corrections can be understood through mirror symmetry \cite{Candelas:1990rm} as the worldsheet instantons that correct the dual K\"ahler moduli space. 

\paragraph{Asymptotic K\"ahler potential and metric.} Let us now put these observations about the asymptotic periods to use in the description of the physical couplings in these regimes. Let us begin with the K\"ahler potential. We define the integer $d_0$ as the integer such that $N^{d_0} \mathbf{a}_0 \neq 0$, $N^{d_0+1}\mathbf{a}_0=0$. Oftentimes it coincides with the nilpotency degree $d=d_0$ of $N$. However, for particular boundaries one can have $d>d_0$, which happens for instance at finite distance points, since these have $d_0=0$. We focus our attention on infinite distance boundaries with $d_0 > 0$, for which we have as K\"ahler potential and metric
\begin{equation}\label{eq:Kasymp}
    K = - (\lambda+d_0) \log s\, , \qquad K_{ss}=K_{aa} = \partial_{\Phi}\partial_{\bar \Phi} K= \frac{\lambda+d_0}{s^2}\, .
\end{equation}
We included here an additional term $-\lambda \log s$ that may come from the volume $\mathcal{V}_b$ in the K\"ahler potential \eqref{eq:K}. For the Type IIB orientifold limit one has $\lambda=3$ \cite{Grimm:2019ixq}, but we choose to keep it arbitrary to encompass also other limit types. In comparison with \eqref{hyper_metric}, notice that the boundary data fixes the leading coefficient of the hyperbolic metric as $C=\lambda+d_0$.

\paragraph{Asymptotic scalar potential.} Let us next turn to the asymptotic form of the scalar potential. Since the periods of the $(4,0)$-form admit the boundary expansion \eqref{eq:piasymp}, a similar expansion holds for all derivatives of the period vector, up to suitable rescalings. In particular, we may therefore expand the F-terms $D_IW$ appearing in the definition \eqref{eq:Vnoscale} of the scalar potential in the same way. This gives rise to a bilinear form for this scalar potential in terms of so-called flux-axion polynomials, which has readily been observed in various limits in \cite{Herraez:2018vae, Grimm:2019ixq,Marchesano:2021gyv}. We define these flux-axion polynomials as
\begin{equation}
    \rho(a) = e^{-a N} q\, , 
\end{equation}
where we expanded the flux quanta in a basis as $\int_{\Gamma_\mathcal{I}}G_4 = q^{\mathcal{I}}$. The scalar potential then takes the bilinear form
\begin{equation}\label{eq:VM}
    V(s,a) = s^{-\lambda} \rho^T(a) \mathcal{M}(a,s) \rho(a)\, ,
\end{equation}
where we explicitly kept track of the factor $s^{-\lambda}$ coming from the additional term in the K\"ahler potential \eqref{eq:Kasymp} that models the dependence of $\mathcal{V}_b$ on the complex structure moduli. The coupling matrix $\mathcal{M}(a,s)$ is periodic in the axions, i.e.~$\mathcal{M}(a+1,s)=\mathcal{M}(a,s)$, while the monodromy behavior of the fluxes under $a \to a +1$ is captured by $\rho(a+1) = e^{-N} \rho(a)$. We can expand the coupling matrix $\mathcal{M}(a,s)$ in the limit $s\to \infty$ as a series of exponential corrections
\begin{equation}
    \mathcal{M}(a,s) = \sum_{n=0}^\infty e^{-2\pi n s} \mathcal{M}_n(a,s)\, ,
\end{equation}
where the terms $\mathcal{M}_n(a,s)$ are algebraic functions in $s$, e.g.~ratios of polynomials. On the other hand, the dependence on $a$ is given by periodic functions $\cos(2\pi m a)$ and $\sin(2\pi m a)$, where the period $m \in \mathbb{N}$ is bounded by the order of the exponential correction as $m\leq n$. In particular, the `perturbative' term $\mathcal{M}_0(a,s)=\mathcal{M}(s)$ is therefore independent of the axion $a$ and its entries are given by algebraic functions in $s$ only.

\paragraph{Sl(2)-approximated scalar potential.} Let us now focus on the perturbative part of the scalar potential given by $\mathcal{M}_0(s)$. In general the dependence on $s$ is still rather involved, but we can systematically break this down through the so-called sl(2)-orbit approximation \cite{CattaniDeligneKaplan}. While we refer the reader to e.g.~\cite{Grimm:2019ixq, Grimm:2020ouv, Grimm:2023lrf} for detailed treatments of this approximation, especially in light of scalar potentials, let us here summarize the main takeaways. The idea is to expand the algebraic functions appearing in $\mathcal{M}_0(s)$ for large $s\gg 1$. However, since we are dealing with a matrix-worth of such functions, and some of the eigenvalues may be parametrically larger in $s$ than the others, we cannot simply take the leading part of each entry. Instead, we have to work in a suitable eigenbasis of the matrix $\mathcal{M}_0(s)$, and this is precisely what is provided by the sl(2)-approximation. It splits the middle cohomology as
\begin{equation}
    H^4_{p}(Y_4,\mathbb{R}) = \bigoplus_{l=-d}^d V_l\, ,
\end{equation}
where the spaces $V_l$ are eigenspaces under the weight operator of the sl(2)-triple. The log-monodromy operator acts as a lowering operator $N V_l \subseteq V_{l-2}$ on these eigenspaces. Let us then decompose the vector of flux-axion polynomials $\rho(a) \in H^4_{p}(Y_4,\mathbb{R})$ in terms of this basis as $\rho(a) = \sum_l \rho_l(a)$. Then the scalar potential in the sl(2)-approximation reads
\begin{equation}\label{eq:Vrho}
    V_{\rm sl(2)}(a,s) = s^{-\lambda} \sum_l s^l \, \rho_l(a) \mathcal{M}_{\infty} \rho_{l}(a)\, .
\end{equation}
where the matrix $\mathcal{M}_\infty$ captures the $\mathcal{O}(1)$-coefficients of the leading part of the scalar potential. In particular, we used that the matrix $\mathcal{M}_\infty$ vanishes for products between elements of different eigenspaces. Let us now make the axion-dependence explicit. To gain some intuition, we specialize first to the case of a single flux component $G_4 = G_4^l$ and turn off all other flux quanta. By expanding the flux-axion polynomials we find
\begin{equation}\label{eq:Vsingle}
    V(s,a) = s^{-\lambda} \sum_{k=0}^{\lfloor \frac{l-d}{2} \rfloor } s^l  \left(\frac{a}{s}\right)^{2k} \, (N^k G_4^l)^T \mathcal{M}_\infty (N^k G_4^l)\, ,
\end{equation}
where the sum runs up to the largest integer $k \leq  \tfrac{l-d}{2}$, beyond which $N^k G_4^l$ has a too low weight. This form of the axion-scalar potential matches precisely with the polynomial expression $V(s,a) = s^{l-\lambda} P(\tfrac{a}{s})$ given in \eqref{eq:onesec} for a single sector. Let us next generalize to the case of multiple flux components $G_4 = \sum_l G_4^l$. We write out the scalar potential as
\begin{equation}\label{eq:Vgen}
    V(s,a) = s^{-\lambda} \sum_{l=-d}^d \sum_{l'=-d}^d \sum_{k=0}^{\lfloor \frac{d+l}{2} \rfloor }\sum_{k'=0}^{\lfloor \frac{d+l'}{2} \rfloor } s^{\frac{l+l'}{2}} \left(\frac{a}{s}\right)^{k+k'}\ \frac{1}{k!k'!} (N^kq_l)^T \mathcal{M}_\infty N^{k'}q_{l'}\, ,
\end{equation}
where we recall that the orthogonality properties of $\mathcal{M}_\infty$ imply that only terms with $l-2k=l'-2k'$ are non-vanishing. In writing \eqref{eq:Vgen} we already chose to describe the dependence on the axions through the ratio $a/s$. Also notice that the scaling in $s$ of the terms is given by $s^{l-2k}$, which matches with \eqref{eq:Vrho} according to the weight of $N^k q_l$. We can encode the dependence on the axion-scalar ratio in terms of the polynomials
\begin{equation}
    P_n\left(\frac{a}{s}\right) = \sum_{l=-d}^d \sum_{k=0}^{\lfloor \frac{d+l}{2} \rfloor }\sum_{k'=0}^{\lfloor \frac{d-l}{2} \rfloor-n} \left(\frac{a}{s}\right)^{k+k'}\ \frac{1}{k!k'!} (N^kq_l)^T \mathcal{M}_\infty N^{k'}q_{-2n-l}\, ,
\end{equation}
where the degree of these polynomials is bounded by $k+k' \leq d-n$.
The scalar potential can be rewritten in terms of the $P_n(\tfrac{a}{s})$ in the simple form
\begin{equation}
    V(a,s) = s^{-\lambda} \sum_{n=-d}^{d} \frac{1}{s^n} P_n\left(\frac{a}{s}\right)\, .
\end{equation}
This expression of the axion-scalar potential in terms of polynomials in $\tfrac{a}{s}$ matches precisely with \eqref{eq:fullp_e} described in Section \ref{sec:gen_action+potential}. 

In the following two subsections we classify this axion-scalar potential for all possible boundaries in complex structure moduli spaces of dimension $h^{3,1}=1$. In the first subsection \ref{ref:classification} we give the leading part of the axion-scalar potential, i.e.~the sl(2)-approximated version \eqref{eq:Vgen}. In the second subsection \ref{sec:corrections} we describe polynomial and exponential corrections to these axion-scalar potentials, i.e.~\eqref{eq:VM} where $\mathcal{M}(a,s)$ is approximated to its full polynomial part $\mathcal{M}_0(s)$ and some particular corrections in $\mathcal{M}_{1,2}(a,s)$. All of these models are obtained by using asymptotic Hodge theory techniques; this systematic analysis is detailed in appendix \ref{app:asympHodge}.

\subsection{Classification of axion-scalar potentials}\label{ref:classification}
In this section we classify the axion-scalar potential for all possible boundaries in complex structure moduli spaces of dimension $h^{3,1}=1$. We refer to appendix \ref{app:asympHodge} for the detailed construction of these asymptotic models. Here we summarize the axion-scalar dependence of the physical couplings across the possible types of boundaries, providing the asymptotic K\"ahler potential and scalar potential. We refer to the next subsection \ref{sec:corrections} for a detailed description of the couplings including perturbative and exponential corrections.

\paragraph{Limit types.} Let us begin by enumerating the possible types of boundaries. These are conveniently labeled by the type of limiting mixed Hodge structure arising at the singularity following \cite{Grimm:2019ixq}. We detailed the correspondence between these singularity types and the Hodge-Deligne diamonds in \eqref{eq:singtypes}. For the purpose of the discussion here, we simply recall the types: $\mathrm{I}_{0,1}$, $\mathrm{I}_{1,1}$, $\mathrm{II}_{0,0}$, $\mathrm{III}_{0,0}$, and $\mathrm{IV}_{1,1}$. The singularities of type $\mathrm{I}$ are of finite distance, while the limit types $\mathrm{II}_{0,0}$, $\mathrm{III}_{0,0}$, and $\mathrm{V}_{1,1}$ are all of infinite distance. 

\paragraph{Type $\mathrm{I}_{0,1}$ limit.} We begin with the finite distance boundary of type $\mathrm{I}_{0,1}$. The K\"ahler potential for this boundary is given by
\begin{equation}
   K = - \log\big( 2 - 4 |A|^2  e^{-4\pi s} s \big)
\end{equation}
The fact that this boundary at $s \to \infty$ is of finite distance can be seen through the exponential dependence on this scalar. The scalar potential is also readily computed, and reads
\begin{equation}\label{eq:V01}
    V_{\rm pol}(a,s)  = \rho(a)^T \begin{pmatrix}
         s & 0 & 0 & -1 \\
         0 & s & 1 & 0 \\
        0 & 1 & 1/s & 0 \\
         -1 & 0 & 0 & 1/s \
    \end{pmatrix}\rho(a)
\end{equation}
where the flux-axion polynomials $\rho(a)$ are given by
\begin{equation}
    \rho(a) = \begin{pmatrix}
        g_3 \\
        g_4 \\
        g_5-ag_3 \\
        g_6-a g_4 \\
    \end{pmatrix}\, ,
\end{equation}
where $g_1,\ldots,g_6 \in \mathbb{R}$ denote the flux quanta. The flux quanta $g_1,g_2$ do not show up in the scalar potential at the polynomial level, but only appear starting with the exponential corrections, cf.~\eqref{eq:VexpI01}.

\paragraph{Type $\mathrm{I}_{1,1}$ limit.} The other one-modulus boundary of finite distance is of type $\mathrm{I}_{1,1}$. The K\"ahler potential for this boundary reads
\begin{equation}
K = - \log\big(2-2 |A|^2 e^{-4 \pi  s} s^2 \big)\, ,
\end{equation}
where again $A\in \mathbb{C}$ denotes a model-dependent parameter. The scalar potential for this boundary is given by
\begin{equation}
    V_{\rm pol}(a,s) = \rho^T(a)\left(\begin{array}{ccc}
\frac{s^2}{4} & 0 & -\frac{1}{2} \\
0 & 1 & 0 \\
-\frac{1}{2} & 0 & \frac{1}{s^2} \\
\end{array}
\right)\rho(a) \, .
\end{equation}
The flux-axion polynomials $\rho(a)$ are given by
\begin{equation}
    \rho(a) = \begin{pmatrix}
        g_3 \\
        g_4 - a g_3 \\
        g_5-a g_4 + \tfrac{1}{2}a^2 g_3\\
    \end{pmatrix}
\end{equation}
where $g_1,\ldots,g_5 \in \mathbb{R}$ denote the flux quanta. Similar to the type $\mathrm{I}_{0,1}$ boundary, the flux quanta $g_1,g_2$ do not appear yet at the polynomial level in the scalar potential. In further comparison to the type $\mathrm{I}_{0,1)}$ boundary, we also find that $\mathrm{I}_{1,1}$ admit quadratic terms in $s$ and $a$ rather than the linear terms appearing in \eqref{eq:V01}. This reflects that $\mathrm{I}_{1,1}$ is a more severe type of singularity.

\paragraph{Type $\mathrm{II}_{0,0}$ limit.} We next move on to the infinite distance boundaries. We start with the weakest type of singularity given by a $\mathrm{II}_{0,0}$ limit. The K\"ahler potential for this boundary is given by
\begin{equation}
   K = - \log(4s)\, .
\end{equation}
From the polynomial behavior in $s$ it follows that this singularity is at infinite distance with the usual hyperbolic metric. The scalar potential for this boundary reads
\begin{equation}\label{eq:VII00}
    V_{\rm pol}(a,s) = \rho^T(a) \left(\begin{array}{cccc}
 s & 0 & 0 & 1  \\
 0 & s & -1 & 0  \\
 0 & -1 & \frac{1}{s} & 0  \\
 1 & 0 & 0 & \frac{1}{s} \\
\end{array}
\right) \rho(a)\, .
\end{equation}
The flux-axion polynomials read
\begin{equation}
    \left(
\begin{array}{c}
 g_1 \\
 g_2 \\
 g_3-a g_1 \\
 g_4-a g_2 \\
\end{array}
\right),
\end{equation}
where $g_1,\ldots, g_4 \in \mathbb{R}$ denote the flux quanta. The types of terms that can arise in $V_{\rm pol}(a,s)$ are similar to the scalar potential \eqref{eq:V01} of the finite distance boundary $\mathrm{I}_{0,1}$. The difference between these boundaries lies in the kinetic terms for the axion-scalar, since we are now dealing with an infinite distance boundary.

\paragraph{Type $\mathrm{III}_{0,0}$ limit.} We next turn to the $\mathrm{III}_{0,0}$ limit. The asymptotic K\"ahler potential is given by
\begin{equation}
    K = - \log\big(4s^2 \big)\, ,
\end{equation}
from which we immediately see again that it is an infinite distance boundary. The polynomial part of the scalar potential is given by
\begin{equation}\label{eq:VIII00}
    V_{\rm pol}(a,s) = \rho^T(a)\left(
\begin{array}{cccccc}
 \frac{s^2}{2} & 0 &  0 & 0 & 1 & 0  \\
 0 & \frac{s^2}{2} &  0 & 0 & 0 & 1  \\
 0 & 0 & 0 & 0 & 0 & 0 \\
 0 & 0 & 0 & 0 & 0 & 0 \\
 1 & 0 & 0 & 0 & \frac{2}{s^2} & 0  \\
 0 & 1 &  0 & 0 & 0 & \frac{2}{s^2} \\
\end{array}
\right)\rho(a)\, ,
\end{equation}
where the flux-axion polynomials are given by
\begin{equation}\label{eq:III00}
\rho(a) = \left(\begin{array}{c}
 g_1 \\
 g_2 \\
 g_3-a g_1 \\
 g_4-a g_2 \\
 \frac{a^2 g_1}{2}-a g_3+g_5 \\
 \frac{a^2 g_2}{2}-a g_4+g_6 \\
\end{array}\right)\, ,
\end{equation}
where $g_1,\ldots,g_6 \in \mathbb{R}$ denote the flux quanta. Due to the vanishing of the third and fourth rows and columns in \eqref{eq:VIII00}, the flux quanta $g_3,g_4$ only appear together multiplied by the axion $a$ in the scalar potential. The fact that this is a stronger singularity type than the $\mathrm{II}_{0,0}$ considered before in \eqref{eq:VII00} also allows for quadratic terms in $a,s$ rather than only linear scalings.

\paragraph{Type $\mathrm{V}_{1,1}$ limit.} We next turn to the strongest type of singularity. This boundary type $\mathrm{V}_{1,1}$ corresponds geometrically to a large complex structure limit for the Calabi--Yau fourfold. The K\"ahler potential depends quartically on the saxion as
\begin{equation}
    K =- \log \big( \tfrac{2}{3}s^4 \big)\, .
\end{equation}
The scalar potential is given by
\begin{equation}\label{eq:VV11}
    V_{\rm pol}(a,s) = \rho^T(a) \left(
\begin{array}{ccccc}
 \frac{y^4}{24} & 0 & 0 & 0 & -1 \\
 0 & \frac{y^2}{6} & 0 & -1 & 0 \\
 0 & 0 & 0 & 0 & 0  \\
 0 & -1 & 0 & \frac{6}{y^2} & 0  \\
 -1 & 0 & 0 & 0 & \frac{24}{y^4}  \\
\end{array}
\right) \rho(a)\, ,
\end{equation}
where the flux-axion polynomials are given by
\begin{equation}\label{eq:V11}
    \rho(a) = \left(
\begin{array}{c}
 g_1 \\
 g_2-a g_1 \\
 \frac{a^2 g_1}{2}-a g_2+g_3 \\
 \frac{a^3 g_1}{6}-\frac{a^2 g_2}{2}+a g_3+g_4 \\
 \frac{a^4 g_1}{24}-\frac{a^3 g_2}{6}+\frac{a^2 g_3}{2}+a g_4+g_5 \\
\end{array}
\right)\, ,
\end{equation}
with flux quanta $g_1,\ldots, g_5 \in \mathbb{R}$. Similar to the $\mathrm{III}_{0,0}$ boundary, due to the vanishing of the third row and column in \eqref{eq:VV11}, the flux quantum $g_3$ now appears only when multiplied by the axion $a$ in the scalar potential.

\subsection{Perturbative and exponential corrections}\label{sec:corrections}
In this section we give the corrections to the scalar potentials from the previous subsection. For the detailed construction of this data we refer to appendix \ref{app:asympHodge}. For the boundaries of type $\mathrm{I}_{0,1}, \mathrm{I}_{1,1}, \mathrm{II}_{0,0}$ there are no perturbative corrections, i.e.~extra polynomial terms in $s$, so we give the first exponential corrections. On the other hand, for the boundaries of type $\mathrm{III}_{0,0}$ and $\mathrm{V}_{1,1}$ perturbative corrections do appear, so for brevity we exclude the exponential corrections. We stress that exponential corrections can also appear for these latter two boundaries, and we characterize them through the periods in appendix \ref{app:asympHodge}.\footnote{In fact, for the boundaries $\mathrm{I}_{0,1}, \mathrm{I}_{1,1}, \mathrm{II}_{0,0}$, and $\mathrm{III}_{0,0}$ the exponential corrections to the periods are essential in the sense of \cite{Bastian:2021eom}: they can never vanish, as otherwise the derivatives of the (4,0)-form cannot span the full middle cohomology.}

\paragraph{Type $\mathrm{I}_{0,1}$ boundary.} In this case, we found that the scalar potential \eqref{eq:V01} at polynomial level did not involve the flux quanta $g_1,g_2 \in \mathbb{R}$. To investigate what sort of scalar potential they induce, we set $g_3=\ldots=g_6=0$ for simplicity. We then find an exponentially suppressed potential
\begin{equation}\label{eq:VexpI01}
    V_{\rm exp}(s,a) = e^{-4\pi s}(g_1^2+g_2^2)\frac{(1+4\pi s)^2}{4\pi^2 s}|A|^2\, ,
\end{equation}
where $A \in \mathbb{C}$. Exponentially suppressed terms for $g_3,\ldots,g_6$, or mixed terms between $g_3,\ldots, g_6$ and $g_1,g_2$, are generically present at order $e^{-2\pi s}$. While we do not give their explicit form here, they may be computed using the period expressions in \ref{app:asympHodge}, and they are readily available in the attached notebook.

\paragraph{Type $\mathrm{I}_{1,1}$ boundary.} Here we found that the scalar potential \eqref{eq:V11} at polynomial level did not involve the flux quanta $g_1,g_2 \in \mathbb{R}$. Similar to the previous example, let us therefore set $g_3=\ldots=g_5=0$, so we can focus on the exponentially suppressed potential that is induced by these fluxes. This scalar potential takes the form
\begin{align}
       V_{\rm exp}(s,a) = &e^{-4\pi s} |A|^2  (g_1^2+g_2^2) \frac{p(s)^2+1}{64 \pi^4 s^2}\\   
   &+ e^{-4\pi s} |A|^2  \frac{p(s)}{32 \pi^4 s^2} \begin{pmatrix}
        g_1 & g_2
    \end{pmatrix} \begin{pmatrix}
        \cos 4\pi a & \sin 4 \pi a\\
        \sin 4 \pi a & -\cos 4\pi a
    \end{pmatrix}\begin{pmatrix}
        g_1 \\
        g_2 
    \end{pmatrix} +\mathcal{O}(e^{-6\pi s})\, ,\nonumber
\end{align}
where we defined the function
\begin{equation}
    p(s) = 1+4\pi s(1+2\pi s)\, .
\end{equation}
Exponentially suppressed terms for products of the flux quanta $g_3,g_4,g_5$ also appear at order $e^{-4\pi s}$, while mixed terms between $g_3,g_4,g_5$ and $g_1,g_2$ are generically present at order $e^{-2\pi s}$. We do not include their explicit form here, but they may be computed from the period expressions in appendix \ref{app:asympHodge} that have been given up to sufficient order; they are also readily available in the attached notebook.

\paragraph{Type $\mathrm{II}_{0,0}$ boundary.} Here we found that the scalar potential \eqref{eq:VII00} at polynomial level did not involve the flux quantum $g_5 \in \mathbb{R}$. Let us therefore set $g_1=\ldots=g_4=0$, so we can focus on the potential induced by $g_5$. It reads
\begin{equation}
    V_{\rm exp}(a,s) = e^{-4\pi s} |A|^2 g_5^2 \frac{(1+4\pi s)^2}{2\pi^2 s}\, .
\end{equation}
Exponentially suppressed terms for products of the flux quanta $g_1,\ldots, g_4$ also appear at order $e^{-4\pi s}$, while mixed terms between $g_1,\ldots , g_4$ and $g_5$ are generically present at order $e^{-2\pi s}$. Their explicit form may be extracted from appendix \ref{app:asympHodge}, and are readily calculated in the attached notebook.

\paragraph{Type $\mathrm{III}_{0,0}$ boundary.} We next consider the corrections to the potential \eqref{eq:VIII00}. In contrast to the previous three boundaries, this boundary does have perturbative corrections. We therefore focus exclusively on these polynomial corrections; we note that the exponential corrections may be extracted from the periods in appendix \ref{app:asympHodge}, and are readily available in the attached notebook. The scalar potential with all polynomial corrections reads
\begin{equation}
    V(s,a) = \frac{1}{s^4-\xi^2}\rho^T(a) \left(
\begin{array}{cccccc}
 \frac{s^6}{2} & 0 & 0 & -s^3\xi  & s^4 & 0 \\
 0 & \frac{s^6}{2} & \xi  s^3 & 0 & 0 & s^4 \\
 0 & \xi  s^3 & 2 \xi ^2 & 0 & 0 & 2 \xi  s \\
 -s^3\xi & 0 & 0 & 2 \xi ^2 & -2 \xi  s & 0 \\
 s^4 & 0 & 0 & -2 \xi  s & 2 s^2 & 0 \\
 0 & s^4 & 2 \xi  s & 0 & 0 & 2 s^2 \\
\end{array}
\right)  \rho(a)\, ,
\end{equation}
where $\xi \in \mathbb{R}$ parametrizes the correction. Indeed, when we set $\xi=0$, note that we recover the leading potential given in \eqref{eq:VIII00}.

\paragraph{Type $\mathrm{V}_{1,1}$ boundary.} Similar to the $\mathrm{III}_{0,0}$ boundary, here we have polynomial corrections, so we exclusively focus on those. The corrected scalar potential takes the form
\begin{equation}
    V(s,a) = \rho^T(a) \scalebox{0.85}{$\left(
\begin{array}{ccccc}
 \frac{s r(s)^2}{96 p(s) q(s)^2} & 0 & -\frac{9 \xi  s^2 r(s)}{8 p(s) q(s)^2} & 0 & -\frac{216 \xi
   ^3+4 s^9+135 \xi ^2 s^3}{4 p(s) q(s)^2} \\
 0 & \frac{p(s)}{6s} & 0 & -1 & 0 \\
 -\frac{9 \xi  s^2 r(s)}{8 p(s) q(s)^2} & 0 & \frac{243 \xi ^2 s^3}{2 p(s) q(s)^2} & 0 & \frac{27
   \xi  s \left(3 \xi +2 s^3\right)}{p(s) q(s)^2} \\
 0 & -1 & 0 & \frac{6 s}{p(s)} & 0 \\
 -\frac{216 \xi ^3+4 s^9+135 \xi ^2 s^3}{4 p(s) q(s)^2} & 0 & \frac{27 \xi  s \left(3 \xi +2
   s^3\right)}{p(s) q(s)^2} & 0 & \frac{6 \left(3 \xi +2 s^3\right)^2}{s p(s) q(s)^2 } \\
\end{array}
\right) $}\rho(a) 
\end{equation}
where we defined the functions
\begin{equation}
    p(s) = s^3+6\xi\, , \quad q(s) = s^3-3\xi\, , \quad r(s) = 2 s^6-3 \xi  s^3+72 \xi ^2\, .
\end{equation}
Alternatively, it may be expanded in $\xi$ up to second order as
\begin{equation}
    V(s,a) = \rho^T(a) \scalebox{0.85}{$\left(
\begin{array}{ccccc}
 \frac{s^4}{24}+\frac{135 \xi ^2}{32 s^2}-\frac{\xi  s}{8} & 0 & \frac{27 \xi ^2}{8 s^4}-\frac{9 \xi
   }{4 s} & 0 & -\frac{243 \xi ^2}{4 s^6}-1 \\
 0 & \frac{s^2}{6}+\frac{\xi }{s} & 0 & -1 & 0 \\
 \frac{27 \xi ^2}{8 s^4}-\frac{9 \xi }{4 s} & 0 & \frac{243 \xi ^2}{2 s^6} & 0 & \frac{81 \xi
   ^2}{s^8}+\frac{54 \xi }{s^5} \\
 0 & -1 & 0 & \frac{216 \xi ^2}{s^8}-\frac{36 \xi }{s^5}+\frac{6}{s^2} & 0 \\
 -\frac{243 \xi ^2}{4 s^6}-1 & 0 & \frac{81 \xi ^2}{s^8}+\frac{54 \xi }{s^5} & 0 & \frac{702 \xi
   ^2}{s^{10}}+\frac{72 \xi }{s^7}+\frac{24}{s^4} \\
\end{array}
\right)$}\rho(a)\, .\nonumber
\end{equation}
The term in the middle row and column corresponds precisely to the correction discussed in Section \ref{sssec:correction} that disrupts the infinite oscillation. In explicit geometrical examples, the parameter $\xi$ is related to the ratio of the integrated third Chern class of the mirror Calabi-Yau fourfold and its intersection number
\begin{equation}
    \xi = \frac{\zeta(3) c_3}{8\pi^3 \kappa}\, .
\end{equation}
For the scalar potential with exponential corrections, we refer to the attached notebook, where the period expansions given in appendix \ref{app:asympHodge} have also been constructed.

\section{Conclusions}\label{sec:con}

In this paper we have studied the cosmological dynamics induced by the evolution of a saxion-axion pair, interacting through a hyperbolic target space metric and a wide class of polynomial scalar potentials. 
While our setup can be motivated phenomenologically, it also provides a controlled arena to explore a dynamical version of the Distance Conjecture. In making this connection, we emphasized that such axion–saxion systems naturally arise in one-modulus asymptotic limits of string theory: the pair can be identified with a complex-structure modulus in type IIB/F-theory flux compactifications, where they form a complex scalar in an $\mathcal{N}=1$ multiplet. This embedding further justifies the class of scalar potentials we considered, as shown in Section \ref{F-theory_embedding}, where we demonstrated that they encompass the one-modulus, asymptotic scalar potentials generated by F-theory with four-form fluxes. The complete classification of such potentials constitutes an additional general result of this work.

 In the context of phenomenology, infinite distance limits have long been considered a hallmark of weakly-coupled physics and large hierarchies, which both seem to play an important role in our observed universe. From a more conceptual perspective, infinite distance limits in String Theory have played a central role in the study of general Quantum Gravity properties such as the distance conjecture. A primary goal of this work was to extend similar considerations to the case of dynamical backgrounds relevant to cosmology. The latter are characterised by explicitly time-dependent configurations, and field-space trajectories that are no longer geodesic. In concrete terms, we verified whether a plausible extension of the distance conjecture that has been suggested in the literature indeed holds in all 1-modulus asymptotic limits. Such a generalisation, which we refer to as the Dynamical Distance Conjecture, postulates that the towers of states become exponentially light in terms of ``traversed" distance, \emph{i.e.}~the length of the actual physical trajectory. 

By reframing the equations of motion as a dynamical system, we were able to provide in Section \ref{sec:Cosm_solutions} a classification of all possible late-time solutions for the scalar field dynamics, which we summarised in Section \ref{ssec:summary}. Aside from the existence of well-known scaling solutions, corresponding to fixed points of the dynamical system, we uncovered a new class of solutions characterised by infinite oscillations of constant amplitude. While the more standard scaling solutions satisfy the dynamical version of the distance conjecture mentioned in the previous paragraph, this is apparently not true for the oscillating ones. However, we presented arguments (in Section \ref{ssec:dynamicaldistances}) as to why in realistic examples the oscillating solutions cannot be the true asymptotic attractor, either from the effect of higher-order corrections (\emph{e.g.}~from the $\alpha'$ expansion) or through more physical mechanisms involving the decay of the oscillating field. 

Our analysis has been carried out for an axion-scalar system, with a well-motivated choice of coupling functions. It would be interesting to investigate whether the dynamical distance conjecture carries over to more general string theoretic settings. We conclude by listing a number of open questions and promising avenues for future research in this direction.

\begin{itemize}
    \item One important open question is to further study oscillating solutions. The most robust argument to exclude such solutions in Section \ref{ssc:ctrex} is based on the existence of higher order $\alpha'$ corrections to the K\"ahler potential. Such corrections can vanish in the case of compactification manifolds with a high degree of symmetry (such as tori), where $c_3(Y_4)=0$. However, we also expect such cases be characterised by a larger amount of supersymmetry, thus implying the existence of other light moduli which we have not taken into account, thus invalidating the analysis. It could be instructive to verify this in explicit examples.

    \item A natural, further step would be to extend our same conclusions in the presence of multiple moduli. For instance, one might consider two-moduli asymptotic limits, studied \emph{e.g.}~in \cite{Grimm:2018ohb,Grimm:2020ouv,vandeHeisteeg:2022gsp}. It has already been shown that (without considering the axions) such set-ups admit richer cosmological dynamics than the one-modulus case, such asymptotic accelerated expansion \cite{Calderon-Infante:2022nxb}. It is therefore plausible that our classification may have to be extended, and contain qualitatively different classes of solutions. Moreover, we have also not included K\"ahler moduli into the picture, which cannot be stabilised by fluxes in F-theory. They will generally give rise to a multiplicative factor of the volume in the scalar potential, and induce a new runaway direction, potentially spoiling our analysis.
    \item On a similar note, it could be interesting to extend our techniques to the ``more singular" cases (both for one and two moduli) where the leading contribution to the K\"ahler potential vanishes, and instanton corrections dominate. This would result in a target space metric that is no longer hyperbolic, with completely different equations of motion. Notice that, while for a single modulus this only happens at a finite distance singularity (\emph{e.g.} Type $\mathrm{I}_{1,1}$ in Section \ref{F-theory_embedding}), for multiple moduli this can also be true for infinite distance ones. Furthermore, if we believe the dynamical version of the distance conjecture discussed in the text, we expect that the finite distance singularities should not admit spiraling trajectories of infinite length. 
    \item An unexpected outcome of our analysis was the observation that the combination of variables
\begin{equation}
    T \propto 2 G_{I \bar{J}}  \frac{{\rm d}}{{\rm d}N} \left( \phi^I \bar{\phi}^{\bar{J}}\right) = \frac{C}{s^2} \frac{{\rm d}}{{\rm d}N} \left( s^2+a^2\right)
\end{equation}
exhibits a universal behaviour as $T \rightarrow \infty$. In particular, it converges to a universal value that depends on the leading term of the scalar potential along the trajectory. We currently have no explanation for this behaviour and it would be interesting if one could come up with a bottom-up explanation. This behavior could signal another peculiar universal pattern that arises at the boundary of moduli space (see \cite{Castellano:2023stg,Castellano:2023jjt} for other examples).
  \item The solutions we have studied also exhibit attractive features which might be amenable to phenomenological applications. Exotic cosmological epochs driven by moduli and axions could be relevant for the very early history of the universe - between inflation and Big Bang Nucleosynthesis (BBN) \cite{Cicoli:2023opf,Apers:2024ffe} - where almost no experimental constraints exist \cite{Allahverdi:2020bys}. A well-studied example is that of kination \cite{Conlon:2022pnx}, dominated by the kinetic energy of a scalar field(s). In presence of a potential for the axion(s), we have found examples where it can be a future attractor.\footnote{In absence of additional sources, such as radiation or matter.} Connecting to one of the points above, an additional future direction could be to investigate consequences for the late universe, for example asymptotic accelerated expansion with multiple moduli and axions, similarly to \cite{Calderon-Infante:2022nxb}. Another curious application concerns the oscillating solutions of subsection \ref{ssec:oscillating} (see also \ref{ssc:ctrex}), in the special case where $\lambda=0$. Their trajectory results in a saxion that is fixed without a minimum in the scalar potential, thanks to the motion of the axion and the resulting Hubble friction. It would be interesting to see whether such a feature can be obtained in realistic models of ``moduli fixing" \cite{Tonioni:2024huw}.\footnote{Fifth forces and the cosmological moduli problem would still need to be addressed for realistic scenarios, which is why we carefully avoided the use of the word stabilisation.} 
    \item One can also study non-oscillating trajectories from an abstract perspective using results from tame geometry. For example, in \cite{rolin_quasianalytic_2006} sufficient criteria on the first-order differential system were given that forbid oscillating (non-tame) trajectories. These criteria can also be applied to the system \eqref{eq:sysNz} and \eqref{eq:sysNgz} in the limit $\hat t = 1/N \rightarrow 0$, to obtain general constraints when the trajectories end on a fix-point. Note, however, that requiring tameness of all trajectories is a strong condition, which rules out any infinite oscillatory behavior.\footnote{Note that while \cite{Grimm:2021vpn,Douglas:2023fcg} conjectures the tameness of effective coupling functions, a differential equation with tame coefficients can have wild (oscillatory) solutions as we have seen in this work.} It is an interesting future direction to explore how far one can exploit the tools of tame geometry to control physical trajectories. 
\end{itemize}

\subsection*{Acknowledgments}

We would like to thank Fien Apers, Thomas van Riet, Cumrun Vafa, and Irene Valenzuela for discussions, as well as David Andriot, José Calderón Infante and Flavio Tonioni for comments on the manuscript. The research of TG and FR was partially supported by the Dutch Research Council (NWO) via a Vici grant, and the work of FR was also supported through a junior postdoctoral fellowship of the Fonds Wetenschappelijk Onderzoek (FWO), project number 12A1Q25N. The research of DH was supported in part by a grant from the Simons Foundation (602883, CV) and the DellaPietra Foundation.

\appendix

\section{Tools for dynamical systems}

In this appendix we give a concise introduction to some techniques from the theory of dynamical systems that are used in the main text. The results will only be justified heuristically, and we refer the interested reader to \cite{Wiggins:2003} for actual proofs. In particular, we are concerned with the study of a $D$-dimensional dynamical system of the type
\begin{eqn}\label{eq:f}
    \dot{x}= f(x),
\end{eqn}where $x$ are the coordinates of some $N-$dimensional manifold and $f(x)$ is a sufficiently regular function, usually taken to be $\mathcal{C}^r$ (with $r \geq 1$) on an open set $U \in \mathbb{R}^D$. Since $f(x)$ does not explicitly depend on time, such a system is said to be \emph{autonomous}.\\

\noindent The solutions to the differential equations \eqref{eq:f} obey important uniqueness and existence properties, summarized by the following theorems.\\

\noindent \textbf{{Theorem}}\\ \textit{Let  $f\in \mathcal{C}^r(U)$ ($r \geq 1$), and $x_i \in U$. Then, there exists a unique (local) solution to \eqref{eq:f} satisfying $x(t_i)=x_i$ for a given initial time $t_i$. Moreover, there exists a compact set $C \subset U \times \mathbb{R}$ allowing the solution to be extended to the boundary of $C$. The solution $x(t;x_i,t_i)$ is a $\mathcal{C}^r$ function of $x_i,t_i$ and $t$ on its domain.} \\ 

\noindent For a fixed initial time $t_i$, the function $x(t;x_i,t_i)$ defines a set of diffeomorphisms from the phase space into itself.\\

\noindent \textbf{{Definition: flow}}\\ \textit{The flow of a map $f$ is given by $\phi(t,x) \equiv x(t;x_i,t_i)$, the solution to \eqref{eq:f} passing through $(x_i,t_i)$. It satisfies the following properties:
\begin{enumerate}[i)]
    \item $\phi(0,x)=x$ 
    \item $\phi(t,x)$ is $ \mathcal{C}^r$ 
    \item $\phi(t+s,x)=\phi(t,\phi(s,x))$
\end{enumerate}\,}\\
\noindent This is known as the \emph{phase flow} generated by \eqref{eq:f}, and naturally provides a geometric perspective to the study of dynamical systems. As an example, it is instrumental in the proof of the Poincaré-Bendixson Theorem cited in the next subsection.\\

\noindent The theory of dynamical systems provides tools and techniques to understand the asymptotic behavior of solutions to the differential equation \eqref{eq:f}, when $t \rightarrow + \infty$. A particularly important concept will be that of an attractor set, that is a particular set of the phase space to which the system tends to evolve. Before we continue, let us give here some definitions relevant to the asymptotic behavior of trajectories.\\

\noindent \textbf{{Definition: limit point}}\\ \textit{Given a solution $x(t)$ to \eqref{eq:f}, a point $P$ is said to be a positive limit point of $x(t)$ if there exists a sequence $\left\{t_n\right\}$ such that $t_n \rightarrow +\infty$ and $x(t_n) \rightarrow P$}.\\

\noindent The set of all limiting points of a trajectory is known as the limiting set.\\

\noindent \textbf{{Definition: limit set}}\\ \textit{The limit set for a curve $x(t)$ is given by the union of all its limiting points.}\\

\noindent Another useful concept is that of invariant sets, \emph{i.e.} sets characterized by the property that any trajectory starting inside the set can never escape it. This formalized by the following definition.\\

\noindent \textbf{{Definition: (positively) invariant set}}\\ \textit{A set $M$ is said to be (positively) invariant with respect to \eqref{eq:f} if, for any solution $x(t)$ and time $t_0$, $x(t_0) \in M$ implies $x(t) \in M$ (for any $t > t_0$).}\\

\noindent This allows us to finally give the definition of an attractor set.\\

\noindent \textbf{{Definition: attractor set}}\\ \textit{A set $\mathcal{A}$ is said to be an attractor set if there exists a neighborhood $U$ of $\mathcal{A}$ such that
\begin{enumerate}[i)]
    \item $\mathcal{A}$ is positively invariant
    \item $\phi\left(t,U\right) \subset U$ for any $t \geq 0$
    \item $\bigcap_{t >0} \phi \left(t,U\right)= \mathcal{A}$
\end{enumerate}\,}\\
\noindent In the following sections, we will present explicit techniques that are used to characterize the attractor set of a given system.

\subsection{Stability}

\noindent An obvious class of solutions to \eqref{eq:f} is given by the points $\bar{x}$ satisfying $f(\bar{x})=0$, also known as critical or equilibrium points. Stable fixed points are perhaps the simplest example of an attractor. Since all the time derivatives vanish at such points, it is obvious that $x=\bar{x}$ is an exact solution to the system. But what happens if we perturb the value of $x$ slightly? To answer this, we need to introduce different notions of (in)stability. Although we are only discussing fixed points for now, we will do this in a way which applies to an arbitrary, limiting trajectory $\bar{x}(t)$. Intuitively, a solution is stable if small perturbations around it cannot deviate too much from the original trajectory. \\

\noindent \textbf{{Definition: (Lyapunov) stability}}\\ \textit{A solution $\bar{x}(t)$ is said to be (Lyapunov) stable if, for any $\epsilon >0$, there exists $\delta > 0$ such that for any other solution $y(t)$ satisfying $ |y(t_0)-x(t_0)| < \delta$, $ |y(t)-x(t)| < \varepsilon $ for any $t > t_0$.}\\

\noindent Furthermore, the stability is said to be asymptotic if all the trajectories that are close enough to the original one converge to it eventually.\\

\noindent \textbf{{Definition: asymptotic stability}}\\ \textit{A (Lyapunov) stable trajectory $\bar{x}(t)$ is said to be asymptotically stable if for any other solution $y(t)$ there exists a $\delta_0$ such that if $|x(t_0)-y(t_0)| < \delta_0$, then $y(t) \rightarrow \bar{x}(t)$ as $t \rightarrow \infty$.}\\

\noindent In a similar way, one can also define stability criteria to approach closed trajectories, also known as \emph{orbits}. It turns out that in 2 dimensions, fixed points and fixed orbits exhaust the set of possibilities. This is formalized by the following result:\\

\noindent \textbf{Theorem: Poincaré-Bendixson}\\ \textit{Consider a 2-dimensional system of the form \eqref{eq:f}, with $f:D \rightarrow \mathbb{R}$ locally Lipschitz and $D$ an open and connected set. For a given solution $x(t)$, let $L^+$ be the positive limit set of the positive semiorbit $\gamma^+= \left\{ x(t), \,\, 0\leq t \leq + \infty \right\}$. If $L^+$ does not contain any fixed points, it must be a closed orbit.}\\

\noindent This means that in dimension $D=2$, every solution must eventually converge to an attractor, which is either a fixed point or a closed orbit. The proof relies on the Jordan curve theorem, and therefore it cannot be extended to higher dimensions. Indeed, for $D>2$, dynamical systems can exhibit much more complicated behavior, including chaos and strange attractors.\\

\noindent In the special case of a fixed point, stability can be studied by linearizing the system around it. Close to $x= \bar{x}$, \eqref{eq:f} can be written as
\begin{eqn}
    \dot{f}= D f \big \lvert_{x=\bar{x}} (x-\bar{x}) + \mathcal{O}\left(|x-\bar{x}|^2\right),
\end{eqn}where $D f \big \lvert_{x=\bar{x}}$ is the Jacobian of $f(x)$ evaluated at the fixed point. Neglecting the $\mathcal{O}\left(|x-\bar{x}|^2\right)$ terms, the equation can then be solved as a matrix exponential in a small neighborhood. This justifies the following:\\

\noindent \textbf{Theorem:}\\ \textit{Given a fixed point $\bar{x}$ of the system \eqref{eq:f}, if all the eigenvalues of $D f \big \lvert_{x=\bar{x}}$ have negative real parts the fixed point is stable.}\\

\noindent The theorem can be proven using Lyapunov functions, which we will introduce below. Moreover, it motivates the following definition. A fixed point is said to be stable, unstable or a saddle point if the real parts of the eigenvalues of $D f \big \lvert_{x=\bar{x}}$ are all negative, all positive or negative and positive respectively. If any eigenvalues have zero real part, the above is not sufficient to determine stability, and one must resort to different techniques. Relevant ones are the Centre Manifold Theorem (not discussed here) and Lyapunov functions, the subject of the next paragraph.\\

\subsection{Lyapunov stability}

\noindent A class of techniques used to prove the stability of and identify the attraction region of non-linear dynamicsl systems makes use fo the so-called \emph{Lyapunov} functions. They can be thought as the generalization of the notion of energy in a mechanical system with losses, where the attractor is usually the lowest energy state. More precisely, they consist of (lower or upper) bounded scalar functions which exhibit certain monotonocity properties along the flow of a dynamical system, \emph{i.e.} when they are evaluated along a generic solution. In such cases, one can show that the Lyapunov function must satisfy certain asymptotic properties (such as converging to a constant), and in turn this can give insight into the asymptotic behaviour of the original system.
To illustrate this in a simple scenario, let us consider a trivial example, before stating the main theorems.\\

\noindent \textbf{Example:}\\ \textit{Consider the equation $\dot{x}=-x$, and define the Lyapunov function $\mathcal{L}(x) \equiv x^2.$ From the equations of motion, $\dot{\mathcal{L}}(t)=- x^2 \leq 0$, where the inequality is saturated only for $x=0$. Since $\mathcal{L}$ is a positive, decreasing function, we conclude $\mathcal{L}(t) \rightarrow 0$, and $x \rightarrow 0$. This is of course consistent with the analytic solution to the system, given by $x(t)=c e ^{-t}$ (for some constant $c$).}\\

\noindent The general result goes by the name of Lyapunov's Theorem, and can be stated as follows.\\

\noindent \textbf{Theorem: Lyapunov}\\ \textit{Consider a system of the form \eqref{eq:f}, with a fixed point $\bar{x}$. Assume there exists a function $\mathcal{L}: U \rightarrow \mathbb{R}$, where $\mathcal{L} \in \mathcal{C}^1 (U)$ in a neighborhood $U$ of $\bar{x}$ and satsifies the following properties:
\begin{enumerate}[i)]
    \item $\mathcal{L}(\bar{x})=0$ and $\mathcal{L}(\bar{x}) >0$ if $x \neq \bar{x}$
    \item $\dot{\mathcal{L}} (x) \leq 0$ for any $x$ in $U-\{ \bar{x}\}$,
\end{enumerate} 
Then $\bar{x}$ is stable. If, in addition to this,
\begin{enumerate}[iii)] 
\item $\dot{\mathcal{L}} (x) \leq 0$ for any $x$ in $U-\{ \bar{x}\}$,
\end{enumerate} 
$\bar{x}$ is asymptotically stable.}
\\

\noindent A function satisfying i) and ii) is known as a Lyapunov function, or a strict Lyapunov function if iii) is also true. A similar theorem can also be used to prove the instability of a fixed point.\\

\noindent \textbf{Theorem: Chetaev}\\ \textit{Let us again start from a system of the form \eqref{eq:f}, with fixed point $\bar{x}$. If there exists a $\mathcal{C}^1$ function $\mathcal{L}: U \rightarrow \mathbb{R}$ and a neighborhood $U$ of $\bar{x}$ satisfying:
\begin{enumerate}[i)]
    \item $\mathcal{L}(\bar{x})=0$ and $\bar{x}$ belongs to the boundary of $G=\left\{x\,\, |\,\, V(x) >0 \right\}$,
    \item  $\dot{\mathcal{L}}(\bar{x}) > 0$ for any $x \in G \cap U$,
\end{enumerate}
then $\bar{x}$ is unstable.}\\

\noindent The corresponding functions are known as Chetaev functions. Although there also exist theorems on the existence of Lyapunov(Chetaev) functions for (un)stable fixed points, their proofs are not constructive. Therefore, finding a Lyapunov function for a given system usually requires some deal of guesswork - a typical ansatz is a manifestly positive quantity, such as a sum of squares. On the other hand, whenever one can be found it provides a very powerful tool to study the associated dynamical system. Notice that the (asymptotic) stability properties are valid for all the points contained in the neighborhood $U$ of $\bar{x}$. Therefore, one can often show the existence of a finite region of attraction, such that any trajectory passing through the region will eventually converge to $\bar{x}$. As a particular case, if $U$ coincided with the whole domain of the dynamical system, the equilibrium is said to be globally asymptotically stable, meaning that any trajectory (irrespectively of initial conditions) will approach $\bar{x}$ asymptotically.\footnote{If the domain is non compact, such as $\mathbb{R}^n$, one must also require the Lyapunov function to be radially unbounded, \emph{i.e.} $\mathcal{L}(x) \rightarrow \infty $ if $|x| \rightarrow \infty$.} This provides a \emph{global} notion of stability, much more powerful than the \emph{local} one that can be inferred from a simple linearization around the fixed point. Moreover, the Lyapunov formulation is more flexible in that it can be applied to fixed loci that are more general than a point, such as manifolds of higher-dimensions. These two points are central to another theorem, that we state below. \\

\noindent \textbf{Theorem: La Salle's invariance principle}\\ \textit{Consider a system of the form \eqref{eq:f}, with a Lyapunov function $\mathcal{L}:  M \rightarrow \mathbb{R}$ that satisfies $\dot{\mathcal{L}} \leq 0$ everywhere on a compact set $M$. Let $A$ be the maximal, positively invariant subset of the set $E= \left \{ x \in M \,\, | \,\, \dot{\mathcal{L}}=0 \right \}$. As $t \rightarrow \infty$, $x(t) \rightarrow A$ for any trajectory starting in $M$.}\\

\noindent Together with that of $\mathcal{L}$, the choice of the set $M$ now allows to study regions of attaction. In particular, it is often useful to consider the level sets of $\mathcal{L}$, which are automatically positive invariant.\\

\noindent 
Finally, let us also state another useful fact. It is most used in the case of non-autonomous (\emph{i.e.} time-dependent) systems, although we also apply it to an autonomous one in the main text. In general, the convergence of a (differentiable) function to a constant does not imply that its derivative should converge to zero (and the converse is also not true). The following lemma clarifies under which circumstances the derivative of a (converging) function vanishes asymptotically. \\

\noindent \textbf{Lemma (Barbalat):}\\ \textit{If $g(t) \in \mathcal{C}^1 (\mathbb{R})$ has a finite limit as $t \rightarrow \infty$, and $\dot{g}(t)$ is uniformly continuous, then $\dot{g}(t) \rightarrow 0$ as $t \rightarrow 0$.}\\

\noindent We recall that, for a function $h(t)$ to be uniformly continuous, it is sufficient (but not necessary) for its first derivative $\dot{h}(t)$ to be a bounded.
\section{Asymptotic Hodge theory}\label{app:asympHodge}
In this section we construct the periods of all types of boundaries in one-dimensional moduli spaces. We recall that there are five types of boundaries that can arise, whose Hodge-Deligne diamonds are given by
\begin{equation}\label{eq:singtypes}
\begin{aligned}
\mathrm{I}_{0,1}: \ \begin{tikzpicture}[baseline={([yshift=-.5ex]current bounding box.center)},scale=0.5,cm={cos(45),sin(45),-sin(45),cos(45),(15,0)}]
  \draw[step = 1, gray, ultra thin] (0, 0) grid (4, 4);
  
\draw[fill] (0,4) circle[radius=0.05];
\draw[fill] (3,2) circle[radius=0.05];
\draw[fill] (2,3) circle[radius=0.05];
\draw[fill] (1,2) circle[radius=0.05];
\draw[fill] (2,1) circle[radius=0.05];
\draw[fill] (4,0) circle[radius=0.05];

\end{tikzpicture},& \quad \mathrm{I}_{1,1}: \ \begin{tikzpicture}[baseline={([yshift=-.5ex]current bounding box.center)},scale=0.5,cm={cos(45),sin(45),-sin(45),cos(45),(15,0)}]
  \draw[step = 1, gray, ultra thin] (0, 0) grid (4, 4);
  
\draw[fill] (0,4) circle[radius=0.05];
\draw[fill] (3,3) circle[radius=0.05];
\draw[fill] (2,2) circle[radius=0.05];
\draw[fill] (1,1) circle[radius=0.05];
\draw[fill] (4,0) circle[radius=0.05];

\end{tikzpicture},& \\
\mathrm{II}_{0,0}: \  \begin{tikzpicture}[baseline={([yshift=-.5ex]current bounding box.center)},scale=0.5,cm={cos(45),sin(45),-sin(45),cos(45),(15,0)}]
  \draw[step = 1, gray, ultra thin] (0, 0) grid (4, 4);
  
\draw[fill] (1,4) circle[radius=0.05];
\draw[fill] (2,2) circle[radius=0.05];
\draw[fill] (4,1) circle[radius=0.05];
\draw[fill] (3,0) circle[radius=0.05];
\draw[fill] (0,3) circle[radius=0.05];
\end{tikzpicture},& \quad  \mathrm{III}_{0,0}: \ \begin{tikzpicture}[baseline={([yshift=-.5ex]current bounding box.center)},scale=0.5,cm={cos(45),sin(45),-sin(45),cos(45),(15,0)}]
  \draw[step = 1, gray, ultra thin] (0, 0) grid (4, 4);
  
\draw[fill] (2,4) circle[radius=0.05];
\draw[fill] (2,2) circle[radius=0.05];
\draw[fill] (4,2) circle[radius=0.05];
\draw[fill] (3,1) circle[radius=0.05];
\draw[fill] (1,3) circle[radius=0.05];
\draw[fill] (2,0) circle[radius=0.05];
\draw[fill] (0,2) circle[radius=0.05];
\end{tikzpicture},& \quad \mathrm{V}_{1,1}: \ \begin{tikzpicture}[baseline={([yshift=-.5ex]current bounding box.center)},scale=0.5,cm={cos(45),sin(45),-sin(45),cos(45),(15,0)}]
  \draw[step = 1, gray, ultra thin] (0, 0) grid (4, 4);
  
\draw[fill] (4,4) circle[radius=0.05];
\draw[fill] (2,2) circle[radius=0.05];
\draw[fill] (3,3) circle[radius=0.05];
\draw[fill] (1,1) circle[radius=0.05];
\draw[fill] (0,0) circle[radius=0.05];
\end{tikzpicture}.  
\end{aligned}
\end{equation}
The first two boundaries of type  $\mathrm{I}_{0,1}$ and $\mathrm{I}_{1,1}$ are at finite distance, while the other three boundaries of type $\mathrm{II}_{0,0}$, $\mathrm{III}_{0,0}$ and $\mathrm{V}_{1,1}$ are at infinite distance. In subsection \ref{ssec:construction} we review the construction of the periods for these boundaries, following the approach laid out in \cite{Bastian:2021eom}. In the subsequent subsections \ref{ssec:I01}-\ref{ssec:V} we apply this approach to each of the above boundary types to construct the periods. 

\subsection{Review of asymptotic period vectors}\label{ssec:construction}
We first briefly review the construction of asymptotic periods. This includes both how to perturbative corrections, i.e.~subleading polynomial terms in the saxion $s$, and exponential corrections in $s$, appear in the periods. Our exposition will be rather short, and we refer the reader interested in a detailed review to \cite{Bastian:2021eom} and the PhD theses \cite{vandeHeisteeg:2022gsp, Monnee:2024gsq}. 

\paragraph{Limiting mixed Hodge structure.} For each boundary, there is a designated splitting of the middle cohomology, known as a limiting mixed Hodge structure
\begin{equation}
H^4_{\rm prim}(Y_4, \mathbb{C}) = \bigoplus_{p,q=0}^4 I^{p,q}\, ,
\end{equation}
where the $I^{p,q}$ are complex subspaces. Usually, this so-called Deligne splitting is obtained from the period data through Hodge filtrations and monodromy weight filtrations; we omit such expressions, as we take the opposite perspective and reconstruct the periods from this data. For a given boundary, as classified by one of the Hodge-Deligne diamonds in \eqref{eq:singtypes}, we will take a convenient real basis that spans these spaces $I^{p,q}$. We will begin from the so-called $\mathbb{R}$-split case, that is $\bar{I}^{p,q} = I^{q,p}$, and then describe the most general rotation away through the so-called phase operator $\delta$, cf.~the discussion surrounding \eqref{eq:delta}. For later reference, it is also helpful to define the operator spaces $\Lambda_{p,q}$ as
\begin{equation}
    O_{p,q} \in \Lambda_{p,q} : \qquad O_{p,q} I^{r,s} \subseteq I^{p+r,q+s}\, .
\end{equation}

\paragraph{Bilinear pairing.} Given such a $\mathbb{R}$-split basis, we write down the most general forms for operators acting on these spaces. We begin with the bilinear pairing $\Sigma$. For the $\mathbb{R}$-split case, it satisfies the orthogonality conditions
\begin{equation}
\omega_{p,q}^T \Sigma \omega_{r,s}  = 0 \quad \text{unless}\quad p+r=q+s=4\, .
\end{equation}
for all $\omega_{p,q} \in I^{p,q}$ and $\omega_{r,s} \in I^{r,s}$. Using the freedom to change basis, we will always bring the pairing to a standard block form associated with SO$(p,q)$.

\paragraph{Log-monodromy matrix.} We next turn to the log-monodromy matrix $N$, which acts on this Deligne splitting as a map in $\Lambda_{-1,-1}$:
\begin{equation}
	N I^{p,q} \subseteq I^{p-1,q-1}\, .
\end{equation}
It must also be an infinitesimal isomorphism of the pairing $N^T \Sigma+\Sigma N = 0$, and satisfy the positivity conditions
\begin{equation}
i^{p-q} \omega_{p,q} \Sigma N^{p+q-4}\overline{\omega_{p,q}} > 0 \, ,
\end{equation}
for an element of one of the primitive parts $\omega_{p,q} \in P^{p,q} \subseteq I^{p,q}$ with respect to $N$. Given a basis for the spaces $I^{p,q}$, we then construct the log-monodromy matrix as the most general $(-1,-1)$-map. In all the examples that we consider in this work, there is only one such map. We normalize its model-dependent coefficient by rescaling the basis vectors to which $N$ maps.

\paragraph{Phase operator.} 
The phase operator $\delta$ describes the complex rotation away from the $\mathbb{R}$-split mixed Hodge structure. It can be decomposed into components as
\begin{equation}\label{eq:delta}
    \delta = \sum_{p,q\geq 1} \delta_{-p,-q}\, .
\end{equation}
It is also an infinitesimal isomorphism $\delta^T \Sigma + \Sigma \delta = 0$, and commutes with the log-monodromy matrix, $[ \delta, N] = 0$. It is accompanied by another operator $\zeta$ to describe the rotation away from the $\mathbb{R}$-split, as described by \eqref{eq:pigen} below. One can compute $\zeta$ componentwise from $\delta$, but in all examples considered in this work it will vanish; for completeness, we refer to appendix B of \cite{Grimm:2021ckh}, where the explicit expressions may be found. For the construction of $\delta$ in examples, we need to write down all possible maps that satisfy \eqref{eq:delta}, are compatible with the pairing, and commute with $N$. The component along $N$ may always be set to zero: based on the expression for the periods \eqref{eq:pigen}, we can adjust it by shifting $t$ by a constant.

\paragraph{Instanton map.} The above data captures the polynomial part in $s$ of the periods, but for the purposes of this work we also want to characterize the exponential corrections. As described in \cite{Cattani_2003, Fernandez_2008}, this can be achieved by introducing a moduli-dependent operator $\Gamma$ acting on the mixed Hodge structure.  This was reviewed for physicists in \cite{Bastian:2021eom}, where it was termed the \textit{instanton map}, since it captures the exponential corrections in the periods; in fact, these exponential corrections were even found to be essential for certain physical couplings such as the K\"ahler metric. Concretely, the periods are encoded as
\begin{equation}\label{eq:pigen}
\Pi(z) = e^{i\delta} e^{-\zeta} e^{t N} e^{\Gamma(z)} \mathbf{a}_0\, .
\end{equation}
where $\mathbf{a}_0 \in I^{4,d}$ in the limiting mixed Hodge structure, and we write $z = e^{2\pi i t}$. The instanton map $\Gamma(z)$ is defined as a matrix of functions that is an infinitesimal isomorphism $\Gamma(z)^T \Sigma + \Sigma \Gamma(z)=0$, and can be expanded into maps that act by lowering the first degree of the Deligne splitting
\begin{equation}\label{eq:Gamma}
    \Gamma(z) = \bigoplus_{p<0} \bigoplus_q \Lambda_{p,q}\, .
\end{equation}
The periods are also known to satisfy horizontality constraints, i.e.~$\Pi \Sigma \Pi = \ldots = \Pi \Sigma \partial_t^3 \Pi = 0$. These conditions can be mapped into differential constraints on the components of $\Gamma$ \cite{Cattani_2003}, see also \cite{Bastian:2021eom}. Explicitly, they read
\begin{equation}\label{eq:diffgamma}
    \partial_t \exp[\Gamma(z)] = [\exp[\Gamma(z)], N] + \exp[\Gamma(z)]\partial_t \Gamma_{-1}(z)\, ,
\end{equation}
where we write $\Gamma_{-p}(z)$ to denote the component in $\Lambda_{-p} = \oplus_q \Lambda_{-p,q}$. These differential constraints fix the components $\Gamma_{-2,-3}(z)$ through $\Gamma_{-1}(z)$. Let us briefly comment on the construction of $\Gamma(z)$ for the asymptotic periods. First, we find all possible maps allowed by \eqref{eq:Gamma} and compatibility with the pairing; note that the component proportional to $N$ may be removed by a change of coordinates. The component $\Gamma_{-1}(z)$ will then be captured by one or two functions as coefficients, which fix the other components uniquely through \eqref{eq:diffgamma}. To get the asymptotic periods, we employ a series ansatz for these functions, and plug the resulting instanton map into \eqref{eq:pigen}.

\subsection{Type $\mathrm{I}_{0,1}$ boundaries}\label{ssec:I01}
In this section we apply the construction to boundaries of type $\mathrm{I}_{0,1}$. We use as boundary data for this singularity
\begin{equation}
\begin{aligned}
I^{4,0} &= (1,\, i,\, 0,\, 0,\, 0, \, 0 )\, , \quad  &I^{3,2} &=(0, \, 0, \, 1, \, i , \, 0, \, 0 )\, , \quad &I^{2,1} &= (0, \, 0, \, 0 , \, 0 , \, 1, \, i)\, , \\
I^{0,4} &= (1,\, -i,\, 0,\, 0,\, 0, \, 0 )\, , \quad  &I^{2,3} &=(0, \, 0, \, 1, \, -i , \, 0, \, 0 )\, , \quad &I^{1,2} &= (0, \, 0, \, 0 , \, 0 , \, 1, \, -i)\, , \\
\end{aligned}
\end{equation}
with as bilinear pairing and log-monodromy matrix 
\begin{equation}
\Sigma = \left(
\begin{array}{cccccc}
 1 & 0 & 0 & 0 & 0 & 0 \\
 0 & 1 & 0 & 0 & 0 & 0 \\
 0 & 0 & 0 & 0 & 0 & 1 \\
 0 & 0 & 0 & 0 & -1 & 0 \\
 0 & 0 & 0 & -1 & 0 & 0 \\
 0 & 0 & 1 & 0 & 0 & 0 \\
\end{array}
\right)\, ,  \quad N = \left(
\begin{array}{cccccc}
 0 & 0 & 0 & 0 & 0 & 0 \\
 0 & 0 & 0 & 0 & 0 & 0 \\
 0 & 0 & 0 & 0 & 0 & 0 \\
 0 & 0 & 0 & 0 & 0 & 0 \\
 0 & 0 & 1 & 0 & 0 & 0 \\
 0 & 0 & 0 & 1 & 0 & 0 \\
\end{array}
\right)\, .
\end{equation}
There is no phase operator $\delta$, as the only possible one is proportional to $N$, so can be removed by a coordinate redefinition.

\paragraph{Instanton map.} The instanton map reads
\begin{equation}
\Gamma(z) =\scalebox{0.75}{$ \frac{1}{2}\left(
\begin{array}{cccccc}
 0 & 0 & -i (d(z)-e(z)) & d(z)+e(z) & i (a(z)-c(z)) & -a(z)-c(z) \\
 0 & 0 & e(z)-d(z) & -i (d(z)+e(z)) & a(z)-c(z) & i (a(z)+c(z)) \\
 a(z)+c(z) & -i (a(z)+c(z)) & b(z) & -i b(z) & 0 & 0 \\
 i (a(z)-c(z)) & a(z)-c(z) & -i b(z) & -b(z) & 0 & 0 \\
 d(z)+e(z) & -i (d(z)+e(z)) & 0 & 0 & b(z) & -i b(z) \\
 i (d(z)-e(z)) & d(z)-e(z) & 0 & 0 & -i b(z) & -b(z) \\
\end{array}
\right)$}\, .
\end{equation}
The functions $a,b$ make up $\Gamma_{-1}$, the functions $c,d$ make up $\Gamma_{-2}$, and the function $e$ makes up $\Gamma_{-3}$. The first two functions $a,b$ fix the other three $c,d,e$ through the differential constraints
\begin{equation}
\begin{aligned}
2\pi i z d'(z) + a(z) = 0\, , \quad -b(z) a'(z) + a(z) b'(z) +2 c'(z)=0\, , \\
-i a(z) b(z)+2i c(z) = 2\pi z (d(z) b'(z)-b(z)d'(z)+2e'(z)\, .
\end{aligned}
\end{equation}
These equations can be solved straightforwardly for a series ansatz of $a,b$. Below we work out a simple example of this that suffices for the purposes of this paper.

\paragraph{Asymptotic periods.} For the asymptotic periods we make the first-order approximation $a(z)=A z$ and $b(z)=B z$. This yields for the other three functions
\begin{equation}
c(z) = 0\, , \qquad d(z) = \frac{i A z}{2\pi}\, , \qquad e(z) = \frac{A B z^2}{8\pi i}\, .
\end{equation}
Plugging these expressions into \eqref{eq:pigen}, we find as period vector
\begin{equation}
\Pi = \left(
\begin{array}{c}
 1+\frac{A^2 B e^{6 i \pi  t}}{8 \pi } \\
 i-\frac{i A^2 B e^{6 i \pi  t}}{8 \pi } \\
 \frac{1}{2} A B e^{4 i \pi  t}+A e^{2 i \pi  t} \\
 i A e^{2 i \pi  t}-\frac{1}{2} i A B e^{4 i \pi  t} \\
 \frac{1}{2} A B e^{4 i \pi  t} t+\frac{i A B e^{4 i \pi  t}}{8 \pi }+A e^{2 i \pi  t} t+\frac{i A
   e^{2 i \pi  t}}{2 \pi } \\
 -\frac{1}{2} i A B e^{4 i \pi  t} t+\frac{A B e^{4 i \pi  t}}{8 \pi }+i A e^{2 i \pi  t} t-\frac{A
   e^{2 i \pi  t}}{2 \pi } \\
\end{array}
\right) + \mathcal{O}(e^{4\pi i t})\,.
\end{equation}
Although there may be corrections to this expression at order $e^{4\pi i t}$, we have included other terms at that order and higher to ensure that the period vector satisfies the horizontality conditions.

\subsection{Type $\mathrm{I}_{1,1}$ boundaries}
In this section we construct the asymptotic periods of the $\mathrm{I}_{1,1}$ boundaries. 

\paragraph{Boundary data.} We use as limiting mixed Hodge structure for the singularity
\begin{equation}
\begin{aligned}
I^{4,0} &= (1,\, i,\, 0,\, 0,\, 0)\, , \quad &I^{0,4} &= (1,\, -i,\, 0,\, 0,\, 0)\, , \\
I^{3,3} &= (0,\,0,\, 1,\, 0,\, 0)\, , \quad &I^{2,2} &= (0,\, 0,\, 0,\, 1,\, 0)\,,  \quad &I^{1,1} &= (0,\, 0,\, 0,\, 0,\, 1)\, .
\end{aligned}
\end{equation}
The bilinear pairing and log-monodromy matrix are given by
\begin{equation}
\Sigma = \left(
\begin{array}{ccccc}
 1 & 0 & 0 & 0 & 0 \\
 0 & 1 & 0 & 0 & 0 \\
 0 & 0 & 0 & 0 & 1 \\
 0 & 0 & 0 & -1 & 0 \\
 0 & 0 & 1 & 0 & 0 \\
\end{array}
\right)\, , \quad N = \left(
\begin{array}{ccccc}
 0 & 0 & 0 & 0 & 0 \\
 0 & 0 & 0 & 0 & 0 \\
 0 & 0 & 0 & 0 & 0 \\
 0 & 0 & 1 & 0 & 0 \\
 0 & 0 & 0 & 1 & 0 \\
\end{array}
\right)
\end{equation}
Again, there is no phase operator $\delta$, as the only possible one is proportional to $N$, and can therefore be removed by a coordinate shift.

\paragraph{Instanton map.} The ansatz for the instanton map reads
\begin{equation}
\Gamma(z) = \frac{1}{2}\left(
\begin{array}{ccccc}
 0 & 0 & -c(z) & b(z) & -a(z) \\
 0 & 0 & i c(z) & -i b(z) & i a(z) \\
 a(z) & -i a(z) & 0 & 0 & 0 \\
 b(z) & -i b(z) & 0 & 0 & 0 \\
 c(z) & -i c(z) & 0 & 0 & 0 \\
\end{array}
\right)\,.
\end{equation}
The function $a(z)$ corresponds to $\Gamma_{-1}(z)$, $b(z)$ to $\Gamma_{-2}(z)$, and $c(z)$ to $\Gamma_{-3}(z)$. The first fixes the other two through the differential equations
\begin{equation}
a(z)+2\pi i z b'(z) = 0\, , \qquad b(z)+2\pi i z c'(z) = 0\, .
\end{equation}

\paragraph{Asymptotic periods.} We make a first-order approximation $a(z)=A z$, which yields for the other two functions
\begin{equation}
b(z) = \frac{iA z}{2\pi} \, , \qquad c(z) =  -\frac{Az}{4\pi^2}\, .
\end{equation}
Plugging these expressions into \eqref{eq:pigen}, we find as period vector
\begin{equation}
\Pi = \left(
\begin{array}{c}
 1+\frac{A^2 e^{4 i \pi  t}}{16 \pi ^2} \\
 i-\frac{i A^2 e^{4 i \pi  t}}{16 \pi ^2} \\
 A e^{2 i \pi  t} \\
 A e^{2 i \pi  t} t+\frac{i A e^{2 i \pi  t}}{2 \pi } \\
 \frac{1}{2} A e^{2 i \pi  t} t^2+\frac{i A e^{2 i \pi  t} t}{2 \pi }-\frac{A e^{2 i \pi  t}}{4 \pi
   ^2} \\
\end{array}
\right)+\mathcal{O}(e^{4\pi i t})\, .
\end{equation}
We have included terms up to order $e^{4\pi it}$ to ensure that the horizontality conditions are satisfied. Nevertheless, we note that there may be corrections that enter at this order as well, as we only did a first-order approximation for the function $a(z)$.

\subsection{Type $\mathrm{II}_{0,0}$ boundaries}
In this section we construct the asymptotic periods for $\mathrm{II}_{0,0}$ boundaries. 

\paragraph{Boundary data.} The limiting mixed Hodge structure is given by
\begin{equation}
\begin{aligned}
I^{4,1}&: \quad (1,\, i,\, 0,\, 0, \, 0), \qquad &I^{1,4}&: \quad (1,\, -i,\, 0,\, 0, \, 0 )\, ,\\
I^{3,0}&: \quad (0,\, 0,\, 1,\, i, \, 0 ), \qquad &I^{0,3}&:\quad (0,\, 0,\, 1,\, -i, \, 0)\, .
\end{aligned}
\end{equation}
The bilinear pairing and log-monodromy matrix read
\begin{equation}
\Sigma = \left(
\begin{array}{ccccc}
 0 & 0 & 0 & -1 & 0 \\
 0 & 0 & 1 & 0 & 0 \\
 0 & 1 & 0 & 0 & 0 \\
 -1 & 0 & 0 & 0 & 0 \\
 0 & 0 & 0 & 0 & 1 \\
\end{array}
\right) \, , \quad N = \left(
\begin{array}{ccccc}
 0 & 0 & 0 & 0 & 0 \\
 0 & 0 & 0 & 0 & 0 \\
 1 & 0 & 0 & 0 & 0 \\
 0 & 1 & 0 & 0 & 0 \\
 0 & 0 & 0 & 0 & 0 \\
\end{array}
\right)\, .
\end{equation}
There is no phase operator $\delta$, since again the only possible one is proportional to $\delta$.

\paragraph{Instanton map.} The instanton map reads
\begin{equation}
\Gamma(z) = \left(
\begin{array}{ccccc}
 \frac{c(z)}{2} & -\frac{1}{2} i c(z) & 0 & 0 & a(z) \\
 -\frac{1}{2} i c(z) & -\frac{c(z)}{2} & 0 & 0 & -i a(z) \\
 0 & 0 & \frac{c(z)}{2} & -\frac{1}{2} i c(z) & b(z) \\
 0 & 0 & -\frac{1}{2} i c(z) & -\frac{c(z)}{2} & -i b(z) \\
 -i b(z) & -b(z) & i a(z) & a(z) & 0 \\
\end{array}
\right)
\end{equation}
The function $a(z)$ makes up $\Gamma_{-1}(z)$, $b(z)$ makes up $\Gamma_{-2}(z)$, and $c(z)$ makes up $\Gamma_{-3}(z)$. These functions satisfy the differential relations
\begin{equation}
a(z) + 2\pi i z b'(z)= 0\, , \qquad a(z)^2 = 2\pi i z (a(z) b'(z) + b(z) a'(z) + i c'(z))\, .
\end{equation}

\paragraph{Asymptotic periods.} For the first-order approximation $a(z) = A z$, we find that the other two functions are given by
\begin{equation}
\b(z) = \frac{i A z}{2\pi}\, , \qquad c(z) = - \frac{A^2 z^2}{4\pi}\, .
\end{equation}
The period expansion at the boundary is given by
\begin{equation}
\Pi = \left(
\begin{array}{c}
 1+\frac{A^2 e^{4 i \pi  t}}{4 \pi } \\
 i-\frac{i A^2 e^{4 i \pi  t}}{4 \pi } \\
 \frac{A^2 e^{4 i \pi  t} t}{4 \pi }+\frac{i A^2 e^{4 i \pi  t}}{4 \pi ^2}+t \\
 -\frac{i A^2 e^{4 i \pi  t} t}{4 \pi }+\frac{A^2 e^{4 i \pi  t}}{4 \pi ^2}+i t \\
 \frac{A e^{2 i \pi  t}}{\pi } \\
\end{array}
\right)+\mathcal{O}(e^{4\pi i t}\, .
\end{equation}
We have included terms up to order $e^{4\pi i t}$ to ensure that the horizontality conditions are satisfied; nevertheless, other corrections may also appear, since we only performed a first-order approximation of $a(z)$.

\subsection{Type $\mathrm{III}_{0,0}$ boundaries}
In this section we construct the asymptotic periods for $\mathrm{III}_{0,0}$ boundaries. 

\paragraph{Boundary data.} The limiting mixed Hodge structure is given by
\begin{equation}
\begin{aligned}
I^{4,2} &= (1,\, i,\, 0,\, 0,\, 0, \, 0 )\, , \quad  &I^{3,1} &=(0, \, 0, \, 1, \, i , \, 0, \, 0 )\, , \quad &I^{2,0} &= (0, \, 0, \, 0 , \, 0 , \, 1, \, i)\, , \\
I^{2,4} &= (1,\, -i,\, 0,\, 0,\, 0, \, 0 )\, , \quad  &I^{1, 3} &=(0, \, 0, \, 1, \, -i , \, 0, \, 0 )\, , \quad &I^{0,2} &= (0, \, 0, \, 0 , \, 0 , \, 1, \, -i)\, , \\
\end{aligned}
\end{equation}
The bilinear pairing and log-monodromy matrix are given by
\begin{equation}
\Sigma = \left(
\begin{array}{cccccc}
 0 & 0 & 0 & 0 & -1 & 0 \\
 0 & 0 & 0 & 0 & 0 & -1 \\
 0 & 0 & 1 & 0 & 0 & 0 \\
 0 & 0 & 0 & 1 & 0 & 0 \\
 -1 & 0 & 0 & 0 & 0 & 0 \\
 0 & -1 & 0 & 0 & 0 & 0 \\
\end{array}
\right)\, , \quad N = \left(
\begin{array}{cccccc}
 0 & 0 & 0 & 0 & 0 & 0 \\
 0 & 0 & 0 & 0 & 0 & 0 \\
 1 & 0 & 0 & 0 & 0 & 0 \\
 0 & 1 & 0 & 0 & 0 & 0 \\
 0 & 0 & 1 & 0 & 0 & 0 \\
 0 & 0 & 0 & 1 & 0 & 0 \\
\end{array}
\right)\, .
\end{equation}
The phase operator reads
\begin{equation}
\delta = \left(
\begin{array}{cccccc}
 0 & 0 & 0 & 0 & 0 & 0 \\
 0 & 0 & 0 & 0 & 0 & 0 \\
 0 & 0 & 0 & 0 & 0 & 0 \\
 0 & 0 & 0 & 0 & 0 & 0 \\
 0 & \xi  & 0 & 0 & 0 & 0 \\
 -\xi  & 0 & 0 & 0 & 0 & 0 \\
\end{array}
\right)\, ,
\end{equation}
where $\xi \in \mathbb{R}$, and we removed the part of $\delta$ proportional to the log-mondromy matrix.

\paragraph{Instanton map.} The instanton map is given by
\begin{equation}
\Gamma(z) = \left(
\begin{array}{cccccc}
 \frac{d(z)}{2} & -\frac{ i d(z) }{2}& \frac{b(z)}{2} & -\frac{i b(z)}{2}  & 0 & 0 \\
 -\frac{1}{2} i d(z) & -\frac{d(z)}{2} & -\frac{i b(z)}{2}  & -\frac{b(z)}{2} & 0 & 0 \\
 \frac{e(z)}{2} & a(z)-\frac{i e(z)}{2}  & 0 & 0 & \frac{b(z)}{2} & -\frac{i b(z)}{2}  \\
 -a(z)-\frac{i e(z)}{2} & -\frac{e(z)}{2} & 0 & 0 & -\frac{i b(z)}{2}  & -\frac{b(z)}{2} \\
 0 & -i c(z) & \frac{e(z)}{2} & -a(z)-\frac{i e(z)}{2}  & -\frac{d(z)}{2} & \frac{i d(z)}{2} \\
 i c(z) & 0 & a(z)-\frac{i e(z)}{2}  & -\frac{e(z)}{2} & \frac{i d(z)}{2}  & \frac{d(z)}{2} \\
\end{array}
\right)\, .
\end{equation}
The functions $a(z),b(z)$ make up $\Gamma_{-1}(z)$, $c(z),d(z)$ make up $\Gamma_{-2}(z)$, and $e(z)$ makes up $\Gamma_{-3}(z)$. These functions satisfy the differential relations
\begin{equation}
\begin{aligned}
a(z) + \pi z c'(z)\, , \qquad b(z) (\pi z a'(z) -1) = \pi z (a(z)b'(z)-2i d'(z))\, ,\\
6 d(z) \left(\pi  z a'(z)+1\right)+6 i \pi  z \left(c(z)
   b'(z)+2 e'(z)\right) = a(z) \left(2 \pi  z d'(z)+i b(z)\right)
\end{aligned}
\end{equation}

\paragraph{Asymptotic periods.} For the ansatz $a(z) = A z$ and $b(z)=B z$, the other three functions are given by
\begin{equation}
c(z) =-  \frac{A z}{\pi}\, , \quad d(z) = \frac{B z}{2\pi i}\, , \qquad e(z) = \frac{z(2+3\pi z A)B}{8\pi^2}\, .
\end{equation}
The period expansion at the boundary is most conveniently written out as
\begin{equation}
\Pi = \left(
\begin{array}{cccccc}
 1 & 0 & 0 & 0 & 0 & 0 \\
 0 & 1 & 0 & 0 & 0 & 0 \\
 t & 0 & 1 & 0 & 0 & 0 \\
 0 & t & 0 & 1 & 0 & 0 \\
 \frac{t^2}{2} & 0 & t & 0 & 1 & 0 \\
 0 & \frac{t^2}{2} & 0 & t & 0 & 1 \\
\end{array}
\right) (a_0 + e^{2\pi i t}a_1 + e^{4\pi i t}a_2 + e^{6\pi i t}a_3 + e^{8\pi i t}a_4+\mathcal{O}(e^{4\pi i t}))
\end{equation}
where the first factor is just $e^{tN}$ written out, and the vectors it act on are given by
\begin{equation}
\begin{aligned}
    a_0 &= (1, i, 0,0,-\xi, -i \xi) \, , \\
    a_1 &= \left(\tfrac{B}{2\pi i}, -\tfrac{B}{2\pi}, iA+\tfrac{B}{4\pi^2}, -A-\tfrac{i B}{4\pi^2}, -\tfrac{A}{\pi}-\tfrac{iB \xi}{2\pi}, -\tfrac{iA}{\pi}-\tfrac{B\xi}{2\pi} \right)\, , \\
    a_2 &= \left(\tfrac{i A B}{2} , \tfrac{A B}{2} , -\tfrac{3 A B}{8 \pi } , \tfrac{3 i A B}{8 \pi } , \tfrac{A^2}{2}+\tfrac{1}{2} i A B \xi -\tfrac{i A B}{4 \pi ^2} , \tfrac{i A^2}{2}+\tfrac{A B \xi
   }{2}-\tfrac{A B}{4 \pi ^2} \right)\, , \\
   a_3 &= \left(0 , 0 , \tfrac{A^2 B}{3} , -\tfrac{1}{3} i A^2 B , \tfrac{3 i A^2 B}{8 \pi } , \tfrac{3 A^2 B}{8 \pi }\right) 
   \\
    a_4 & = (0 , 0 , 0 , 0 , \tfrac{1}{12} i A^3 B , \tfrac{A^3 B}{12})\, .
\end{aligned}
\end{equation}
We included terms up to order $e^{8\pi i t}$ to guarantee that the period vector satisfies the orthogonality conditions. Nonetheless, we note that there may be quadratic terms that go beyond our first order ansatz for $a(z)$, and thereby produce terms at order $e^{4\pi i t}$ in the period vector.

\subsection{Type $\mathrm{V}_{1,1}$ boundaries}\label{ssec:V}
In this section we construct the asymptotic periods for $\mathrm{V}_{1,1}$ boundaries. The boundary data is given by
\begin{equation}
\begin{aligned}
I^{4,4}&: \quad (1,\, 0,\, 0,\, 0,\, 0,\, 0),\\
I^{3,3}&: \quad (0,\, 1,\, 0,\, 0,\, 0,\, 0)\, ,\\
I^{2,2}&: \quad (0,\, 0,\, 1,\, 0,\, 0,\, 0)\,  \\
I^{1,1}&:\quad (0,\, 0,\, 0,\, 1,\, 0,\, 0)\, , \\
I^{0,0}&:\quad (0,\, 0,\, 0,\, 0,\, 1,\, 0)\, .
\end{aligned}
\end{equation}
The bilinear pairing and log-monodromy matrix are given by
\begin{equation}
\Sigma = \left(
\begin{array}{ccccc}
 0 & 0 & 0 & 0 & 1 \\
 0 & 0 & 0 & 1 & 0 \\
 0 & 0 & 1 & 0 & 0 \\
 0 & 1 & 0 & 0 & 0 \\
 1 & 0 & 0 & 0 & 0 \\
\end{array}
\right)\, , \quad N = \left(
\begin{array}{ccccc}
 0 & 0 & 0 & 0 & 0 \\
 1 & 0 & 0 & 0 & 0 \\
 0 & 1 & 0 & 0 & 0 \\
 0 & 0 & -1 & 0 & 0 \\
 0 & 0 & 0 & -1 & 0 \\
\end{array}
\right)
\end{equation}
The phase operator reads
\begin{equation}
\delta = \left(
\begin{array}{ccccc}
 0 & 0 & 0 & 0 & 0 \\
 0 & 0 & 0 & 0 & 0 \\
 0 & 0 & 0 & 0 & 0 \\
 \xi  & 0 & 0 & 0 & 0 \\
 0 & -\xi  & 0 & 0 & 0 \\
\end{array}
\right)\, , 
\end{equation}
where $\xi \in \mathbb{R}$, and we removed the part of $\delta$ proportional to the log-monodromy matrix by a coordinate shift.

\paragraph{Instanton map.} The instanton map reads
\begin{equation}
\Gamma(z) = \left(
\begin{array}{ccccc}
 0 & 0 & 0 & 0 & 0 \\
 a(z) & 0 & 0 & 0 & 0 \\
 b(z) & -a(z) & 0 & 0 & 0 \\
 c(z) & 0 & a(z) & 0 & 0 \\
 0 & -c(z) & -b(z) & -a(z) & 0 \\
\end{array}
\right)\, ,
\end{equation}
where $a(z)$ corresponds to the $\Gamma_{-1}(z)$ component, $b(z)$ to $\Gamma_{-2}(z)$, and $c(z)$ to $\Gamma_{-3}(z)$. These functions satisfy the differential relations
\begin{equation}
a(z) + i \pi z b'(z) = 0\, , \qquad i \pi z(b(z) a'(z)+2c'(z)) = b(z)\, .
\end{equation}

\paragraph{Asymptotic periods.} We make the leading approximation $a(z)=A z$, for which we find the other two functions to be
\begin{equation}
b(z) = \frac{i A z}{\pi}\, , \qquad c(z) = \frac{Az}{4\pi^2} (2-\pi i A z)\, , .
\end{equation}
The resulting period vector is most conveniently expressed as
\begin{equation}
    \Pi = \left(
\begin{array}{ccccc}
 1 & 0 & 0 & 0 & 0 \\
 t & 1 & 0 & 0 & 0 \\
 \frac{t^2}{2} & t & 1 & 0 & 0 \\
 -\frac{t^3}{6} & -\frac{t^2}{2} & -t & 1 & 0 \\
 \frac{t^4}{24} & \frac{t^3}{6} & \frac{t^2}{2} & -t & 1 \\
\end{array}
\right) (a_0 + e^{2\pi i t}a_1 + e^{4\pi i t}a_2 + e^{6\pi i t}a_3 + e^{8\pi i t}a_4 +\mathcal{O}(e^{4\pi i t}))\, ,
\end{equation}
where we wrote out $e^{tN}$ in matrix form, which acts on the vectors
\begin{equation}
\begin{aligned}
    a_0 &= (1, 0,0,i\xi,0)\, , \quad &a_1 &= (0,A,\tfrac{i A}{\pi },\tfrac{A}{2 \pi ^2},-i A \xi ) \, , \\
    a_2 &= (0,0,-\tfrac{A^2}{2},\tfrac{i A^2}{4 \pi },0)\, , \quad &a_3 &= ( 0,0,0,-\tfrac{A^3}{6},\tfrac{i A^3}{4 \pi })\, , \\
    a_4 &= ( 0,0,0,0,\tfrac{A^4}{24})\, .
\end{aligned}
\end{equation}
We included terms up to order $e^{8\pi i t}$ to guarantee that the period vector satisfies the orthogonality conditions. Nonetheless, we note that there may be quadratic terms that correct the ansatz for $a(z)$, and thereby produce terms at order $e^{4\pi i t}$ in the period vector.

\bibliography{bibl}

@article{rolin_quasianalytic_2006,
	title = {Quasianalytic solutions of differential equations and o-minimal structures},
	eprint = "math/0505073",
    archivePrefix = "arXiv",
    primaryClass = "math.CA",
 	author = {Rolin, J.-P. and Sanz, F. and Schaefke, R.},
	month = jun,
	year = {2006}
}

@article{Douglas:2023fcg,
    author = "Douglas, Michael R. and Grimm, Thomas W. and Schlechter, Lorenz",
    title = "{The Tameness of Quantum Field Theory, Part II -- Structures and CFTs}",
    eprint = "2302.04275",
    archivePrefix = "arXiv",
    primaryClass = "hep-th",
    month = "2",
    year = "2023"
}

@article{Schimmrigk:2018gch,
    author = "Schimmrigk, Rolf",
    title = "{The Swampland Spectrum Conjecture in Inflation}",
    eprint = "1810.11699",
    archivePrefix = "arXiv",
    primaryClass = "hep-th",
    month = "10",
    year = "2018"
}

@article{Rahimy:2025iyj,
    author = "Rahimy, Saba and Teixeira, Elsa M. and Zavala, Ivonne",
    title = "{Deciphering coupled scalar dark sectors}",
    eprint = "2503.01961",
    archivePrefix = "arXiv",
    primaryClass = "hep-th",
    doi = "10.1103/8hgv-l6ph",
    journal = "Phys. Rev. D",
    volume = "112",
    number = "4",
    pages = "043512",
    year = "2025"
}

@article{Ferreira:1997hj,
    author = "Ferreira, Pedro G. and Joyce, Michael",
    title = "{Cosmology with a primordial scaling field}",
    eprint = "astro-ph/9711102",
    archivePrefix = "arXiv",
    reportNumber = "CFPA-97-TH-20",
    doi = "10.1103/PhysRevD.58.023503",
    journal = "Phys. Rev. D",
    volume = "58",
    pages = "023503",
    year = "1998"
}

@article{Licciardello:2025fhx,
    author = "Licciardello, Daniele and Rahimy, Saba and Zavala, Ivonne",
    title = "{Extending the Dynamical Systems Toolkit: Coupled Fields in Multiscalar Dark Energy}",
    eprint = "2509.02539",
    archivePrefix = "arXiv",
    primaryClass = "hep-th",
    month = "9",
    year = "2025"
}

@article{Mohseni:2024njl,
    author = "Mohseni, Amineh and Montero, Miguel and Vafa, Cumrun and Valenzuela, Irene",
    title = "{On measuring distances in the quantum gravity landscape}",
    eprint = "2407.02705",
    archivePrefix = "arXiv",
    primaryClass = "hep-th",
    reportNumber = "IFT-24-097, CERN-TH-2024-101",
    doi = "10.1007/JHEP12(2024)168",
    journal = "JHEP",
    volume = "12",
    pages = "168",
    year = "2024"
}

@article{Payeur:2024kyy,
    author = "Payeur, Guillaume and McDonough, Evan and Brandenberger, Robert",
    title = "{Swampland conjectures constraints on dark energy from a highly curved field space}",
    eprint = "2405.05304",
    archivePrefix = "arXiv",
    primaryClass = "hep-th",
    doi = "10.1103/PhysRevD.110.106011",
    journal = "Phys. Rev. D",
    volume = "110",
    number = "10",
    pages = "106011",
    year = "2024"
}

@article{Andriot:2025cyi,
    author = "Andriot, David and Cribiori, Niccol\`o and Van Riet, Thomas",
    title = "{Scale separation, rolling solutions and entropy bounds}",
    eprint = "2504.08634",
    archivePrefix = "arXiv",
    primaryClass = "hep-th",
    month = "4",
    year = "2025"
}

@article{vandeHeisteeg:2023uxj,
    author = "van de Heisteeg, Damian and Vafa, Cumrun and Wiesner, Max and Wu, David H.",
    title = "{Bounds on field range for slowly varying positive potentials}",
    eprint = "2305.07701",
    archivePrefix = "arXiv",
    primaryClass = "hep-th",
    doi = "10.1007/JHEP02(2024)175",
    journal = "JHEP",
    volume = "02",
    pages = "175",
    year = "2024"
}

@article{Bedroya:2025ris,
    author = "Bedroya, Alek and Lee, Hayden and Steinhardt, Paul",
    title = "{A species scale-driven breakdown of effective field theory in time-dependent string backgrounds}",
    eprint = "2504.13260",
    archivePrefix = "arXiv",
    primaryClass = "hep-th",
    month = "4",
    year = "2025"
}

@article{Palti:2024voy,
    author = "Palti, Eran and Petri, Nicol\`o",
    title = "{A positive metric over DGKT vacua}",
    eprint = "2405.01084",
    archivePrefix = "arXiv",
    primaryClass = "hep-th",
    doi = "10.1007/JHEP06(2024)019",
    journal = "JHEP",
    volume = "06",
    pages = "019",
    year = "2024"
}

@article{Basile:2023rvm,
    author = "Basile, Ivano and Montella, Carmine",
    title = "{Domain walls and distances in discrete landscapes}",
    eprint = "2309.04519",
    archivePrefix = "arXiv",
    primaryClass = "hep-th",
    doi = "10.1007/JHEP02(2024)227",
    journal = "JHEP",
    volume = "02",
    pages = "227",
    year = "2024"
}

@article{Shiu:2023bay,
    author = "Shiu, Gary and Tonioni, Flavio and Van Hemelryck, Vincent and Van Riet, Thomas",
    title = "{Connecting flux vacua through scalar field excursions}",
    eprint = "2311.10828",
    archivePrefix = "arXiv",
    primaryClass = "hep-th",
    reportNumber = "UUITP-32/23",
    doi = "10.1103/PhysRevD.109.066017",
    journal = "Phys. Rev. D",
    volume = "109",
    number = "6",
    pages = "066017",
    year = "2024"
}

@article{Kehagias:2019akr,
    author = {Kehagias, Alex and L\"ust, Dieter and L\"ust, Severin},
    title = "{Swampland, Gradient Flow and Infinite Distance}",
    eprint = "1910.00453",
    archivePrefix = "arXiv",
    primaryClass = "hep-th",
    reportNumber = "MPP-2019-198, LMU-ASC 32/19, IPhT-T19/133, CPHT-RR055.092019",
    doi = "10.1007/JHEP04(2020)170",
    journal = "JHEP",
    volume = "04",
    pages = "170",
    year = "2020"
}

@article{Li:2023gtt,
    author = "Li, Yixuan and Palti, Eran and Petri, Nicol\`o",
    title = "{Towards AdS distances in string theory}",
    eprint = "2306.02026",
    archivePrefix = "arXiv",
    primaryClass = "hep-th",
    doi = "10.1007/JHEP08(2023)210",
    journal = "JHEP",
    volume = "08",
    pages = "210",
    year = "2023"
}

@article{Demulder:2024glx,
    author = "Demulder, Saskia and Lust, Dieter and Raml, Thomas",
    title = "{Navigating string theory field space with geometric flows}",
    eprint = "2412.10364",
    archivePrefix = "arXiv",
    primaryClass = "hep-th",
    month = "12",
    year = "2024"
}

@article{Palti:2025ydz,
    author = "Palti, Eran and Petri, Nicol\`o",
    title = "{Metrics over multi-parameter AdS vacua}",
    eprint = "2504.01316",
    archivePrefix = "arXiv",
    primaryClass = "hep-th",
    month = "4",
    year = "2025"
}

@article{Shiu:2022oti,
    author = "Shiu, Gary and Tonioni, Flavio and Van Hemelryck, Vincent and Van Riet, Thomas",
    title = "{AdS scale separation and the distance conjecture}",
    eprint = "2212.06169",
    archivePrefix = "arXiv",
    primaryClass = "hep-th",
    doi = "10.1007/JHEP05(2023)077",
    journal = "JHEP",
    volume = "05",
    pages = "077",
    year = "2023"
}

@article{Lust:2019zwm,
    author = {L\"ust, Dieter and Palti, Eran and Vafa, Cumrun},
    title = "{AdS and the Swampland}",
    eprint = "1906.05225",
    archivePrefix = "arXiv",
    primaryClass = "hep-th",
    doi = "10.1016/j.physletb.2019.134867",
    journal = "Phys. Lett. B",
    volume = "797",
    pages = "134867",
    year = "2019"
}

@article{Ooguri:2018wrx,
    author = "Ooguri, Hirosi and Palti, Eran and Shiu, Gary and Vafa, Cumrun",
    title = "{Distance and de Sitter Conjectures on the Swampland}",
    eprint = "1810.05506",
    archivePrefix = "arXiv",
    primaryClass = "hep-th",
    doi = "10.1016/j.physletb.2018.11.018",
    journal = "Phys. Lett. B",
    volume = "788",
    pages = "180--184",
    year = "2019"
}

@article{Hebecker:2018vxz,
    author = "Hebecker, Arthur and Wrase, Timm",
    title = "{The Asymptotic dS Swampland Conjecture - a Simplified Derivation and a Potential Loophole}",
    eprint = "1810.08182",
    archivePrefix = "arXiv",
    primaryClass = "hep-th",
    doi = "10.1002/prop.201800097",
    journal = "Fortsch. Phys.",
    volume = "67",
    number = "1-2",
    pages = "1800097",
    year = "2019"
}

@article{Scalisi:2018eaz,
    author = "Scalisi, Marco and Valenzuela, Irene",
    title = "{Swampland distance conjecture, inflation and $\alpha$-attractors}",
    eprint = "1812.07558",
    archivePrefix = "arXiv",
    primaryClass = "hep-th",
    doi = "10.1007/JHEP08(2019)160",
    journal = "JHEP",
    volume = "08",
    pages = "160",
    year = "2019"
}

@article{Hebecker:2023qke,
    author = "Hebecker, Arthur and Schreyer, Simon and Venken, Victoria",
    title = "{No asymptotic acceleration without higher-dimensional de Sitter vacua}",
    eprint = "2306.17213",
    archivePrefix = "arXiv",
    primaryClass = "hep-th",
    doi = "10.1007/JHEP11(2023)173",
    journal = "JHEP",
    volume = "11",
    pages = "173",
    year = "2023"
}

@article{Allahverdi:2010xz,
    author = "Allahverdi, Rouzbeh and Brandenberger, Robert and Cyr-Racine, Francis-Yan and Mazumdar, Anupam",
    title = "{Reheating in Inflationary Cosmology: Theory and Applications}",
    eprint = "1001.2600",
    archivePrefix = "arXiv",
    primaryClass = "hep-th",
    doi = "10.1146/annurev.nucl.012809.104511",
    journal = "Ann. Rev. Nucl. Part. Sci.",
    volume = "60",
    pages = "27--51",
    year = "2010"
}

@article{Amin:2014eta,
    author = "Amin, Mustafa A. and Hertzberg, Mark P. and Kaiser, David I. and Karouby, Johanna",
    title = "{Nonperturbative Dynamics Of Reheating After Inflation: A Review}",
    eprint = "1410.3808",
    archivePrefix = "arXiv",
    primaryClass = "hep-ph",
    doi = "10.1142/S0218271815300037",
    journal = "Int. J. Mod. Phys. D",
    volume = "24",
    pages = "1530003",
    year = "2014"
}

@article{Cicoli:2020cfj,
    author = "Cicoli, Michele and Dibitetto, Giuseppe and Pedro, Francisco G.",
    title = "{New accelerating solutions in late-time cosmology}",
    eprint = "2002.02695",
    archivePrefix = "arXiv",
    primaryClass = "gr-qc",
    doi = "10.1103/PhysRevD.101.103524",
    journal = "Phys. Rev. D",
    volume = "101",
    number = "10",
    pages = "103524",
    year = "2020"
}

@article{Cicoli:2020noz,
    author = "Cicoli, Michele and Dibitetto, Giuseppe and Pedro, Francisco G.",
    title = "{Out of the Swampland with Multifield Quintessence?}",
    eprint = "2007.11011",
    archivePrefix = "arXiv",
    primaryClass = "hep-th",
    doi = "10.1007/JHEP10(2020)035",
    journal = "JHEP",
    volume = "10",
    pages = "035",
    year = "2020"
}

@article{Cicoli:2023opf,
    author = "Cicoli, Michele and Conlon, Joseph P. and Maharana, Anshuman and Parameswaran, Susha and Quevedo, Fernando and Zavala, Ivonne",
    title = "{String cosmology: From the early universe to today}",
    eprint = "2303.04819",
    archivePrefix = "arXiv",
    primaryClass = "hep-th",
    doi = "10.1016/j.physrep.2024.01.002",
    journal = "Phys. Rept.",
    volume = "1059",
    pages = "1--155",
    year = "2024"
}

@article{Andriot:2024sif,
    author = "Andriot, David",
    title = "{Quintessence: an analytical study, with theoretical and observational applications}",
    eprint = "2410.17182",
    archivePrefix = "arXiv",
    primaryClass = "hep-th",
    month = "10",
    year = "2024"
}

@article{Sonner:2006yn,
    author = "Sonner, Julian and Townsend, Paul K.",
    title = "{Recurrent acceleration in dilaton-axion cosmology}",
    eprint = "hep-th/0608068",
    archivePrefix = "arXiv",
    reportNumber = "DAMTP-2006-63",
    doi = "10.1103/PhysRevD.74.103508",
    journal = "Phys. Rev. D",
    volume = "74",
    pages = "103508",
    year = "2006"
}

@article{Russo:2022pgo,
    author = "Russo, J. G. and Townsend, P. K.",
    title = "{A dilaton-axion model for string cosmology}",
    eprint = "2203.09398",
    archivePrefix = "arXiv",
    primaryClass = "hep-th",
    doi = "10.1007/JHEP06(2022)001",
    journal = "JHEP",
    volume = "06",
    pages = "001",
    year = "2022"
}

@article{Shiu:2024sbe,
    author = "Shiu, Gary and Tonioni, Flavio and Tran, Hung V.",
    title = "{Analytic bounds on late-time axion-scalar cosmologies}",
    eprint = "2406.17030",
    archivePrefix = "arXiv",
    primaryClass = "hep-th",
    doi = "10.1007/JHEP09(2024)158",
    journal = "JHEP",
    volume = "09",
    pages = "158",
    year = "2024"
}

@article{Copeland:1997et,
    author = "Copeland, Edmund J. and Liddle, Andrew R and Wands, David",
    title = "{Exponential potentials and cosmological scaling solutions}",
    eprint = "gr-qc/9711068",
    archivePrefix = "arXiv",
    reportNumber = "SUSX-TH-97-022, SUSSEX-AST-97-11-1, PU-RCG-97-20",
    doi = "10.1103/PhysRevD.57.4686",
    journal = "Phys. Rev. D",
    volume = "57",
    pages = "4686--4690",
    year = "1998"
}

@article{Conlon:2022pnx,
    author = "Conlon, Joseph P. and Revello, Filippo",
    title = "{Catch-me-if-you-can: the overshoot problem and the weak/inflation hierarchy}",
    eprint = "2207.00567",
    archivePrefix = "arXiv",
    primaryClass = "hep-th",
    doi = "10.1007/JHEP11(2022)155",
    journal = "JHEP",
    volume = "11",
    pages = "155",
    year = "2022"
}

@article{Apers:2022cyl,
    author = "Apers, Fien and Conlon, Joseph P. and Mosny, Martin and Revello, Filippo",
    title = "{Kination, meet Kasner: on the asymptotic cosmology of string compactifications}",
    eprint = "2212.10293",
    archivePrefix = "arXiv",
    primaryClass = "hep-th",
    doi = "10.1007/JHEP08(2023)156",
    journal = "JHEP",
    volume = "08",
    pages = "156",
    year = "2023"
}

@article{Andriot:2024jsh,
    author = "Andriot, David and Parameswaran, Susha and Tsimpis, Dimitrios and Wrase, Timm and Zavala, Ivonne",
    title = "{Exponential quintessence: curved, steep and stringy?}",
    eprint = "2405.09323",
    archivePrefix = "arXiv",
    primaryClass = "hep-th",
    doi = "10.1007/JHEP08(2024)117",
    journal = "JHEP",
    volume = "08",
    pages = "117",
    year = "2024"
}

@article{Shiu:2023fhb,
    author = "Shiu, Gary and Tonioni, Flavio and Tran, Hung V.",
    title = "{Late-time attractors and cosmic acceleration}",
    eprint = "2306.07327",
    archivePrefix = "arXiv",
    primaryClass = "hep-th",
    doi = "10.1103/PhysRevD.108.063528",
    journal = "Phys. Rev. D",
    volume = "108",
    number = "6",
    pages = "063528",
    year = "2023"
}

@article{Cicoli:2018kdo,
    author = "Cicoli, Michele and De Alwis, Senarath and Maharana, Anshuman and Muia, Francesco and Quevedo, Fernando",
    title = "{De Sitter vs Quintessence in String Theory}",
    eprint = "1808.08967",
    archivePrefix = "arXiv",
    primaryClass = "hep-th",
    doi = "10.1002/prop.201800079",
    journal = "Fortsch. Phys.",
    volume = "67",
    number = "1-2",
    pages = "1800079",
    year = "2019"
}

@article{Bedroya:2019snp,
    author = "Bedroya, Alek and Vafa, Cumrun",
    title = "{Trans-Planckian Censorship and the Swampland}",
    eprint = "1909.11063",
    archivePrefix = "arXiv",
    primaryClass = "hep-th",
    doi = "10.1007/JHEP09(2020)123",
    journal = "JHEP",
    volume = "09",
    pages = "123",
    year = "2020"
}

@article{Montero:2019ekk,
    author = "Montero, Miguel and Van Riet, Thomas and Venken, Victoria",
    title = "{Festina Lente: EFT Constraints from Charged Black Hole Evaporation in de Sitter}",
    eprint = "1910.01648",
    archivePrefix = "arXiv",
    primaryClass = "hep-th",
    doi = "10.1007/JHEP01(2020)039",
    journal = "JHEP",
    volume = "01",
    pages = "039",
    year = "2020"
}

@article{Montero:2021otb,
    author = "Montero, Miguel and Vafa, Cumrun and Van Riet, Thomas and Venken, Victoria",
    title = "{The FL bound and its phenomenological implications}",
    eprint = "2106.07650",
    archivePrefix = "arXiv",
    primaryClass = "hep-th",
    reportNumber = "UUITP-26/2",
    doi = "10.1007/JHEP10(2021)009",
    journal = "JHEP",
    volume = "10",
    pages = "009",
    year = "2021"
}

@article{Bedroya:2019tba,
    author = "Bedroya, Alek and Brandenberger, Robert and Loverde, Marilena and Vafa, Cumrun",
    title = "{Trans-Planckian Censorship and Inflationary Cosmology}",
    eprint = "1909.11106",
    archivePrefix = "arXiv",
    primaryClass = "hep-th",
    doi = "10.1103/PhysRevD.101.103502",
    journal = "Phys. Rev. D",
    volume = "101",
    number = "10",
    pages = "103502",
    year = "2020"
}

@article{Andriot:2023wvg,
    author = "Andriot, David and Tsimpis, Dimitrios and Wrase, Timm",
    title = "{Accelerated expansion of an open universe and string theory realizations}",
    eprint = "2309.03938",
    archivePrefix = "arXiv",
    primaryClass = "hep-th",
    doi = "10.1103/PhysRevD.108.123515",
    journal = "Phys. Rev. D",
    volume = "108",
    number = "12",
    pages = "123515",
    year = "2023"
}

@article{Cicoli:2021fsd,
    author = "Cicoli, Michele and Cunillera, Francesc and Padilla, Antonio and Pedro, Francisco G.",
    title = "{Quintessence and the Swampland: The Parametrically Controlled Regime of Moduli Space}",
    eprint = "2112.10779",
    archivePrefix = "arXiv",
    primaryClass = "hep-th",
    doi = "10.1002/prop.202200009",
    journal = "Fortsch. Phys.",
    volume = "70",
    number = "4",
    pages = "2200009",
    year = "2022"
}

@article{Shiu:2023nph,
    author = "Shiu, Gary and Tonioni, Flavio and Tran, Hung V.",
    title = "{Accelerating universe at the end of time}",
    eprint = "2303.03418",
    archivePrefix = "arXiv",
    primaryClass = "hep-th",
    doi = "10.1103/PhysRevD.108.063527",
    journal = "Phys. Rev. D",
    volume = "108",
    number = "6",
    pages = "063527",
    year = "2023"
}

@article{Gerhardus:2016iot,
    author = "Gerhardus, Andreas and Jockers, Hans",
    title = "{Quantum periods of Calabi\textendash{}Yau fourfolds}",
    eprint = "1604.05325",
    archivePrefix = "arXiv",
    primaryClass = "hep-th",
    reportNumber = "BONN-TH-2016-02",
    doi = "10.1016/j.nuclphysb.2016.09.021",
    journal = "Nucl. Phys. B",
    volume = "913",
    pages = "425--474",
    year = "2016"
}

@article{Apers:2024ffe,
    author = "Apers, Fien and Conlon, Joseph P. and Copeland, Edmund J. and Mosny, Martin and Revello, Filippo",
    title = "{String theory and the first half of the universe}",
    eprint = "2401.04064",
    archivePrefix = "arXiv",
    primaryClass = "hep-th",
    doi = "10.1088/1475-7516/2024/08/018",
    journal = "JCAP",
    volume = "08",
    pages = "018",
    year = "2024"
}

@article{Brinkmann:2022oxy,
    author = "Brinkmann, Max and Cicoli, Michele and Dibitetto, Giuseppe and Pedro, Francisco G.",
    title = "{Stringy multifield quintessence and the Swampland}",
    eprint = "2206.10649",
    archivePrefix = "arXiv",
    primaryClass = "hep-th",
    doi = "10.1007/JHEP11(2022)044",
    journal = "JHEP",
    volume = "11",
    pages = "044",
    year = "2022"
}

@article{Lozanov:2019jxc,
    author = "Lozanov, Kaloian D.",
    title = "{Lectures on Reheating after Inflation}",
    eprint = "1907.04402",
    archivePrefix = "arXiv",
    primaryClass = "astro-ph.CO",
    month = "7",
    year = "2019"
}

@article{Tonioni:2024huw,
    author = "Tonioni, Flavio",
    title = "{A mechanism for freezing moduli into Minkowski spacetime}",
    eprint = "2407.21104",
    archivePrefix = "arXiv",
    primaryClass = "hep-th",
    doi = "10.1016/j.physletb.2025.139455",
    journal = "Phys. Lett. B",
    volume = "865",
    pages = "139455",
    year = "2025"
}

@article{Calderon-Infante:2022nxb,
    author = "Calder\'on-Infante, Jos\'e and Ruiz, Ignacio and Valenzuela, Irene",
    title = "{Asymptotic accelerated expansion in string theory and the Swampland}",
    eprint = "2209.11821",
    archivePrefix = "arXiv",
    primaryClass = "hep-th",
    reportNumber = "CERN-TH-2022-153, IFT-UAM/CSIC-22-110",
    doi = "10.1007/JHEP06(2023)129",
    journal = "JHEP",
    volume = "06",
    pages = "129",
    year = "2023"
}

@article{Rudelius:2021oaz,
    author = "Rudelius, Tom",
    title = "{Dimensional reduction and (Anti) de Sitter bounds}",
    eprint = "2101.11617",
    archivePrefix = "arXiv",
    primaryClass = "hep-th",
    doi = "10.1007/JHEP08(2021)041",
    journal = "JHEP",
    volume = "08",
    pages = "041",
    year = "2021"
}

@article{Rudelius:2021azq,
    author = "Rudelius, Tom",
    title = "{Asymptotic observables and the swampland}",
    eprint = "2106.09026",
    archivePrefix = "arXiv",
    primaryClass = "hep-th",
    doi = "10.1103/PhysRevD.104.126023",
    journal = "Phys. Rev. D",
    volume = "104",
    number = "12",
    pages = "126023",
    year = "2021"
}

@article{Becker:2006ks,
    author = "Becker, Katrin and Becker, Melanie and Vafa, Cumrun and Walcher, Johannes",
    title = "{Moduli Stabilization in Non-Geometric Backgrounds}",
    eprint = "hep-th/0611001",
    archivePrefix = "arXiv",
    reportNumber = "HUTP-06-A044",
    doi = "10.1016/j.nuclphysb.2007.01.034",
    journal = "Nucl. Phys. B",
    volume = "770",
    pages = "1--46",
    year = "2007"
}

@article{Rajaguru:2024emw,
    author = "Rajaguru, Muthusamy and Sengupta, Anindya and Wrase, Timm",
    title = "{Fully stabilized Minkowski vacua in the 2$^{6}$ Landau-Ginzburg model}",
    eprint = "2407.16756",
    archivePrefix = "arXiv",
    primaryClass = "hep-th",
    doi = "10.1007/JHEP10(2024)095",
    journal = "JHEP",
    volume = "10",
    pages = "095",
    year = "2024"
}

@article{Becker:2024ayh,
    author = "Becker, Katrin and Brady, Nathan and Gra\~na, Mariana and Morros, Miguel and Sengupta, Anindya and You, Qi",
    title = "{Tadpole conjecture in non-geometric backgrounds}",
    eprint = "2407.16758",
    archivePrefix = "arXiv",
    primaryClass = "hep-th",
    doi = "10.1007/JHEP10(2024)021",
    journal = "JHEP",
    volume = "10",
    pages = "021",
    year = "2024"
}

@article{Chen:2025rkb,
    author = "Chen, Shi and van de Heisteeg, Damian and Vafa, Cumrun",
    title = "{Symmetries and M-theory-like Vacua in Four Dimensions}",
    eprint = "2503.16599",
    archivePrefix = "arXiv",
    primaryClass = "hep-th",
    month = "3",
    year = "2025"
}

@article{Gukov:1999ya,
    author = "Gukov, Sergei and Vafa, Cumrun and Witten, Edward",
    title = "{CFT's from Calabi-Yau four folds}",
    eprint = "hep-th/9906070",
    archivePrefix = "arXiv",
    reportNumber = "HUTP-99-A034, IASSNS-HEP-99-52, PUPT-1864",
    doi = "10.1016/S0550-3213(00)00373-4",
    journal = "Nucl. Phys. B",
    volume = "584",
    pages = "69--108",
    year = "2000",
    note = "[Erratum: Nucl.Phys.B 608, 477--478 (2001)]"
}

@article{Haack:2001jz,
    author = "Haack, Michael and Louis, Jan",
    title = "{M theory compactified on Calabi-Yau fourfolds with background flux}",
    eprint = "hep-th/0103068",
    archivePrefix = "arXiv",
    doi = "10.1016/S0370-2693(01)00464-6",
    journal = "Phys. Lett. B",
    volume = "507",
    pages = "296--304",
    year = "2001"
}

@article{Sethi:1996es,
    author = "Sethi, S. and Vafa, C. and Witten, Edward",
    title = "{Constraints on low dimensional string compactifications}",
    eprint = "hep-th/9606122",
    archivePrefix = "arXiv",
    reportNumber = "HUTP-96-A025, IASSNS-HEP-96-60",
    doi = "10.1016/S0550-3213(96)00483-X",
    journal = "Nucl. Phys. B",
    volume = "480",
    pages = "213--224",
    year = "1996"
}

@article{vandeHeisteeg:2024lsa,
    author = "van de Heisteeg, Damian",
    title = "{Charting the Complex Structure Landscape of F-theory}",
    eprint = "2404.03456",
    archivePrefix = "arXiv",
    primaryClass = "hep-th",
    month = "4",
    year = "2024"
}

@article{Schmid,
	Author = {Wilfried Schmid},
	Journal = {Invent. Math. , 22:211--319, 1973},
	year = {1973},
	Title = "{Variation of Hodge structure: the singularities of the period mapping}"
}

@article{CattaniDeligneKaplan,
 ISSN = {08940347, 10886834},
 URL = {http://www.jstor.org/stable/2152824},
 author = {Eduardo Cattani and Pierre Deligne and Aroldo Kaplan},
 journal = {Journal of the American Mathematical Society},
 number = {2},
 pages = {483--506},
 publisher = {American Mathematical Society},
 title = {On the Locus of Hodge Classes},
 urldate = {2022-04-06},
 volume = {8},
 year = {1995}
}

@article{Cattani_2003,
   title={Frobenius Modules and Hodge Asymptotics},
   volume={238},
   ISSN={1432-0916},
   url={http://dx.doi.org/10.1007/s00220-003-0848-y},
   DOI={10.1007/s00220-003-0848-y},
   number={3},
   journal={Communications in Mathematical Physics},
   publisher={Springer Science and Business Media LLC},
   author={Cattani, Eduardo and Fernandez, Javier},
   year={2003},
   month=jul, pages={489–504} }

@article{Fernandez_2008,
   title={Infinitesimal variations of Hodge structure at infinity},
   volume={139},
   ISSN={1572-9168},
   url={http://dx.doi.org/10.1007/s10711-008-9330-5},
   DOI={10.1007/s10711-008-9330-5},
   number={1},
   journal={Geometriae Dedicata},
   publisher={Springer Science and Business Media LLC},
   author={Fernandez, Javier and Cattani, Eduardo},
   year={2008},
   month=nov, pages={299–312} }

@article{Castellano:2023stg,
    author = "Castellano, Alberto and Ruiz, Ignacio and Valenzuela, Irene",
    title = "{Universal Pattern in Quantum Gravity at Infinite Distance}",
    eprint = "2311.01501",
    archivePrefix = "arXiv",
    primaryClass = "hep-th",
    reportNumber = "CERN-TH-2023-203",
    doi = "10.1103/PhysRevLett.132.181601",
    journal = "Phys. Rev. Lett.",
    volume = "132",
    number = "18",
    pages = "181601",
    year = "2024"
}

@article{Castellano:2023jjt,
    author = "Castellano, Alberto and Ruiz, Ignacio and Valenzuela, Irene",
    title = "{Stringy evidence for a universal pattern at infinite distance}",
    eprint = "2311.01536",
    archivePrefix = "arXiv",
    primaryClass = "hep-th",
    reportNumber = "CERN-TH-2023-204",
    doi = "10.1007/JHEP06(2024)037",
    journal = "JHEP",
    volume = "06",
    pages = "037",
    year = "2024"
}

@article{EDMUNDS201247,
title = {Properties of generalized trigonometric functions},
journal = {Journal of Approximation Theory},
volume = {164},
number = {1},
pages = {47-56},
year = {2012},
issn = {0021-9045},
doi = {https://doi.org/10.1016/j.jat.2011.09.004},
url = {https://www.sciencedirect.com/science/article/pii/S0021904511001560},
author = {David E. Edmunds and Petr Gurka and Jan Lang},
}

@article{Ooguri:2006in,
    author = "Ooguri, Hirosi and Vafa, Cumrun",
    title = "{On the Geometry of the String Landscape and the Swampland}",
    eprint = "hep-th/0605264",
    archivePrefix = "arXiv",
    reportNumber = "CALT-68-2600, HUTP-06-A017",
    doi = "10.1016/j.nuclphysb.2006.10.033",
    journal = "Nucl. Phys. B",
    volume = "766",
    pages = "21--33",
    year = "2007"
}

@article{Grimm:2019wtx,
    author = "Grimm, Thomas W. and Van De Heisteeg, Damian",
    title = "{Infinite Distances and the Axion Weak Gravity Conjecture}",
    eprint = "1905.00901",
    archivePrefix = "arXiv",
    primaryClass = "hep-th",
    doi = "10.1007/JHEP03(2020)020",
    journal = "JHEP",
    volume = "03",
    pages = "020",
    year = "2020"
}

@article{Lanza:2024uis,
    author = "Lanza, Stefano and Westphal, Alexander",
    title = "{Uplifts in the penumbra: features of the moduli potential away from infinite-distance boundaries}",
    eprint = "2412.12253",
    archivePrefix = "arXiv",
    primaryClass = "hep-th",
    reportNumber = "DESY-24-199",
    doi = "10.1007/JHEP05(2025)071",
    journal = "JHEP",
    volume = "05",
    pages = "071",
    year = "2025"
}

@article{Gendler:2020dfp,
    author = "Gendler, Naomi and Valenzuela, Irene",
    title = "{Merging the weak gravity and distance conjectures using BPS extremal black holes}",
    eprint = "2004.10768",
    archivePrefix = "arXiv",
    primaryClass = "hep-th",
    doi = "10.1007/JHEP01(2021)176",
    journal = "JHEP",
    volume = "01",
    pages = "176",
    year = "2021"
}

@article{Marchesano:2019ifh,
    author = "Marchesano, Fernando and Wiesner, Max",
    title = "{Instantons and infinite distances}",
    eprint = "1904.04848",
    archivePrefix = "arXiv",
    primaryClass = "hep-th",
    reportNumber = "IFT-UAM/CSIC-19-049",
    doi = "10.1007/JHEP08(2019)088",
    journal = "JHEP",
    volume = "08",
    pages = "088",
    year = "2019"
}

@article{Font:2019cxq,
    author = "Font, Anamar\'\i{}a and Herr\'aez, Alvaro and Ib\'a\~nez, Luis E.",
    title = "{The Swampland Distance Conjecture and Towers of Tensionless Branes}",
    eprint = "1904.05379",
    archivePrefix = "arXiv",
    primaryClass = "hep-th",
    reportNumber = "IFT-UAM/CSIC-19-50, IFT-UAM-CSIC-19-50",
    doi = "10.1007/JHEP08(2019)044",
    journal = "JHEP",
    volume = "08",
    pages = "044",
    year = "2019"
}

@book{Baumann:2014nda,
    author = "Baumann, Daniel and McAllister, Liam",
    title = "{Inflation and String Theory}",
    eprint = "1404.2601",
    archivePrefix = "arXiv",
    primaryClass = "hep-th",
    doi = "10.1017/CBO9781316105733",
    isbn = "978-1-107-08969-3, 978-1-316-23718-2",
    publisher = "Cambridge University Press",
    series = "Cambridge Monographs on Mathematical Physics",
    month = "5",
    year = "2015"
}

@article{vanBeest:2021lhn,
    author = "van Beest, Marieke and Calder\'on-Infante, Jos\'e and Mirfendereski, Delaram and Valenzuela, Irene",
    title = "{Lectures on the Swampland Program in String Compactifications}",
    eprint = "2102.01111",
    archivePrefix = "arXiv",
    primaryClass = "hep-th",
    doi = "10.1016/j.physrep.2022.09.002",
    journal = "Phys. Rept.",
    volume = "989",
    pages = "1--50",
    year = "2022"
}

@article{Palti:2019pca,
    author = "Palti, Eran",
    title = "{The Swampland: Introduction and Review}",
    eprint = "1903.06239",
    archivePrefix = "arXiv",
    primaryClass = "hep-th",
    reportNumber = "MPP-2019-53",
    doi = "10.1002/prop.201900037",
    journal = "Fortsch. Phys.",
    volume = "67",
    number = "6",
    pages = "1900037",
    year = "2019"
}

@article{CaboBizet:2014ovf,
    author = "Cabo Bizet, Nana and Klemm, Albrecht and Vieira Lopes, Daniel",
    title = "{Landscaping with fluxes and the E8 Yukawa Point in F-theory}",
    eprint = "1404.7645",
    archivePrefix = "arXiv",
    primaryClass = "hep-th",
    reportNumber = "BONN-TH-2013-21",
    month = "4",
    year = "2014"
}

@article{Klaewer:2016kiy,
    author = "Klaewer, Daniel and Palti, Eran",
    title = "{Super-Planckian Spatial Field Variations and Quantum Gravity}",
    eprint = "1610.00010",
    archivePrefix = "arXiv",
    primaryClass = "hep-th",
    doi = "10.1007/JHEP01(2017)088",
    journal = "JHEP",
    volume = "01",
    pages = "088",
    year = "2017"
}

@article{McAllister:2008hb,
    author = "McAllister, Liam and Silverstein, Eva and Westphal, Alexander",
    title = "{Gravity Waves and Linear Inflation from Axion Monodromy}",
    eprint = "0808.0706",
    archivePrefix = "arXiv",
    primaryClass = "hep-th",
    reportNumber = "SLAC-PUB-13357, SU-ITP-08-15",
    doi = "10.1103/PhysRevD.82.046003",
    journal = "Phys. Rev. D",
    volume = "82",
    pages = "046003",
    year = "2010"
}

@article{Silverstein:2008sg,
    author = "Silverstein, Eva and Westphal, Alexander",
    title = "{Monodromy in the CMB: Gravity Waves and String Inflation}",
    eprint = "0803.3085",
    archivePrefix = "arXiv",
    primaryClass = "hep-th",
    reportNumber = "SU-ITP-08-07, SLAC-PUB-13183",
    doi = "10.1103/PhysRevD.78.106003",
    journal = "Phys. Rev. D",
    volume = "78",
    pages = "106003",
    year = "2008"
}

@article{Calderon-Infante:2020dhm,
    author = "Calder\'on-Infante, Jos\'e and Uranga, Angel M. and Valenzuela, Irene",
    title = "{The Convex Hull Swampland Distance Conjecture and Bounds on Non-geodesics}",
    eprint = "2012.00034",
    archivePrefix = "arXiv",
    primaryClass = "hep-th",
    reportNumber = "IFT-UAM/CSIC-20-169",
    doi = "10.1007/JHEP03(2021)299",
    journal = "JHEP",
    volume = "03",
    pages = "299",
    year = "2021"
}

@article{Lee:2019xtm,
    author = "Lee, Seung-Joo and Lerche, Wolfgang and Weigand, Timo",
    title = "{Emergent strings, duality and weak coupling limits for two-form fields}",
    eprint = "1904.06344",
    archivePrefix = "arXiv",
    primaryClass = "hep-th",
    reportNumber = "CERN-TH-2019-044",
    doi = "10.1007/JHEP02(2022)096",
    journal = "JHEP",
    volume = "02",
    pages = "096",
    year = "2022"
}

@article{Grimm:2018ohb,
    author = "Grimm, Thomas W. and Palti, Eran and Valenzuela, Irene",
    title = "{Infinite Distances in Field Space and Massless Towers of States}",
    eprint = "1802.08264",
    archivePrefix = "arXiv",
    primaryClass = "hep-th",
    doi = "10.1007/JHEP08(2018)143",
    journal = "JHEP",
    volume = "08",
    pages = "143",
    year = "2018"
}

@article{Allahverdi:2020bys,
    author = "Allahverdi, Rouzbeh and others",
    title = "{The First Three Seconds: a Review of Possible Expansion Histories of the Early Universe}",
    eprint = "2006.16182",
    archivePrefix = "arXiv",
    primaryClass = "astro-ph.CO",
    reportNumber = "FERMILAB-PUB-20-242-A, KCL-PH-TH/2020-33, KEK-Cosmo-257,
  KEK-TH-2231, IPMU20-0070, PI/UAN-2020-674FT, RUP-20-22",
    doi = "10.21105/astro.2006.16182",
    journal = "Open J. Astrophys.",
    volume = "4",
    pages = "astro.2006.16182",
    year = "2021"
}

@article{Ghoshal:2025tlk,
    author = "Ghoshal, Anish and Revello, Filippo and Villa, Gonzalo",
    title = "{Cosmic superstrings in large volume compactifications: PTAs, LISA and time-varying tension}",
    eprint = "2504.20994",
    archivePrefix = "arXiv",
    primaryClass = "astro-ph.CO",
    month = "4",
    year = "2025"
}

@article{Revello:2024gwa,
    author = "Revello, Filippo and Villa, Gonzalo",
    title = "{Cosmic (super)strings with a time-varying tension}",
    eprint = "2411.04186",
    archivePrefix = "arXiv",
    primaryClass = "hep-ph",
    doi = "10.1088/1475-7516/2025/04/049",
    journal = "JCAP",
    volume = "04",
    pages = "049",
    year = "2025"
}

@article{Brunelli:2025ems,
    author = "Brunelli, Luca and Cicoli, Michele and Pedro, Francisco G.",
    title = "{Growth of cosmic strings beyond kination}",
    eprint = "2503.11293",
    archivePrefix = "arXiv",
    primaryClass = "hep-th",
    doi = "10.1088/1475-7516/2025/07/010",
    journal = "JCAP",
    volume = "07",
    pages = "010",
    year = "2025"
}

@article{SanchezGonzalez:2025uco,
    author = "S{\'a}nchez Gonz{\'a}lez, Noelia and Conlon, Joseph P. and Copeland, Edmund J. and Hardy, Edward",
    title = "{Dynamical Systems and Superstring Phases in the Early Universe}",
    eprint = "2505.14187",
    archivePrefix = "arXiv",
    primaryClass = "hep-ph",
    month = "5",
    year = "2025"
}

@article{Conlon:2024uob,
    author = "Conlon, Joseph P. and Copeland, Edmund J. and Hardy, Edward and Gonz{\'a}lez, Noelia S{\'a}nchez",
    title = "{Percolating cosmic string networks from kination}",
    eprint = "2406.12637",
    archivePrefix = "arXiv",
    primaryClass = "hep-ph",
    doi = "10.1103/PhysRevD.110.083537",
    journal = "Phys. Rev. D",
    volume = "110",
    number = "8",
    pages = "083537",
    year = "2024"
}

@article{Apers:2024dtn,
    author = "Apers, Fien and Conlon, Joseph P. and Mosny, Martin",
    title = "{A note on 4d kination and higher-dimensional uplifts}",
    eprint = "2409.08049",
    archivePrefix = "arXiv",
    primaryClass = "hep-th",
    doi = "10.1140/epjc/s10052-025-14030-2",
    journal = "Eur. Phys. J. C",
    volume = "85",
    number = "3",
    pages = "337",
    year = "2025"
}

@article{Mosny:2025cyd,
    author = "Mosny, Martin and Conlon, Joseph P. and Copeland, Edmund J.",
    title = "{Self-Tracking Solutions for Asymptotic Scalar Fields}",
    eprint = "2507.04161",
    archivePrefix = "arXiv",
    primaryClass = "hep-th",
    month = "7",
    year = "2025"
}

@phdthesis{Monnee:2024gsq,
    author = "Monnee, Jeroen",
    title = "{Asymptotic Hodge Theory in String Compactifications and Integrable Systems}",
    eprint = "2409.06794",
    archivePrefix = "arXiv",
    primaryClass = "hep-th",
    doi = "10.33540/2451",
    school = "Universiteit Utrecht, Utrecht U.",
    year = "2024"
}

@article{Corvilain:2018lgw,
    author = "Corvilain, Pierre and Grimm, Thomas W. and Valenzuela, Irene",
    title = {{The Swampland Distance Conjecture for K\"ahler moduli}},
    eprint = "1812.07548",
    archivePrefix = "arXiv",
    primaryClass = "hep-th",
    doi = "10.1007/JHEP08(2019)075",
    journal = "JHEP",
    volume = "08",
    pages = "075",
    year = "2019"
}

@article{Grimm:2018cpv,
    author = "Grimm, Thomas W. and Li, Chongchuo and Palti, Eran",
    title = "{Infinite Distance Networks in Field Space and Charge Orbits}",
    eprint = "1811.02571",
    archivePrefix = "arXiv",
    primaryClass = "hep-th",
    reportNumber = "MPP-2018-260",
    doi = "10.1007/JHEP03(2019)016",
    journal = "JHEP",
    volume = "03",
    pages = "016",
    year = "2019"
}

@article{Grimm:2023lrf,
    author = "Grimm, Thomas W. and Monnee, Jeroen",
    title = "{Finiteness theorems and counting conjectures for the flux landscape}",
    eprint = "2311.09295",
    archivePrefix = "arXiv",
    primaryClass = "hep-th",
    doi = "10.1007/JHEP08(2024)039",
    journal = "JHEP",
    volume = "08",
    pages = "039",
    year = "2024"
}

@article{Grimm:2019bey,
    author = "Grimm, Thomas W. and Ruehle, Fabian and van de Heisteeg, Damian",
    title = "{Classifying Calabi\textendash{}Yau Threefolds Using Infinite Distance Limits}",
    eprint = "1910.02963",
    archivePrefix = "arXiv",
    primaryClass = "hep-th",
    doi = "10.1007/s00220-021-03972-9",
    journal = "Commun. Math. Phys.",
    volume = "382",
    number = "1",
    pages = "239--275",
    year = "2021"
}

@article{Grimm:2019ixq,
    author = "Grimm, Thomas W. and Li, Chongchuo and Valenzuela, Irene",
    title = "{Asymptotic Flux Compactifications and the Swampland}",
    eprint = "1910.09549",
    archivePrefix = "arXiv",
    primaryClass = "hep-th",
    doi = "10.1007/JHEP06(2020)009",
    journal = "JHEP",
    volume = "06",
    pages = "009",
    year = "2020",
    note = "[Erratum: JHEP 01, 007 (2021)]"
}

@article{Bastian:2020egp,
    author = "Bastian, Brice and Grimm, Thomas W. and van de Heisteeg, Damian",
    title = "{Weak gravity bounds in asymptotic string compactifications}",
    eprint = "2011.08854",
    archivePrefix = "arXiv",
    primaryClass = "hep-th",
    doi = "10.1007/JHEP06(2021)162",
    journal = "JHEP",
    volume = "06",
    pages = "162",
    year = "2021"
}

@article{Grimm:2020cda,
    author = "Grimm, Thomas W.",
    title = "{Moduli space holography and the finiteness of flux vacua}",
    eprint = "2010.15838",
    archivePrefix = "arXiv",
    primaryClass = "hep-th",
    doi = "10.1007/JHEP10(2021)153",
    journal = "JHEP",
    volume = "10",
    pages = "153",
    year = "2021"
}

@article{Grimm:2021ikg,
    author = "Grimm, Thomas W. and Monnee, Jeroen and van de Heisteeg, Damian",
    title = "{Bulk reconstruction in moduli space holography}",
    eprint = "2103.12746",
    archivePrefix = "arXiv",
    primaryClass = "hep-th",
    doi = "10.1007/JHEP05(2022)010",
    journal = "JHEP",
    volume = "05",
    pages = "010",
    year = "2022"
}

@article{Bastian:2021eom,
    author = "Bastian, Brice and Grimm, Thomas W. and van de Heisteeg, Damian",
    title = "{Modeling General Asymptotic Calabi-Yau Periods}",
    eprint = "2105.02232",
    archivePrefix = "arXiv",
    primaryClass = "hep-th",
    month = "5",
    year = "2021"
}

@article{Bastian:2021hpc,
    author = "Bastian, Brice and Grimm, Thomas W. and van de Heisteeg, Damian",
    title = "{Engineering small flux superpotentials and mass hierarchies}",
    eprint = "2108.11962",
    archivePrefix = "arXiv",
    primaryClass = "hep-th",
    doi = "10.1007/JHEP02(2023)149",
    journal = "JHEP",
    volume = "02",
    pages = "149",
    year = "2023"
}

@article{Grimm:2021ckh,
    author = "Grimm, Thomas W. and Plauschinn, Erik and van de Heisteeg, Damian",
    title = "{Moduli stabilization in asymptotic flux compactifications}",
    eprint = "2110.05511",
    archivePrefix = "arXiv",
    primaryClass = "hep-th",
    doi = "10.1007/JHEP03(2022)117",
    journal = "JHEP",
    volume = "03",
    pages = "117",
    year = "2022"
}

@article{Bakker:2021uqw,
    author = "Bakker, Benjamin and Grimm, Thomas W. and Schnell, Christian and Tsimerman, Jacob",
    title = "{Finiteness for self-dual classes in integral variations of Hodge structure}",
    eprint = "2112.06995",
    archivePrefix = "arXiv",
    primaryClass = "math.AG",
    reportNumber = "MPIM-Bonn-2022",
    doi = "10.46298/epiga.2023.specialvolumeinhonourofclairevoisin.9626",
    month = "12",
    year = "2021"
}

@article{Grimm:2021vpn,
    author = "Grimm, Thomas W.",
    title = "{Taming the landscape of effective theories}",
    eprint = "2112.08383",
    archivePrefix = "arXiv",
    primaryClass = "hep-th",
    doi = "10.1007/JHEP11(2022)003",
    journal = "JHEP",
    volume = "11",
    pages = "003",
    year = "2022"
}

@article{Grimm:2024fip,
    author = "Grimm, Thomas W. and van de Heisteeg, Damian",
    title = "{Exact flux vacua, symmetries, and the structure of the landscape}",
    eprint = "2404.12422",
    archivePrefix = "arXiv",
    primaryClass = "hep-th",
    doi = "10.1007/JHEP01(2025)005",
    journal = "JHEP",
    volume = "01",
    pages = "005",
    year = "2025"
}

@phdthesis{vandeHeisteeg:2022gsp,
    author = "van de Heisteeg, Damian Theodorus Engelbert",
    title = "{Asymptotic String Compactifications: Periods, flux potentials, and the swampland}",
    eprint = "2207.00303",
    archivePrefix = "arXiv",
    primaryClass = "hep-th",
    doi = "10.33540/1380",
    school = "Utrecht U.",
    year = "2022"
}

@article{Cota:2017aal,
    author = "Cota, Cesar Fierro and Klemm, Albrecht and Schimannek, Thorsten",
    title = "{Modular Amplitudes and Flux-Superpotentials on elliptic Calabi-Yau fourfolds}",
    eprint = "1709.02820",
    archivePrefix = "arXiv",
    primaryClass = "hep-th",
    doi = "10.1007/JHEP01(2018)086",
    journal = "JHEP",
    volume = "01",
    pages = "086",
    year = "2018"
}

@article{Palti:2021ubp,
    author = "Palti, Eran",
    title = "{Stability of BPS states and weak coupling limits}",
    eprint = "2107.01539",
    archivePrefix = "arXiv",
    primaryClass = "hep-th",
    doi = "10.1007/JHEP08(2021)091",
    journal = "JHEP",
    volume = "08",
    pages = "091",
    year = "2021"
}

@article{Candelas:1990rm,
    author = "Candelas, Philip and De La Ossa, Xenia C. and Green, Paul S. and Parkes, Linda",
    editor = "Yau, Shing-Tung",
    title = "{A Pair of Calabi-Yau manifolds as an exactly soluble superconformal theory}",
    reportNumber = "UTTG-25-90",
    doi = "10.1016/0550-3213(91)90292-6",
    journal = "Nucl. Phys. B",
    volume = "359",
    pages = "21--74",
    year = "1991"
}

@article{Marchesano:2021gyv,
    author = "Marchesano, Fernando and Prieto, David and Wiesner, Max",
    title = "{F-theory flux vacua at large complex structure}",
    eprint = "2105.09326",
    archivePrefix = "arXiv",
    primaryClass = "hep-th",
    reportNumber = "IFT-UAM/CSIC-21-58",
    doi = "10.1007/JHEP08(2021)077",
    journal = "JHEP",
    volume = "08",
    pages = "077",
    year = "2021"
}

@article{Herraez:2018vae,
    author = "Herraez, Alvaro and Ibanez, Luis E. and Marchesano, Fernando and Zoccarato, Gianluca",
    title = "{The Type IIA Flux Potential, 4-forms and Freed-Witten anomalies}",
    eprint = "1802.05771",
    archivePrefix = "arXiv",
    primaryClass = "hep-th",
    reportNumber = "IFT-UAM/CSIC-18-016, IFT-UAM-CSIC-18-016",
    doi = "10.1007/JHEP09(2018)018",
    journal = "JHEP",
    volume = "09",
    pages = "018",
    year = "2018"
}

@article{Grana:2022dfw,
    author = "Gra\~na, Mariana and Grimm, Thomas W. and van de Heisteeg, Damian and Herraez, Alvaro and Plauschinn, Erik",
    title = "{The tadpole conjecture in asymptotic limits}",
    eprint = "2204.05331",
    archivePrefix = "arXiv",
    primaryClass = "hep-th",
    doi = "10.1007/JHEP08(2022)237",
    journal = "JHEP",
    volume = "08",
    pages = "237",
    year = "2022"
}

@article{Bastian:2023shf,
    author = "Bastian, Brice and van de Heisteeg, Damian and Schlechter, Lorenz",
    title = "{Beyond large complex structure: quantized periods and boundary data for one-modulus singularities}",
    eprint = "2306.01059",
    archivePrefix = "arXiv",
    primaryClass = "hep-th",
    doi = "10.1007/JHEP07(2024)151",
    journal = "JHEP",
    volume = "07",
    pages = "151",
    year = "2024"
}

@article{Lee:2019wij,
    author = "Lee, Seung-Joo and Lerche, Wolfgang and Weigand, Timo",
    title = "{Emergent strings from infinite distance limits}",
    eprint = "1910.01135",
    archivePrefix = "arXiv",
    primaryClass = "hep-th",
    reportNumber = "CERN-TH-2019-159",
    doi = "10.1007/JHEP02(2022)190",
    journal = "JHEP",
    volume = "02",
    pages = "190",
    year = "2022"
}

@book{Wiggins:2003,
    author = "Wiggins",
    title = "{Introduction to Applied Nonlinear Dynamical Systems and Chaos}",
    doi = "10.1017/9781108937092",
    isbn = "978-0-387-00177-7, 978-1-4419-1807-9",
    publisher = "Springer New York, NY",
    month = "10",
    year = "2003"
}

@article{Revello:2023hro,
    author = "Revello, Filippo",
    title = "{Attractive (s)axions: cosmological trackers at the boundary of moduli space}",
    eprint = "2311.12429",
    archivePrefix = "arXiv",
    primaryClass = "hep-th",
    doi = "10.1007/JHEP05(2024)037",
    journal = "JHEP",
    volume = "05",
    pages = "037",
    year = "2024"
}

@article{Debusschere:2024rmi,
    author = "Debusschere, C\'edric and Tonioni, Flavio and Van Riet, Thomas",
    title = "{A distance conjecture beyond moduli?}",
    eprint = "2407.03715",
    archivePrefix = "arXiv",
    primaryClass = "hep-th",
    month = "7",
    year = "2024"
}

@article{Grimm:2020ouv,
    author = "Grimm, Thomas W. and Li, Chongchuo",
    title = "{Universal axion backreaction in flux compactifications}",
    eprint = "2012.08272",
    archivePrefix = "arXiv",
    primaryClass = "hep-th",
    doi = "10.1007/JHEP06(2021)067",
    journal = "JHEP",
    volume = "06",
    pages = "067",
    year = "2021"
}

@article{Landete:2018kqf,
    author = "Landete, Aitor and Shiu, Gary",
    title = "{Mass Hierarchies and Dynamical Field Range}",
    eprint = "1806.01874",
    archivePrefix = "arXiv",
    primaryClass = "hep-th",
    reportNumber = "MAD-TH-18-03",
    doi = "10.1103/PhysRevD.98.066012",
    journal = "Phys. Rev. D",
    volume = "98",
    number = "6",
    pages = "066012",
    year = "2018"
}

@article{Buratti:2018xjt,
    author = "Buratti, Ginevra and Calder\'on, Jos\'e and Uranga, Angel M.",
    title = "{Transplanckian axion monodromy!?}",
    eprint = "1812.05016",
    archivePrefix = "arXiv",
    primaryClass = "hep-th",
    reportNumber = "IFT-UAM/CSIC-18-121",
    doi = "10.1007/JHEP05(2019)176",
    journal = "JHEP",
    volume = "05",
    pages = "176",
    year = "2019"
}

@article{Blumenhagen:2018nts,
    author = {Blumenhagen, Ralph and Kl\"awer, Daniel and Schlechter, Lorenz and Wolf, Florian},
    title = "{The Refined Swampland Distance Conjecture in Calabi-Yau Moduli Spaces}",
    eprint = "1803.04989",
    archivePrefix = "arXiv",
    primaryClass = "hep-th",
    reportNumber = "MPP-2018-34",
    doi = "10.1007/JHEP06(2018)052",
    journal = "JHEP",
    volume = "06",
    pages = "052",
    year = "2018"
}

@article{Baume:2016psm,
    author = "Baume, Florent and Palti, Eran",
    title = "{Backreacted Axion Field Ranges in String Theory}",
    eprint = "1602.06517",
    archivePrefix = "arXiv",
    primaryClass = "hep-th",
    doi = "10.1007/JHEP08(2016)043",
    journal = "JHEP",
    volume = "08",
    pages = "043",
    year = "2016"
}
\bibliographystyle{utphys}

\end{document}